\documentclass[%
 amsmath,amssymb,
 aps, 
 physrev,
 twocolumn
]{revtex4-2}
\renewcommand{\P}{\mathbb{P}}
\usepackage{graphicx}
\usepackage{bm}
\usepackage{amssymb,amsmath,hyperref,esint,multirow,subfig,subfloat,amsthm,thmtools,amsopn,float}

\declaretheoremstyle[
  spaceabove=\topsep,
  spacebelow=\topsep,
  headfont=\normalfont\bfseries,
  notefont=\normalfont\bfseries,
  bodyfont=\normalfont\itshape,
  headpunct=,
]{myplain}

\theoremstyle{myplain}
\newtheorem{algorithm}{Algorithm}

\usepackage{xcolor}

\begin{document}

\preprint{APS/123-QED}

\title{\textbf{Inference of Hierarchical Core--Periphery Structure in Temporal Networks}}

\author{Theodore Y. Faust}
\email{Contact author: tyfaust@umich.edu}
 \affiliation{Department of Mathematics, University of California, Los Angeles, Los Angeles, CA, USA}
 \affiliation{Institute for Social Research, University of Michigan, Ann Arbor, MI, USA}
\author{Mason A. Porter}
\affiliation{Department of Mathematics, University of California, Los Angeles, Los Angeles, CA, USA;}
\affiliation{Department of Sociology, University of California, Los Angeles, Los Angeles, CA, USA;}
\affiliation{Nordita, KTH Royal Institute of Technology, Stockholm University, and Uppsala University, Stockholm, Sweden;}
\affiliation{Department of Mathematics, Stockholm University, Stockholm, Sweden;}
\affiliation{Santa Fe Institute, Santa Fe, NM, USA}


\date{\today}
\begin{abstract}

Networks can have various types of mesoscale structures. One such structure is core--periphery structure, which consists of sets of densely-connected core nodes and sets of sparsely-connected peripheral nodes. Core nodes are connected densely to other core nodes and are potentially connected densely to peripheral nodes, which are connected sparsely to other nodes.
There has been much research on core--periphery structure in time-independent networks, but few core--periphery detection methods have been developed for time-dependent (i.e., ``temporal") networks.
Using a multilayer-network representation of temporal networks and an inference approach that employs stochastic block models, we generalize a recent method for detecting hierarchical core--periphery structure \cite{Polanco23} from time-independent networks to temporal networks. In contrast to ``onion-like'' nested core--periphery structures (where each node is assigned to a group according to how deeply it is nested in a network's core), hierarchical core--periphery structures encompass networks with nested structures, tree-like structures (where any two groups must either be disjoint or have one group as a strict subset of the other group), and general non-nested mesoscale structures (where the group assignments of nodes do not have to be nested in any way).
To perform statistical inference and thereby identify hierarchical core--periphery structure, we use a Markov-chain Monte Carlo (MCMC) approach. We illustrate our method for detecting hierarchical core--periphery structure in one synthetic temporal network and two real-world temporal networks, and we discuss the structures that we identify in these networks. 

\end{abstract}

\maketitle




\section{Introduction} \label{cpintro}

In network analysis, it is common to study various types of mesoscale structures \cite{Newman18}. The best-known type of mesoscale structure is community structure, in which densely-connected sets of nodes are connected sparsely to other sets of densely-connected nodes \cite{Fortunato16}. Another well-known type of mesoscale structure is role structure, in which one seeks to assign nodes to roles based on similarities between the local network structures in their neighborhoods~\cite{rossi2015}. A third prominent type of mesoscale structure is core--periphery structure, in which well-connected nodes (so-called ``core" nodes) are connected densely to other well-connected nodes and potentially densely connected to other nodes (so-called ``peripheral" nodes), which are connected sparsely to other nodes~\cite{csermely2013,Rombach17,yan2022}.

In core--periphery structure, well-connected nodes (so-called ``core" nodes) are connected densely to each other and potentially connected densely to other nodes (so-called ``peripheral" nodes), which are connected sparsely to other nodes~\cite{csermely2013,Rombach17,yan2022}.
In the past few decades, core--periphery structure has been studied in many time-independent networks and has yielded insights into topics such as
social networks \cite{Granovetter83,White76}, academic networks \cite{Doreian85, Willis90}, economic networks \cite{Roy83, Williams77}, transportation networks \cite{lee2014}, and many other areas. In many situations --- including in the analysis of the spread of diseases through face-to-face contacts \cite{Fournet14}, transportation systems \cite{Morer20}, and legislation cosponsorships \cite{lee2016, Neal20} --- it is important to consider relationships and/or interactions that change with time.  In such situations, one can study a time-dependent network (i.e., ``temporal network")~\cite{Holme12,Holme15,Holme19}. 
In a temporal network, the entities (i.e., nodes) of a network and/or the ties (i.e., edges) between those entities can change with time. 
Researchers have developed a variety of distinct approaches to identify community structure in temporal networks \cite{Holme15}, and many of the ideas in these investigations will surely also be insightful for studying core--periphery structure and other mesoscale structures. For example, as emphasized by Palla et al.~\cite{Palla07} in their clique-percolation approach, densely-connected groups of nodes in a network can merge, split, appear, and disappear over time. Other approaches to examine community structure in temporal networks include statistical inference using generative models from Markov chains~\cite{Peixoto17}, nonnegative-tensor factorization~\cite{Gauvin14}, and optimization of objective functions~\cite{Mucha10}.
See \cite{Rossetti18} for a review of community structure in temporal networks.

One way to examine a temporal network is as a sequence of ordinary time-independent networks (i.e., monolayer networks) in which each network in the sequence encodes the relationships between entities during one time point or time period. One can study such a temporal network using a multilayer-network representation~\cite{kivela2014,aleta2026}. 
Many approaches have been developed to study core--periphery structure in time-independent networks~\cite{Zhang15, Rombach17, Gallagher21, Polanco23}, but there are only a few studies of core--periphery structure in multilayer networks.
Bergermann et al.~\cite{Bergermann1024,Bergermann24} used spectral methods to identify core--periphery structure in multilayer networks, Nie et al.~\cite{Nie25} used a rich-club approach to identify core--periphery structure in multiplex networks, Hashemi and Behrouz~\cite{Hashemi24} generalized the concept of $k$-cores to multiplex networks, and Mosca and Mazzarisi~\cite{mosca2026} optimized a likelihood function to identify core--periphery structure in temporal networks.

In the present paper, we study a hierarchical notion of core--periphery structure in temporal networks (which we represent as multilayer networks).
We generalize the hierarchical core--periphery structure of Polanco and Newman \cite{Polanco23} from ordinary networks (i.e., graphs) to temporal networks. This notion of hierarchical core--periphery structure encompasses a rich variety of possible mesoscale structures, including ones that are not nested.
In a nested (i.e., onion-like) core--periphery structure \cite{Gallagher21}, each node of a network is part of exactly one group,
with higher-numbered groups signifying nodes that are deeper in a core.
In such a setting, the probability that there is
an edge between two nodes depends on the lower of the two group assignments of those
nodes. By contrast, in the hierarchical core--periphery structure in \cite{Polanco23}, each node of a network can be in
several groups simultaneously. One determines the probability of an edge between two nodes using
the highest-numbered group that includes both nodes.
This hierarchical formulation allows one to generate networks with nested structures, tree-like structures (where any two groups must either be disjoint or have one group be a strict subset of the other group), and general 
non-nested mesoscale structures (where the group assignments of nodes do not have to be nested in any way).\footnote{For illustrations of the types of structures that one can generate using this hierarchical notion of core--periphery structure, see Figure 1 of \cite{Polanco23}.}

To identify hierarchical core--periphery structure in temporal networks, we use statistical inference.
This choice makes our approach more computationally costly than spectral methods to detect core--periphery structure, but it also yields several important benefits \cite{peixoto2023}. 
For example, the use of statistical inference guarantees that posterior-sampling methods yield samples precisely from the specified posterior distribution in the limit of infinitely many nodes.
Indeed, we prove in this limit that the stable distribution of a slightly-modified version of the Markov-chain Monte Carlo (MCMC) approach that
we use for statistical inference is the same as the desired posterior distribution. Additionally, in contrast to many other approaches, such as non-inferential optimization approaches,
statistical-inference methods avoid identifying mesoscale structures in completely random networks, which (by construction) arise from models without any mesoscale structure~\cite{peixoto2023}.

After we discuss our approach and give reasons for various choices that we make, we use it to study a synthetic network with known hierarchical core--periphery structure and two real-world temporal networks. The latter two networks are a network of ties between terrorist organizations in the Indian states of Jammu and Kashmir \cite{Saxena04} and a network of co-appearances of individuals in the \emph{Gospel of Luke}~\cite{Holanda19}. For both real-world examples, we illustrate that the identified core--periphery structure is plausible.

Our paper proceeds as follows. In Section \ref{Notation}, we introduce our main notation. In Section \ref{GenerativeModel}, we present the details of the generative model that we use for statistical inference of hierarchical core--periphery structure. 
In Section \ref{StatInfApproach}, we present the MCMC approach that we use for statistical inference.
We also discuss variants of this approach and examine a variety of properties (including convergence guarantees and run-time complexity) of our main approach.
In Section \ref{GroupAssignmentDiscussion}, we discuss the consequences of the choice of generative model on the performance of our MCMC approach. 
In Section \ref{ComparisonSICP}, we use our statistical-inference approach to identify hierarchical core--periphery structure in one synthetic temporal network and two real-world temporal networks. In Section \ref{FinalConclusionsCP}, we conclude and discuss several future directions. Our code and the outputs of our computational experiments are available at \cite{faust-code}. 


\section{Notation}\label{Notation}

In this section, we introduce notation for both monolayer networks (i.e., ordinary graphs) and temporal networks, which we represent as multilayer networks in which each layer corresponds to one time step (which can represent a time point or a time period) \cite{kivela2014}.

We first discuss our notation for monolayer networks. For simplicity, we consider unweighted and undirected networks without self-edges or multi-edges. 
A monolayer network is a graph $G = (V,E)$, which consists of a set $V = \{1, \ldots, n\}$ of $n$ nodes (i.e., vertices) and a set $E \subseteq V \times V$ of edges. 
We denote an undirected edge by $(i,j)$. We represent a monolayer network $G$ using an adjacency matrix $A \in \{0,1\}^{n \times n}$, where $A_{ij} = 1$ if nodes $i$ and $j$ are connected directly by an edge (i.e., they are adjacent) and $A_{ij} = 0$ otherwise.

To represent a temporal network, we use a multilayer network in which each layer encodes the adjacencies between the nodes at its associated time step. We model a temporal network as a sequence of network layers (i.e., times)
$\ell \in \{1,\ldots,L\}$. Each time $\ell \in \{1,\ldots,L\}$ includes all nodes $i \in \{1,\ldots,n\}$. We refer to an instantiation of a node in a given layer as a node-layer $(i,\ell) \in \{1,\ldots,n\} \times \{1,\ldots,L\}$. We again use an adjacency representation, so we have a sequence $(A^{(1)},\ldots,A^{(L)})$, with $A^{(\ell)} \in \{0,1\}^{n \times n}$ for each layer $\ell$. We assume that our temporal networks are unweighted and undirected, so $A^{(\ell)}_{ij} = 1$ if node-layers $(i,\ell)$ and $(j,\ell)$ are adjacent and $A^{(\ell)}_{ij} = 0$ if they are not adjacent. For notational convenience, we let $A$ denote the sequence $(A^{(1)},\ldots,A^{(L)})$. Technically, this is an abuse of notation, as we already used $A$ to refer to a single adjacency matrix for a monolayer network, but we always clearly state whether we are considering a monolayer network or a temporal network. As general terminology, we refer to $A$ as an \emph{adjacency structure}. For convenience (and despite the additional associated abuse of notation), we also sometimes refer to $A$ as a ``network".


\section{A Hierarchical Generative Model for Core--Periphery Structure in Temporal Networks} \label{GenerativeModel}

As in \cite{Polanco23}, we think of core--periphery structure as a hierarchy with groups of nodes that may or may not be nested.
We allow each node-layer to be a member of each of $k$ groups, which we label with the indices $0,1,\ldots,k - 1$. 
We assume that the group labels correspond to the same groups across all layers.
The number $k$ of groups is the same in each layer, but it is possible for one or more groups to have no nodes in a layer. This allows groups to appear or disappear over time. In our hierarchical core--periphery setting, the probability of an edge between two nodes is determined by the highest-numbered group that includes both nodes \cite{Polanco23}.
By contrast, in a nested core--periphery setting \cite{Gallagher21}, researchers typically require each node of a network to be a member of exactly one group with indices in $0,1,\ldots,k - 1$. In such a nested core--periphery setting, the probability that there is an edge between two nodes depends on the lower of the two group assignments of the nodes.

To ensure that the hierarchical structure has a ``base level'', we assume (as in \cite{Polanco23}) that every node is in group $0$.
We define a set of indicator variables $g^r_{(i,\ell)}$, where $g^r_{(i,\ell)} = 1$ if node-layer $(i,\ell)$ is in group $r$ and $g^r_{(i,\ell)} = 0$ otherwise. 
The ``group-indicator array" $g$ is an array of indicator variables $g^r_{(i,\ell)}$ for all groups $r \in \{0,1,\ldots,k - 1\}$, nodes $i \in \{1,\ldots, n\}$, and layers $\ell \in \{1,\ldots, L\}$. For convenience, we often refer to $g$ as simply ``group assignments''.

Given group assignments $g$, we generate a temporal network $A$ using the model in \cite{Polanco23} independently for each layer. 
Although many real-world temporal networks have persistent edges \cite{Mingueza23}, as the existence of an edge between nodes $i$ and $j$ in layer $\ell + 1$ is positively correlated with the existence of an edge between nodes $i$ and $j$ in layer $\ell$, we generate edges independently in each layer (i.e., at each time) to ensure that our MCMC approach (see Section \ref{StatInfApproach}) is tractable.
We place an edge between each pair of node-layers, $(i,\ell)$ and $(j,\ell)$, in the same layer $\ell$ independently with probability $\omega_{h((i,\ell),(j,\ell))}^{(\ell)} \in [0,1]$, where $h((i,\ell),(j,\ell))$ is the highest common group that includes both node-layers (i.e., the largest $r$ such that $g^r_{(i,\ell)} = g^r_{(j,\ell)} = 1$). 
Applying the argument in Eqs.~(1)--(3) in Section II of \cite{Polanco23}, we then have
\begin{equation*}
	P(A|\omega,k,g) = \prod_{\ell = 1}^L \prod_{r = 0}^{k - 1}\left [ (\omega_r^{(\ell)})^{m_r^{(\ell)}} (1 - \omega_r^{(\ell)})^{t_r^{(\ell)} - m_r^{(\ell)}}\right] \,,
\end{equation*}	
where
\begin{equation*}
	t_r^{(\ell)} = \sum_{1 \le i < j \le n} \delta_{r,h((i,\ell),(j,\ell))}
\end{equation*}	
is the number of node-layer pairs $(i,\ell), (j,\ell)$ in layer $\ell$ with highest common group $r$ and 
\begin{equation*}
	m_r^{(\ell)} = \sum_{1 \le i < j \le n} A^{(\ell)}_{ij} \delta_{r,h((i,\ell),(j,\ell))}
\end{equation*}	
is the number of such pairs that are adjacent to each other.

To avoid dependence on the parameters $\omega_r^{(\ell)}$, we follow the approach in \cite{Polanco23}. For all groups $r$ and layers $\ell$, we assume that there is a uniform prior $\P(\omega_r^{(\ell)}) = 1$. 
The computations in Eq.~(4) of \cite{Polanco23} show that marginalizing according to these prior distributions yields
\begin{equation}\label{ProbAdjacencyMarginal}
	\P(A|k,g) = \prod_{\ell = 1}^L \prod_{r = 0}^{k - 1} \frac{m_r^{(\ell)}! (t_r^{(\ell)} - m_r^{(\ell)})!}{(t_r^{(\ell)} + 1)!} \,.
\end{equation}	

We break the discussion of our model into two parts. In Section \ref{FixedK}, we consider a variant of our model in which we fix the number $k$ of groups.
In Section \ref{VariableK}, we remove this assumption and introduce our main model, in which we do not specify the number $k$ of groups.


\subsection{Fixed Number of Groups}\label{FixedK}

Suppose that we fix the number $k$ of groups. To perform Bayesian inference on a temporal network $A$ to yield a posterior distribution $\P(g|A)$, we need a prior distribution $\P(g|k)$ of the group assignments of nodes. There are many possible choices, such as a uniform distribution over group assignments (i.e., $\P(g|k) \propto 1$), for such a prior distribution. However, Faust et al.~\cite{FaustThesis,faust2026} showed in the context of community structure that many common prior distributions, such as a uniform distribution and generating $g$ via a discrete-time Markov process, make it prohibitively unlikely to obtain large or small groups, which typically is an undesirable situation. This, in turn, impacts the accuracy of statistical inference of small and large groups.

To mitigate the group-size bias, we determine group assignments using the approach in \cite{FaustThesis,faust2026}. 
This approach yields group assignments $g$ that are correlated with time. When node-layer $(i,\ell)$ is in group $r$, node-layer $(i,\ell + 1)$ has an augmented probability to also be in group $r$.
We start by uniformly randomly selecting the group sizes in the first layer of a network.
Given these group sizes, we uniformly randomly choose the group assignments of nodes from all group assignments with the chosen group sizes.
In other words, we choose the group assignments $g_{(1)}$ of the nodes in the first layer according to the probability distribution
\begin{equation}\label{gsubone}
	P(g_{(1)} | k) = \prod_{r = 1}^{k - 1} \frac{(n_{1}(g^r_{(1)}))! \times (n - n_{1}(g^r_{(1)}))!}{(n + 1)!} \,,
\end{equation}
where $n_{1}(g^r_{(1)})$ is the number of nodes in layer $1$ that have group-$r$ indicator variables of $1$.

For each layer beyond the first one, we use the group assignments from the previous layer to generate group assignments via the probability distribution
\begin{widetext}
\begin{align} \label{prob}
	\P(g^r_{(\ell)} | g^r_{(\ell - 1)}) = \prod_{s = 0}^1 \left[ \frac{1}{\begin{pmatrix} n_{s}(g^r_{(\ell - 1)}) \\[0.6em] c_{r;ss}^{(\ell)} \end{pmatrix}} \times \int_0^1 {p_{r;s,\ell}}^{n_{s}(g^r_{(\ell - 1)}) - c_{r;ss}^{(\ell)}}\frac{p_{r;s,\ell} - 1}{{p_{r;s,\ell}}^{n_{s}(g^r_{(\ell - 1)}) + 1} - 1}  \; dp_{r;s,\ell} \right]\, ,
\end{align}
\end{widetext}
where $g^r_{(\ell)}$ is the set of group-$r$ indicator variables for all nodes in layer $\ell$, the quantity $n_{s}(g^r_{(\ell - 1)})$ is the number of nodes in layer $\ell - 1$ with a group-$r$ indicator variable of $s$, the quantity $c_{r;ss}^{(\ell)}$ is the number of nodes $i$ with $g^r_{(i,\ell - 1)} = s$ and $g^r_{(i,\ell)} = s$, and each $p_{r;s,\ell}$ is the parameter of an independent geometric distribution \footnote{For a complete discussion of the approach that we use to generate $g^r_{(\ell)}$ from $g^r_{(\ell - 1)}$, see \cite[Section 4.2.3.3 and Appendix B.2]{FaustThesis} and \cite[Section 3.3.3 and Appendices A.2 and A.3]{faust2026}.}. 
For notational convenience, we define
\begin{equation*}
	J(k_1,k_2) = \int_0^1 x^{k_1} \frac{x - 1}{x^{k_2 + 1} - 1} \; dx \, .
\end{equation*}
Using this notation, we write Eq.~\eqref{prob} as
\begin{widetext}
\begin{equation} \label{probJ}
	\P(g^r_{(\ell)} | g^r_{(\ell - 1)}) = \prod_{s = 0}^1 \left[ \frac{1}{\begin{pmatrix} n_{s}(g^r_{(\ell - 1)}) \\[0.6em] c_{r;ss}^{(\ell)} \end{pmatrix}} \times J(n_{s}(g^r_{(\ell - 1)}) - c_{r;ss}^{(\ell)}, n_{s}(g^r_{(\ell - 1)}))\right]\, .
\end{equation}
\end{widetext}
Finally, we set
\begin{equation}\label{ggivenkprod}
	\P(g|k) = \P(g_{(1)}) \prod_{\ell = 2}^L \prod_{r = 1}^{k - 1} \P(g^r_{(\ell)} | g^r_{(\ell - 1)}) \,,
\end{equation}
and we thereby obtain the prior distribution $\P(g|k)$. 
Using the notation
\begin{equation}\label{Fdef}
	F(g|k) = \prod_{\ell = 2}^L \prod_{r = 1}^{k - 1} \P(g^r_{(\ell)} | g^r_{(\ell - 1)}) \,,
\end{equation}	
we write Eq.~\eqref{ggivenkprod} as
\begin{equation}\label{Fdefinition}
	\P(g|k) = \P(g_{(1)} | k) F(g|k) \,.
\end{equation}
By Bayes' rule,
\begin{equation}\label{bayesposterior}
	\P(g|A,k) = \frac{\P(A|g,k)\P(g|k)}{\P(A|k)} \,.
\end{equation}
Because we have expressions for $\P(A|g,k)$ and $\P(g|k)$ (and because $\P(A|k) \propto 1$), we can use Eq.~\eqref{bayesposterior} to sample from the posterior distribution $\P(g|A,k)$.


\subsection{Main Model: Unspecified Number of Groups} \label{VariableK}

In our main model for hierarchical core--periphery structure, we suppose that the number $k$ of groups is unspecified.
As in \cite{Polanco23}, we use a Poisson prior distribution on $k$ with mean $1$. Namely,
\begin{equation} \label{above-prior}
	\P(k) = \frac{e^{-1}}{(k - 1)!} \,.
\end{equation}	
We then have
\begin{equation}\label{bayesposteriorkvariable}
	\P(g,k|A) = \P(k)\P(g|A,k) \,.
\end{equation}	
In Section \ref{StatInfApproach}, we derive our main MCMC algorithm to sample from the posterior distribution \eqref{bayesposteriorkvariable} and we prove that the stationary distribution of a slightly-modified version of this algorithm is the same as Eq.~\eqref{bayesposteriorkvariable}. In Section \ref{MultiNodeMoves}, we discuss the required modifications of our main MCMC algorithm.


\section{Our Markov-Chain Monte Carlo (MCMC) Approach for Statistical Inference} \label{StatInfApproach}

To sample from the posterior distribution \eqref{bayesposteriorkvariable}, we use an MCMC method that is similar to the ones in \cite{Polanco23}. 
In Sections \ref{MNMoves}, \ref{AcceptanceProb}, and \ref{MainMethod}, we introduce our MCMC method (see Algorithm \ref{FinalAlg}). To prove that the stable distribution of a slightly-modified version of Algorithm \ref{FinalAlg} is the same as the desired posterior distribution \eqref{bayesposteriorkvariable}, we introduce versions of Algorithm \ref{FinalAlg} that do not include certain types of moves and prove intermediate results involving their stable distributions. 
In Section \ref{MCMCKFixed}, we consider Algorithm \ref{KFixedAlg}, which is an MCMC algorithm for a fixed number of groups.
In Section \ref{MCMCKVariable}, we consider Algorithm \ref{KVariableAlg}, which is similar to Algorithm \ref{FinalAlg} but (1) allows the number of groups to vary and (2) does not permit one class of moves that we consider
in Algorithm \ref{FinalAlg}.
In Section \ref{MNMoves2}, we prove that the stable distribution of a slightly-modified version of Algorithm \ref{FinalAlg} is the same as the desired posterior distribution \eqref{bayesposteriorkvariable}. 
In Section \ref{AcceptanceProbCalc}, we explain how we compute the quantities that we need to execute Algorithm \ref{FinalAlg}.
In Section \ref{FinalAlgComplexity}, we discuss the run-time complexity of Algorithm \ref{FinalAlg}.


\subsection{Our Main MCMC Algorithm} \label{MNMoves}

We begin by presenting the three types of MCMC moves in Algorithm \ref{FinalAlg}. 


\subsubsection{Standard Moves}\label{StandardMoves}

The first type of MCMC move, which we call a \emph{standard move} (see Algorithm \ref{StandardProposal}), adds a node-layer to a group, removes a node-layer from a group, or removes an empty group (i.e., a group that has no nodes).
In this type of move, we first uniformly randomly select a group $r$ and a layer $\ell$.
The specific move that we propose is different for the first layer $\ell = 1$ and the other layers $\ell \ge 2$.

First consider layer $\ell = 1$. With probability $1/2$, we propose a move that adds a node-layer in layer $1$ to group $r$. With probability $1/2$, we propose a move that removes a node-layer in layer $1$ from group $r$.
If we choose to add a node-layer to a group, we select a node-layer $(i,1)$ in layer $1$ that is not in group $r$ (i.e., $g^r_{(i,1)} = 0$) uniformly at random from the set of all node-layers $(i,1)$ in layer $1$ that are not in group $r$. If all node-layers in layer $1$ are already in group $r$, then we do nothing. Analogously, if we choose to remove a node-layer from a group, we select a node-layer $(i,1)$ in layer $1$ that is currently in group $r$ (i.e., $g^r_{(i,1)} = 1$) uniformly at random from the set of all node-layers $(i,1)$ in layer $1$ that are in group $r$. If no node-layers are in group $r$, we reduce the number $k$ of groups by $1$ by removing group $r$ and shifting the labels of all groups above $r$ down by $1$. If group $r$ has no node-layers in layer $1$ but at least one node-layer in another layer, then we do nothing.

If layer $\ell \ge 2$, we select a node $i$ uniformly at random. If node-layer $(i,\ell)$ is in group $r$, then we propose a move that removes it from group $r$. If node-layer $(i,\ell)$ is not in group $r$, then we propose a move that adds it to group $r$. For $\ell \ge 2$, a proposed move cannot remove a group.

\medskip
\begin{algorithm}[Proposal of a standard move]
\label{StandardProposal} 
\ \\
\vspace{-5 mm}
\begin{enumerate}
\item Select a layer $\ell$ uniformly at random from $1, \ldots, L$.
\item Select a group $r$ uniformly at random from $1, \ldots, k - 1$.
\item If $\ell = 1$:
\begin{enumerate}
\item With probability $1/2$, we choose to 
add a node-layer in layer $1$ to group $r$. With probability $1/2$, we choose to 
remove a node-layer in layer $1$ from group $r$.
\item If we choose to add a node-layer in layer $1$ to group $r$:
\begin{enumerate}
\item If all node-layers $(i,1)$ in layer $1$ are in group $r$:
\begin{enumerate}
\item Do nothing.
\end{enumerate}
\item Otherwise (i.e., if there is at least one node-layer in layer $1$ that is not in group $r$):
\begin{enumerate}
\item Select a node-layer $(i,1)$ that is not in group $r$ uniformly at random from all node-layers $(i,1)$ that are not in group $r$.
\item Propose a move that adds node-layer $(i,1)$ to group $r$.
\end{enumerate}
\end{enumerate}
\item If we choose to remove a node-layer in layer $1$ from group $r$:
\begin{enumerate}
\item If no node-layers $(i,\ell)$ are in
 group $r$:
\begin{enumerate}
\item Propose a move that removes group $r$ and shifts the labels of all groups above $r$ down by $1$.
\end{enumerate}
\item Otherwise, if no node-layers $(i,1)$ in layer $1$ are in group $r$ but group $r$ has at least one node-layer from another layer:
\begin{enumerate}
\item Do nothing.
\end{enumerate}
\item Otherwise (i.e., if group $r$ has at least one node-layer from layer $1$):
\begin{enumerate}
\item Select a node-layer $(i,1)$ from group $r$ uniformly at random from all node-layers $(i,1)$ in group $r$.
\item Propose a move that removes node-layer $(i,1)$ from group $r$.
\end{enumerate}
\end{enumerate}
\end{enumerate}
\item Otherwise (i.e., if $\ell \ge 2$):
\begin{enumerate}
\item Select a node $i$ uniformly at random from $1, \ldots, n$.
\begin{enumerate}
\item If node-layer $(i,\ell)$ is in group $r$:
\begin{enumerate}
\item Propose a move that removes node-layer $(i,\ell)$ from group $r$.
\end{enumerate}
\item If node-layer $(i,\ell)$ is not in group $r$:
\begin{enumerate}
\item Propose a move that adds node-layer $(i,\ell)$ to group $r$.
\end{enumerate}
\end{enumerate}
\end{enumerate}
\end{enumerate}
\end{algorithm}


\subsubsection{Group-Addition Moves} \label{GroupAdditionMoves}

Our second type of MCMC move is a \emph{group-addition move} (see Algorithm \ref{GroupAdditionProposal}). This type of move increases the number $k$ of groups by $1$. 

\medskip
\begin{algorithm}[Proposal of a group-addition move]
 \label{GroupAdditionProposal} 
 \ \\
 \vspace{-5 mm}
\begin{enumerate}
\item Select a group $r$ uniformly at random from $1, \ldots, k$.
\item Propose a move that increases the labels of all groups $r$ and higher by $1$, creates a new empty group with the label $r$, and increases the value of $k$ by $1$.
\end{enumerate}
\end{algorithm}


\subsubsection{Multi-Node Moves} \label{MultiNodeMoves}

Our third type of MCMC move is a \emph{multi-node move} (see Algorithm \ref{MultiNodeProposal}). This type of move changes more than one group-assignment indicator variable $g_{(i,\ell)}^r$ at a time. For a layer $\ell$ and two subsets $\mathcal{G}_1$ and 
$\mathcal{G}_2$ of the groups $\{1,\ldots, k - 1\}$, we propose new group assignments $g'$ that satisfies
\begin{equation}\label{gprimedefinition}
	(g')^r_{(i,\ell')} = \begin{cases} 
				\delta_{\mathcal{G}_2,r}\,, & \ell' = \ell \, \text{ and } \, \delta_{\mathcal{G}_1,(i,\ell')} = 1 \\ 
				\delta_{\mathcal{G}_1,r}\,, & \ell' = \ell \, \text{ and } \, \delta_{\mathcal{G}_2,(i,\ell')} = 1 \\ 
				g^r_{(i,\ell')}\,, & \text{otherwise} \, , \end{cases}
\end{equation}	
where
\begin{equation*} 
	\delta_{\mathcal{G},r} = \begin{cases} 1\,, & r \in \mathcal{G} \\ 
	0\,, & \text{otherwise} \end{cases}
\end{equation*}	
and
\begin{equation*} 
	\delta_{\mathcal{G},(i,\ell)} =  \begin{cases} 1\,, & g_{(i,\ell)}^{r} = \delta_{\mathcal{G},r} \text{ for all } r \in \{1,\ldots,k - 1\} \\ 
							0\,, & \text{otherwise} \, . \end{cases}
\end{equation*}	

\medskip
\begin{algorithm}[Proposal of a multi-node move]
\label{MultiNodeProposal} 
\ \\
\vspace{-5 mm}
\begin{enumerate}
\item Select subsets $\mathcal{G}_1$ and $\mathcal{G}_2$ uniformly at random from the groups
$\{1,\ldots,k - 1\}$.
\item Select a layer $\ell$ uniformly at random from $1,\ldots,L$.
\item Propose a move with new group assignments $g'$ using Eq.~\eqref{gprimedefinition}.
\end{enumerate}
\end{algorithm}

As we discussed in Section \ref{VariableK}, we need to slightly modify our main MCMC algorithm (see Algorithm \ref{FinalAlg}) so that the stationary distribution precisely equals the desired posterior distribution \eqref{bayesposteriorkvariable}. Specifically, we need to modify Algorithm \ref{MultiNodeProposal} by choosing the layer $\ell$ uniformly at random from $2,\ldots,L$ instead of from $1,\ldots,L$.
 In Section \ref{MNMoves2}, we prove that the stationary distribution of the modified version of Algorithm \ref{FinalAlg} matches the desired posterior distribution \eqref{bayesposteriorkvariable} and discuss why we nevertheless use the unmodified version of Algorithm \ref{FinalAlg} to infer core--periphery structure.


\subsection{Acceptance Probability} \label{AcceptanceProb}

In Sections \ref{StandardMoves}, \ref{GroupAdditionMoves}, and \ref{MultiNodeMoves}, we discussed three types of MCMC moves. In each step of our MCMC algorithm (see Algorithm \ref{FinalAlg}), we propose a move of one of these three types.
After proposing a move from $g,k$ to $g',k'$, we need to determine whether to accept or reject the move.
Due to the design of our MCMC algorithm, using the standard Metropolis--Hastings acceptance probability \cite{Robert04}
\begin{equation}\label{MH}
	\min\left\{1,\frac{\P(A|g',k')\P(g'|k')}{\P(A|g,k)\P(g|k)}\right\}
\end{equation}
would cause the stationary distribution of our Markov chain to differ from the desired posterior distribution \eqref{bayesposteriorkvariable}. Therefore, we slightly modify \eqref{MH} and instead use the acceptance probability
\begin{equation}\label{acceptanceprob}
	\alpha(g,k \to g',k') = \min\left\{1,\frac{\P(A|g',k')F(g'|k')}{\P(A|g,k)F(g|k)}\right\} \,,
\end{equation}
where we recall that $F(g|k)$ is given by Eq.~\eqref{Fdef}.


\subsection{Statement of our Main MCMC Algorithm} \label{MainMethod}

Now that we have discussed our three types of MCMC moves (see Sections \ref{StandardMoves}, \ref{GroupAdditionMoves}, and \ref{MultiNodeMoves}) and the acceptance probability (see Section \ref{AcceptanceProb}) of a move, we now state Algorithm \ref{FinalAlg}, which gives the MCMC approach that we use to identify hierarchical core--periphery structure in temporal networks. 

Let $p \in [0,1]$ denote the probability that a move is a multi-node move. We use Algorithm \ref{FinalAlg} to sample from the posterior distribution \eqref{bayesposteriorkvariable} and identify hierarchical core--periphery structure. Our implementation of this algorithm is available at \cite{faust-code}.

\medskip
\begin{algorithm}[Main MCMC algorithm]
 \label{FinalAlg}
\ \\
 \vspace{-5 mm}
\begin{enumerate}
\item Propose a move: 
\begin{enumerate}
\item With probability $p$, propose a multi-node move (see Algorithm \ref{MultiNodeProposal}).
\item Otherwise, with probability $1 - \frac{1}{2k(n + 1)}$, propose a standard move (see Algorithm \ref{StandardProposal}).
\item Otherwise, propose a group-addition move (see Algorithm \ref{GroupAdditionProposal}).
\end{enumerate}
\item Accept the proposed move from $g, k$ to $g', k'$ with probability
\begin{equation*}
	\alpha(g,k \to g',k') := \min\left\{1,\frac{\P(A|g',k')F(g'|k')}{\P(A|g,k)F(g|k)}\right\} \,.
\end{equation*}
\item Otherwise, reject the proposed move.
\end{enumerate}
\end{algorithm}


\subsection{MCMC Algorithm for a Fixed Number of Groups} \label{MCMCKFixed}

As we discussed in the introduction of Section \ref{StatInfApproach}, we prove our main theoretical result, which is that the stationary distribution of a slightly-modified version of Algorithm \ref{FinalAlg} is the same as the desired posterior distribution \eqref{bayesposteriorkvariable}. 
We prove this result by proving similar results for intermediate algorithms that do not include certain types of moves. 
We use arguments in the proofs of these intermediate results in the proof of our main theoretical result.
In this subsection, we consider a fixed number $k$ of groups. In this case, we use only standard moves (see Section \ref{StandardMoves}) in our MCMC algorithm.

Because we fix the number $k$ of groups, we slightly modify our algorithm for proposing a standard move (see Algorithm \ref{StandardProposal}). In this modified version of the algorithm, if we choose to remove a node-layer in layer $1$ and no node-layers are in group $r$, we do nothing (instead of removing group $r$, as we did before). In Algorithm \ref{StandardProposalKFixed}, we give our modified algorithm for proposing a standard move.

\medskip
\begin{algorithm}[Proposal of a standard move when we fix the number of groups]
\label{StandardProposalKFixed}
\ \\ 
\vspace{-5 mm}
\begin{enumerate}
\item Select a layer $\ell$ uniformly at random from $1, \ldots, L$.
\item Select a group $r$ uniformly at random from $1, \ldots, k - 1$.
\item If $\ell = 1$:
\begin{enumerate}
\item With probability $1/2$, we choose to add a node-layer in layer $1$ to group $r$. With probability $1/2$, we choose to remove a node-layer in layer $1$ from group $r$.
\item If we choose to add a node-layer in layer $1$ to group $r$:
\begin{enumerate}
\item If all node-layers $(i,1)$ in layer $1$ are in group $r$:
\begin{enumerate}
\item Do nothing.
\end{enumerate}
\item Otherwise (i.e., if there is at least one node-layer in layer $1$ that is not in group $r$):
\begin{enumerate}
\item Select a node-layer $(i,1)$ that is not in group $r$ uniformly at random from all node-layers $(i,1)$ that are not in group $r$.
\item Propose a move that adds node-layer $(i,1)$ to group $r$.
\end{enumerate}
\end{enumerate}
\item If we choose to remove a node-layer in layer $1$ from group $r$:
\begin{enumerate}
\item If no node-layers $(i,1)$ in layer $1$ are in group $r$:
\begin{enumerate}
\item Do nothing.
\end{enumerate}
\item Otherwise (i.e., if group $r$ has at least one node-layer from layer $1$):
\begin{enumerate}
\item Select a node-layer $(i,1)$ that is in group $r$ uniformly at random from all node-layers $(i,1)$ that are in group $r$.
\item Propose a move that removes node-layer $(i,1)$ from group $r$.
\end{enumerate}
\end{enumerate}
\end{enumerate}
\item Otherwise (i.e., if $\ell \ge 2$):
\begin{enumerate}
\item Select a node $i \in \{1,\ldots,n\}$.
\begin{enumerate}
\item If node-layer $(i,\ell)$ is in group $r$:
\begin{enumerate}
\item Propose a move that removes node-layer $(i,\ell)$ from group $r$.
\end{enumerate}
\item If node-layer $(i,\ell)$ is not in group $r$:
\begin{enumerate}
\item Propose a move that adds node-layer $(i,\ell)$ to group $r$.
\end{enumerate}
\end{enumerate}
\end{enumerate}
\end{enumerate}
\end{algorithm}

Because we fix the number $k$ of groups, we slightly modify the acceptance probability \eqref{acceptanceprob}.
The acceptance probability of a move from group assignments $g$ to group assignments $g'$ for this algorithm is
\begin{equation}\label{acceptanceprobkfixed}
	\alpha(g \to g') = \min\left\{1, \frac{\P(A|g',k)F(g'|k)}{\P(A|g,k)F(g|k)}\right\} \,.
\end{equation}
In Algorithm \ref{KFixedAlg}, we give this algorithm for sampling from the posterior distribution \eqref{bayesposterior}.

\medskip
\begin{algorithm}[MCMC algorithm when there is a fixed number of groups]
\label{KFixedAlg}
\ \\
\vspace{-5 mm}
\begin{enumerate}
\item Propose a move: 
\begin{enumerate}
\item Propose a standard move (see Algorithm \ref{StandardProposalKFixed}).
\end{enumerate}
\item Accept the proposed move from group assignments $g$ to group assignments $g'$ with probability
\begin{equation*}
	\alpha(g \to g') := \min\left\{1,\frac{\P(A|g',k)F(g'|k)}{\P(A|g,k)F(g|k)}\right\} \,.
\end{equation*}
\item Otherwise, reject the proposed move.
\end{enumerate}
\end{algorithm}

We now show that the equilibrium distribution of this MCMC process is the same as the desired posterior distribution $\P(g|A,k)$ in Eq.~\eqref{bayesposterior}. To verify this, it is sufficient to show that the MCMC process satisfies ergodicity and detailed balance \cite{Robert04}. To prove ergodicity, we need to show that one can access every state in the system from every other state using a finite sequence of moves. To prove detailed balance, we need to show that the mean rate of $g \to g'$ moves equals the mean rate of $g' \to g$ moves at equilibrium. 
That is, we need to verify that
\begin{equation} \label{eq1}
	\P(g|A,k) \P(g\to g') = \P(g'|A,k)\P(g' \to g)\,.
\end{equation}	
In Algorithm \ref{KFixedAlg}, ergodicity clearly holds because we can first remove all node-layers from all groups and then re-add 
node-layers to attain any desired group assignments $g$. 
We thus only need to prove that the MCMC process satisfies detailed balance. 

To prove detailed balance, we need to verify Eq.~\eqref{eq1}.
We write $\P(g \to g') = \pi(g \to g')\alpha(g \to g')$, where $\pi$ is the probability of proposing a move and $\alpha$ is the probability of accepting it.
This implies that \eqref{eq1} is equivalent to
\begin{equation}\label{eq2}
	\frac{\P(g'|A,k)}{\P(g|A,k)} = \frac{\pi(g \to g') \alpha(g \to g')}{\pi(g' \to g) \alpha(g' \to g)}\,.
\end{equation}
Because $\P(g|A,k) = \frac{\P(A|g,k)\P(g|k)}{\P(A|k)} =  \frac{\P(A|g,k)F(g|k) \P(g_{(1)}|k)}{\P(A|k)}$ by Eqs.~\eqref{bayesposterior} and \eqref{Fdefinition} and $\alpha(g \to g') = \min\left\{1, \frac{\P(A|g',k)F(g'|k)}{\P(A|g,k)F(g|k)}\right\}$ by Eq.~\eqref{acceptanceprob}, we obtain that Eq.~\eqref{eq2} is equivalent to
\begin{equation*}
	\frac{\P(A|g',k)F(g'|k) \P(g'_{(1)}|k)}{\P(A|g,k)F(g|k) \P(g_{(1)}|k)} = \frac{\pi(g \to g') \P(A|g',k)F(g'|k)}{\pi(g' \to g) \P(A|g,k)F(g|k)}\, ,
\end{equation*}
which in turn is equivalent to
\begin{equation}\label{detailedbalancefinal}
	\frac{\P(g'_{(1)}|k)}{\P(g_{(1)}|k)} = \frac{\pi(g \to g')}{\pi(g' \to g)}\, .
\end{equation} 

We now show that Eq.~\eqref{detailedbalancefinal} holds for a standard move $g \to g'$ that adds node-layer $(i,1)$ to group $r$. We omit the proofs of Eq.~\eqref{detailedbalancefinal} for the cases in which 
$g \to g'$ adds or removes a node-layer in layer $\ell \geq 1$, as they use extremely similar arguments.

First, because $\P(g_{(1)}|k) = \prod_{s = 1}^{k - 1} \frac{n_s^{(1)}!(n - n_s^{(1)})!}{(n + 1)!}$, we have
\begin{equation*}
	\frac{\P(g_{(1)}'|k)}{\P(g_{(1)}|k)} = \prod_{s = 1}^{k - 1} \frac{{n_s'}^{(1)}!(n - {n_s'}^{(1)})!}{n_s^{(1)}!(n - n_s^{(1)})!}\, .
\end{equation*}	
By adding a node-layer in layer $1$ to group $r$, we obtain ${n_r'}^{(1)} = n_r^{(1)} + 1$ and ${n_s'}^{(1)} = n_s^{(1)}$, which then yields
\begin{equation*}
	\frac{\P(g_{(1)}'|k)}{\P(g_{(1)}|k)} = \frac{n_r^{(1)} + 1}{n - n_r^{(1)}} \, .
\end{equation*}	

From Algorithms \ref{StandardProposalKFixed} and \ref{KFixedAlg}, the probability of proposing the move $g \to g'$ is
\begin{equation*}
	\pi(g \to g') = \frac{1}{L} \times \frac{1}{k - 1} \times \frac{1}{2} \times \frac{1}{n - n_r^{(1)}} \, .
\end{equation*}	
Similarly, $\pi(g' \to g) = \frac{1}{L} \times \frac{1}{k - 1}\times \frac{1}{2} \times \frac{1}{n_r^{(1)} + 1}$. Therefore,
\begin{equation*}
	\frac{\pi(g \to g')}{\pi(g' \to g)} =  \frac{n_r^{(1)} + 1}{n - n_r^{(1)}}= \frac{\P(g_{(1)}'|k)}{\P(g_{(1)}|k)} \, .
\end{equation*}	
The detailed-balance equation \eqref{detailedbalancefinal} is thus satisfied in this instance.

We now verify Eq.~\eqref{detailedbalancefinal} for a standard move $g \to g'$ that adds node-layer $(i,\ell)$ to group $r$ when $\ell \ge 2$. First, ${n_r'}^{(1)} = n_r^{(1)}$ for all $r$, as adding a node-layer in layer $\ell \ge 2$ does not affect the group sizes in layer $1$. This yields 
\begin{equation*}
	\frac{\P(g_{(1)}'|k)}{\P(g_{(1)}|k)} = 1 \,.
\end{equation*}	

From Algorithms \ref{StandardProposalKFixed} and \ref{KFixedAlg}, the probability of proposing the move $g \to g'$ is
\begin{equation*}
	\pi(g \to g') = \frac{1}{L} \times \frac{1}{k - 1} \times \frac{1}{n} \, .
\end{equation*}	
Similarly, $\pi(g' \to g) = \frac{1}{L} \times \frac{1}{k - 1} \times \frac{1}{n}$, so
\begin{equation*}
	\frac{\pi(g \to g')}{\pi(g' \to g)} = 1 = \frac{\P(g_{(1)}'|k)}{\P(g_{(1)}|k)} \, .
\end{equation*}	
Consequently, the detailed-balance equation \eqref{detailedbalancefinal} is satisfied.

Now that we have verified both ergodicity and detailed balance, we know that the equilibrium distribution of Algorithm \ref{KFixedAlg} is the same as the desired posterior distribution \eqref{bayesposterior}.


\subsection{Intermediate Algorithm for a Variable Number of Groups} \label{MCMCKVariable}

We now allow the number $k$ of groups to vary, and we consider an MCMC algorithm for sampling from the desired posterior distribution \eqref{bayesposteriorkvariable}. In contrast to our main algorithm (see Algorithm \ref{FinalAlg}), this algorithm does not use multi-node moves (see Section \ref{MultiNodeMoves}).
Because the number of groups can vary, we use our main algorithms for proposing a standard move (see Algorithm \ref{StandardProposal}) and for proposing a group-addition move (see Algorithm \ref{GroupAdditionProposal}). We also again use the acceptance probability
\begin{equation*}
	\alpha(g,k \to g',k') = \min\left\{1,\frac{\P(A|g',k')F(g'|k')}{\P(A|g,k)F(g|k)}\right\}
\end{equation*}
in Eq.~\eqref{acceptanceprob}. In Algorithm \ref{KVariableAlg}, we give our MCMC algorithm that samples from the desired posterior distribution \eqref{bayesposteriorkvariable} without using multi-node moves.

\medskip
\begin{algorithm}[MCMC algorithm with no multi-node moves when the number of groups can vary]
\label{KVariableAlg}
\ \\
\vspace{-5 mm}
\begin{enumerate}
\item Propose a move: 
\begin{enumerate}
\item With probability $1 - \frac{1}{2k(n + 1)}$, propose a standard move (see Algorithm \ref{StandardProposal}).
\item Otherwise, propose a group-addition move (see Algorithm \ref{GroupAdditionProposal}).
\end{enumerate}
\item Accept the proposed move from $g, k$ to $g', k'$ with probability 
\begin{equation*}
	\alpha(g,k \to g',k') := \min\left\{1,\frac{\P(A|g',k')F(g'|k')}{\P(A|g,k)F(g|k)}\right\} \,.
\end{equation*}
\item Otherwise, reject the proposed move.
\end{enumerate}
\end{algorithm}

Algorithm \ref{KVariableAlg} is the same as Algorithm \ref{FinalAlg} except that it does not allow multi-node moves.

We now show that the equilibrium distribution of this MCMC process is the same as the desired posterior distribution $\P(g,k|A)$ (see Eq.~\eqref{bayesposteriorkvariable}). As in the fixed-$k$ case, ergodicity clearly holds for this MCMC procedure. Therefore, it suffices to prove detailed balance. However, in contrast to the fixed-$k$ case, detailed balance holds only in the limit $n \to \infty$. 
Following similar logic to the logic for Eqs.~\eqref{eq1}--\eqref{detailedbalancefinal} for the fixed-$k$ case, to prove detailed balance for the variable-$k$ case, we need to show that
\begin{equation}\label{detailedbalancefinalKVariable}
	\frac{\P(g_{(1)}',k')}{\P(g_{(1)},k)}  = \frac{\pi(g,k \to g',k')}{\pi(g',k' \to g,k)}\, .
\end{equation}

First suppose that $k = k'$, which occurs when $g, k \to g', k'$ is a standard move that does not change the number of groups.
Algorithms \ref{StandardProposalKFixed} and \ref{StandardProposal} have the same probability of proposing a specific standard move, so the proposed-move probabilities $\pi(g,k \to g',k')$ and $\pi(g',k' \to g,k)$ in Algorithm \ref{KVariableAlg} are (aside from a factor of $1 - \frac{1}{2k(n + 1)}$) the same as the probabilities $\pi(g \to g')$ and $\pi(g \to g')$, respectively, in Algorithm \ref{KFixedAlg}.
Additionally, by the definition of conditional probability, $\P(g,k) \propto \P(g|k)$ and $\P(g',k') \propto \P(g'|k')$ with the same proportionality constant. The detailed-balance equation \eqref{detailedbalancefinal} for the fixed-$k$ case then implies that 
\begin{equation*}
	\frac{\P(g_{(1)}',k')}{\P(g_{(1)},k)}  = \frac{\pi(g,k \to g',k')}{\pi(g',k' \to g,k)}\, ,
\end{equation*}	
so the detailed-balance equation \eqref{detailedbalancefinalKVariable} holds for this case.

Now suppose that $k' \ne k$. We will show that detailed balance holds when $g, k \to g', k'$ is a group-addition move. 
Because $\P(g_{(1)}|k) = \prod_{r = 1}^{k - 1} \frac{n_r^{(1)}!(n - n_r^{(1)})!}{(n + 1)!}$ and $\P(k) = \frac{e^{-1}}{(k - 1)!}$, we obtain
\begin{equation*}
	\frac{\P(g_{(1)}',k')}{\P(g_{(1)},k)} = \frac{(k - 1)! \prod_{r = 1}^k \frac{{n_r'}^{(1)}!(n - {n_r'}^{(1)})!}{(n + 1)!}}{k!\prod_{r = 1}^{k - 1} \frac{n_r^{(1)}!(n - n_r^{(1)})!}{(n + 1)!}} = \frac{1}{k(n + 1)} \, .
\end{equation*}	
From Algorithms \ref{StandardProposal}, \ref{GroupAdditionProposal}, and \ref{KVariableAlg}, the proposed-move probabilities $\pi(g,k \to g',k')$ and $\pi(g',k' \to g,k) $ are
\begin{align*}
	\pi(g,k \to g',k') &= \frac{1}{L} \times \frac{1}{2k(n + 1)} \times \frac{1}{k} \, , \\
	\pi(g',k' \to g,k) &= \frac{1}{L} \times \left(1 - \frac{1}{2(k + 1)(n + 1)}\right) \times \frac{1}{k} \times \frac{1}{2} \, ,
\end{align*}	
where the expression for $\pi(g',k' \to g,k)$ follows from the fact that $g',k' \to g,k$ must be a standard move that removes a group.
Therefore,
\begin{equation*}
	\frac{\pi(g,k \to g',k')}{\pi(g',k' \to g,k)} = \frac{1}{k(n + 1)} + \Theta\left(\frac{1}{n^2}\right) \, ,
\end{equation*}	
where $\Theta(\cdot)$ denotes a tight asymptotic bound on the growth rate of its argument.
Because $\frac{\P(g',k')}{\P(g,k)} = \frac{1}{k(n + 1)}$, we have 
\begin{equation*}
	\frac{\P(g',k')}{\P(g,k)} = \frac{\pi(g,k \to g',k')}{\pi(g',k' \to g,k)}
\end{equation*}	
as $n \to \infty$. Therefore, the detailed-balance equation \eqref{detailedbalancefinalKVariable} holds in the limit $n \to \infty$. 

As with the fixed-$k$ case (see Section \ref{MCMCKFixed}), we omit the proof of detailed balance when $g,k \to g',k'$ is a standard move that removes a group, as it is extremely similar to the proof of detailed balance when $g,k \to g',k'$ is a group-addition move. 


\subsection{Revisiting our Main MCMC Algorithm \ref{FinalAlg}}\label{MNMoves2}

\begin{figure}[H]
\centering
\subfloat[]{\includegraphics[width=0.5\textwidth]{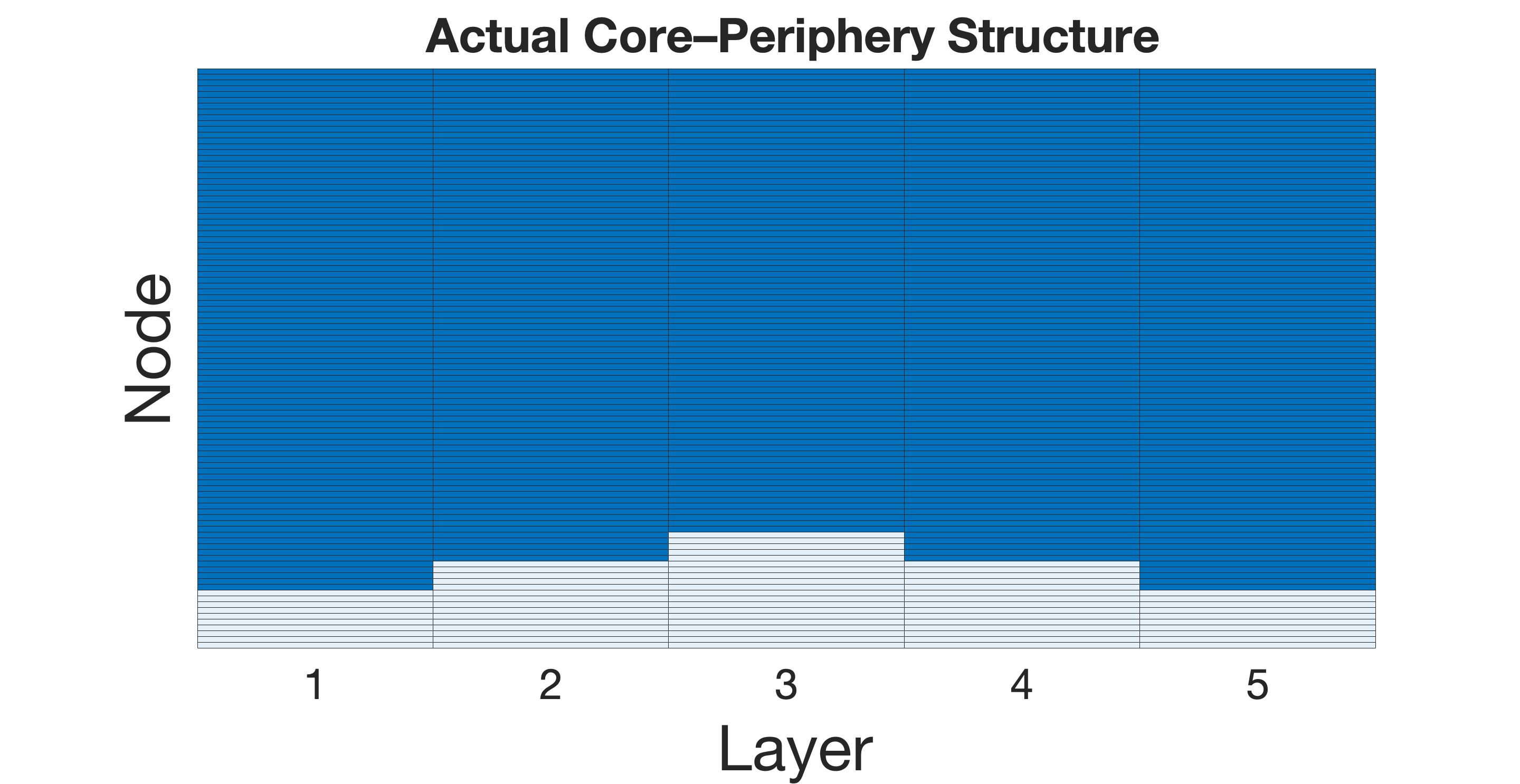}}\\
\subfloat[]{\includegraphics[width=0.5\textwidth]{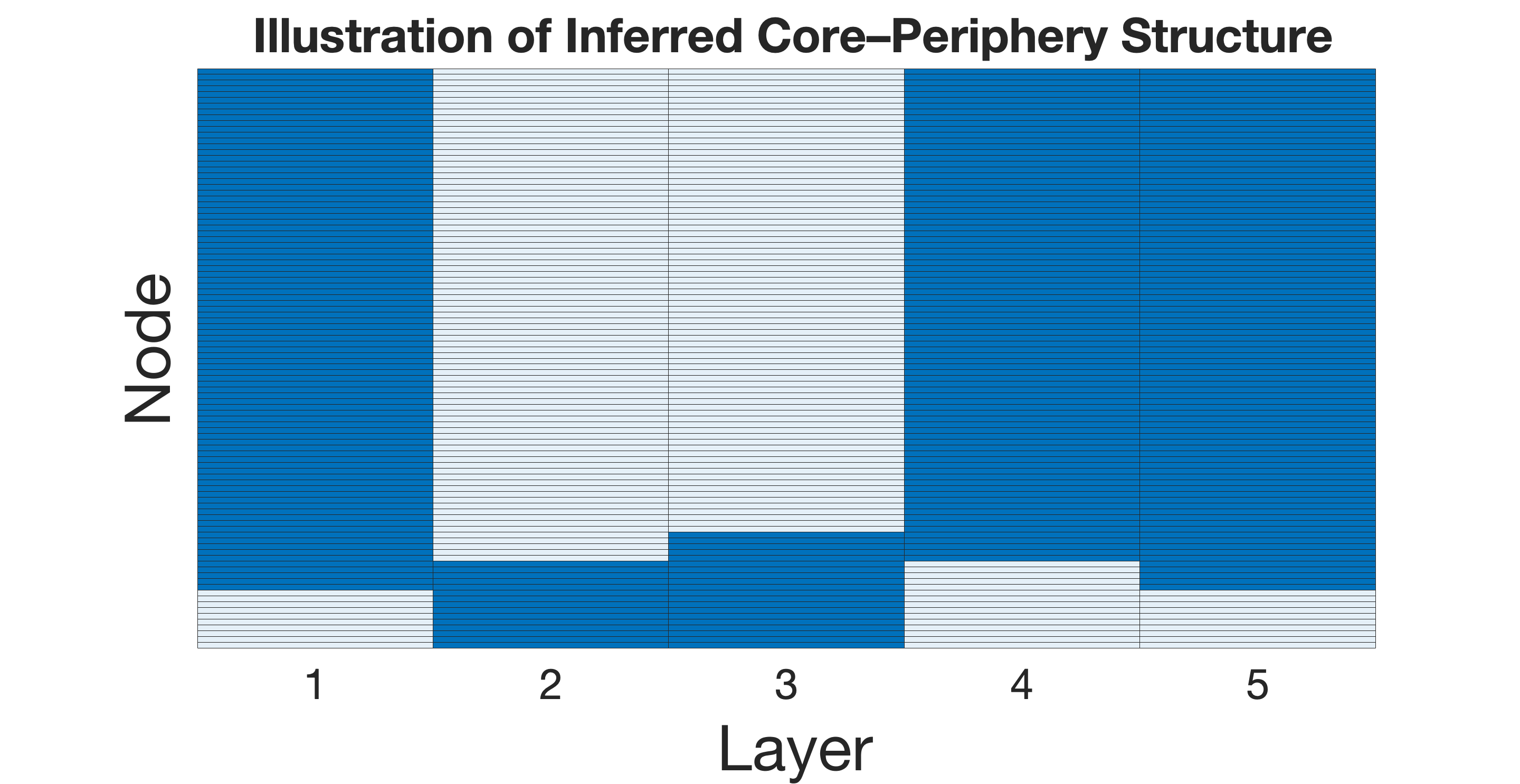}}
\caption{Heat maps of (a) an example of actual core--periphery structure and (b) an illustration of permuted group structure that we often observe at local maxima of the posterior distribution $\P(g|A)$ for a 100-node network with 5 layers. 
Each rectangle in a heat map corresponds to one node-layer $(i,\ell)$. The dark blue rectangles signify the value $g^1_{(i,\ell)} = 1$, and the light blue rectangles signify the value $g^1_{(i,\ell)} = 0$.}
\label{localextrema}
\end{figure}

In Section \ref{MCMCKVariable}, we proved that the stationary distribution of the MCMC approach in Algorithm \ref{KVariableAlg} is the same as the desired posterior distribution \eqref{bayesposteriorkvariable} in the limit $n \to \infty$. However, in practice, this approach often does not lead to the correct identification of core--periphery structure in temporal networks. 
The reason for this is that the posterior distribution $\P(g|A)$ from which we sample using this approach has many local maxima, so the algorithm often becomes stuck at suboptimal maxima.  
In Figure \ref{localextrema}, we illustrate the behavior that we commonly see at such suboptimal maxima.

In this example, we consider a network with 100 nodes and 5 layers. In some layers, we observe that the group structure of the local maximum is a permutation of the actual core--periphery structure. Specifically, for the nodes in layers $\ell = 2$ and $\ell = 3$, the group assignments $g$ match the group assignments of the actual core--periphery structure if we swap the values $0$ and $1$ of the group-assignment indicator variables $g^1_{(i,\ell)}$. If $g^1_{(i,\ell)} = 1$ initially, then $g^1_{(i,\ell)}$ becomes $0$ after the swap; if $g^1_{(i,\ell)} = 0$ initially, then $g^1_{(i,\ell)}$ becomes $1$ after the swap.

To mitigate this issue, we incorporate multi-node moves into our main MCMC algorithm (see Algorithm \ref{FinalAlg}). Faust et al.~\cite{faust2026,FaustThesis} showed that including multi-node moves in posterior sampling methods for community detection in temporal networks greatly reduces the frequency that such methods become stuck at suboptimal local maxima where the inferred group assignments are permutations of the correct group assignments in some layers. This observation motivates our choice to include multi-node moves in our main MCMC algorithm.

We now prove that the stationary distribution of our main algorithm (see Algorithm \ref{FinalAlg}) matches the desired posterior distribution \eqref{bayesposteriorkvariable}. 
Recall that Algorithm \ref{FinalAlg} is the same as Algorithm \ref{KVariableAlg} except for the addition of multi-node moves. Including this additional type of move does not impact the detailed-balance calculations in Section \ref{MCMCKVariable} (aside from an additional factor of $1 - p$ in all terms), so it suffices to verify the detailed-balance equation 
\begin{equation}\label{detailedBalanceMN}
	\frac{\P(g_{(1)}',k')}{\P(g_{(1)},k)}  = \frac{\pi(g,k \to g',k')}{\pi(g',k' \to g,k)}
\end{equation}
when $g,k \to g',k'$ and $g',k' \to g,k$ are multi-node moves. 
From Algorithms \ref{MultiNodeProposal} and \ref{FinalAlg}, we have
\begin{align*}
	\pi(g,k \to g',k') &= \frac{1}{L} \times p \times \frac{1}{2^{k - 1}} \times \frac{1}{2^{k - 1}} \,, \\
	\pi(g',k' \to g,k) &= \frac{1}{L} \times p \times \frac{1}{2^{k' - 1}} \times \frac{1}{2^{k' - 1}} \,.
\end{align*}	
A multi-node move does not change the number of groups, so $k' = k$. Therefore, $\pi(g,k \to g',k')  = \pi(g',k' \to g,k)$. 
Recall that we determine the group assignment $g'$ using Eq.~\eqref{gprimedefinition} for a multi-node move $g,k \to g',k'$. 
Therefore, for a multi-node move $g,k \to g',k'$ in layer $\ell \in \{2,\ldots,L\}$, we know that $g$ and $g'$ have the same group assignments in the first layer. Consequently,
\begin{equation*}
	\P(g_{(1)}',k') = \P(g_{(1)},k) \, ,
\end{equation*}
which implies that the detailed-balance equation \eqref{detailedBalanceMN} holds for multi-node moves $g,k \to g',k'$ in layers $\ell \in \{2,\ldots,L\}$. However, because a multi-node move in layer $\ell = 1$ almost always changes the group sizes in layer $1$, the detailed-balance equation \eqref{detailedBalanceMN} does not hold for such a move. If we choose to restrict Algorithm \ref{MultiNodeProposal} to allow multi-node moves only in layers $\ell \in \{2,\ldots,k\}$ (as we discussed previously in Section \ref{MultiNodeMoves}), the stationary distribution of our main algorithm (see Algorithm \ref{FinalAlg}) is then the same as the desired posterior distribution \eqref{bayesposteriorkvariable}. However, this choice causes our MCMC algorithm to become stuck at local extrema with permuted group assignments in the first layer. This 
would significantly decrease the performance of our approach, so we do not do this, and we instead allow multi-node moves in all layers.


\medskip

\subsection{Computation of the Acceptance Probability} \label{AcceptanceProbCalc}

We now discuss how we compute the acceptance probability \eqref{acceptanceprob}. Recall from Eq.~\eqref{acceptanceprob} that we must compute  
\begin{equation*}
	\alpha(g,k \to g',k') = \min\left\{1,\frac{\P(A|g',k)F(g'|k)}{\P(A|g,k)F(g|k)}\right\} \,.
\end{equation*}
It is straightforward to compute $\P(A|g,k)$ from Eq.~\eqref{ProbAdjacencyMarginal}. 
To compute $F(g|k)$, recall from Eqs.~\eqref{probJ} and \eqref{Fdef} that
\begin{widetext}
\begin{equation*}
	\P(g^r_{(\ell)} | g^r_{(\ell - 1)}) = \prod_{s = 0}^1 \left[ \frac{1}{\begin{pmatrix} n_{r;s}^{(\ell - 1)} \\[0.6em] c_{r;ss}^{(\ell)} \end{pmatrix}} \times J(n_{s}(g^r_{(\ell - 1)}) - c_{r;ss}^{(\ell)}, n_{s}(g^r_{(\ell - 1)}))\right]\, ,
\end{equation*}
\end{widetext}
where
\begin{equation*}
	F(g|k) = \prod_{\ell = 2}^L \prod_{r = 1}^{k - 1} \P(g^r_{(\ell)} | g^r_{(\ell - 1)}) 
\end{equation*}	
and
\begin{equation*}
	J(k_1,k_2) = \int_0^1 x^{k_1} \frac{x - 1}{x^{k_2 + 1} - 1} \; dx \, .
\end{equation*}	

To minimize computational cost, we precompute
\begin{equation*}
	J(k_1,k_2) = \int_0^1 x^{k_1} \frac{x - 1}{x^{k_2 + 1} - 1} \; dx
\end{equation*}	
for all $k_1, k_2 \in \{0, \ldots, n\}$ with $k_1 \leq k_2$ using the procedure in Section B.2.1 of \cite{FaustThesis} (that is also discussed in Appendix A.3 of \cite{faust2026}). 
This precomputation allows us to avoid needing to repeatedly recompute $J(k_1,k_2)$ when computing Eq.~\eqref{probJ} \footnote{If $n$ is sufficiently large, the integral $J(k_1,k_2)$ can become very small, which causes finite-precision issues and thereby leads to inaccurate results when computing $J(k_1,k_2)$ using the procedure in \cite[Appendix B.2.1]{FaustThesis} and \cite[Appendix A.3]{faust2026}. 
To mitigate this problem, we use the approximation
\begin{equation*}
	\frac{J(k_1 + 1,k_2)}{J(k_1,k_2)}  \approx 1
\end{equation*}
for large $k_1$ and $k_2$. In particular, for fixed $k_2$, we set the computed values of $J(k_1,k_2)$ to $\exp(-16)$ for all $k_1 \ge k_1'$, where $k_1'$ is the smallest $k_1'$ such that $J(k_1',k_2) < \exp(-16)$.}.
We also use analogous arguments to those in Eqs.~(4.34)--(4.36) of \cite{FaustThesis} (also see Section 5.1.1 of \cite{faust2026}) to avoid directly calculating terms in Eq.~\eqref{Fdef} that cancel out in the expression \eqref{acceptanceprob} for the acceptance probability $\alpha(g,k \to g',k')$. This reduces the number of times that we need to evaluate $J(k_1,k_2)$. Our implementation of Algorithm \ref{FinalAlg} is available at \cite{faust-code}.


\subsection{Run-Time Complexity of Algorithm \ref{FinalAlg}} \label{FinalAlgComplexity}

We now discuss the run-time complexity of Algorithm \ref{FinalAlg} (i.e., our main algorithm). The space complexity of Algorithm \ref{FinalAlg} grows much more slowly than its run-time complexity, so we discuss only run-time complexity.
Each component of Algorithm \ref{FinalAlg} takes at most $O(kn^2)$ operations, where $k$ is the number of groups and $n$ is the number of nodes, so the complete algorithm has a run-time complexity of $O(kn^2)$.

In Section \ref{StandardComplexity}, we discuss the run-time complexity of standard moves (see Algorithm \ref{StandardProposal}). In Section \ref{GroupAdditionComplexity}, we discuss the run-time complexity of group-addition 
moves (see Algorithm \ref{GroupAdditionProposal}). In Section \ref{MultiNodeComplexity}, we discuss the run-time complexity of multi-node moves. Finally, in Section \ref{summary-complexity}, we summarize the overall run-time complexity of Algorithm \ref{FinalAlg}.


\subsubsection{Run-Time Complexity of Standard Moves} \label{StandardComplexity}

In this subsubsection, we explain that it takes $O(kn^2)$ operations to propose a standard move, compute the acceptance probability, and perform the move if it is accepted.

In Algorithm \ref{StandardProposal}, it takes at most $O(n)$ operations to propose a standard move.
To add a node-layer from group $r$ in layer $1$, we need to construct the set of node-layers in layer $1$ that are not in group $r$. It requires $O(n)$ operations to do this. By an analogous argument, it takes $O(n)$ operations to remove a node-layer from group $r$ in layer $1$.
For layer $\ell \ge 2$, the move proposal and implementation requires $O(1)$ operations, as we choose a node $i$ uniformly at random.

We now discuss the run-time complexity of computing the acceptance probability of a standard move. We first examine a standard move that removes an empty group, and we then examine a standard move that changes the group assignment of one node-layer.

Consider a standard move that removes an empty group $r$. Recall from Eq.~\eqref{GADiscAcceptance} that 
\begin{equation*}
	\alpha(g,k \to g',k') = \min\left\{1,\frac{\P(A|g',k)F(g'|k)}{\P(A|g,k)F(g|k)}\right\} \,,
\end{equation*}
where $\P(A|g,k)$ is given by Eq.~\eqref{ProbAdjacencyMarginal} and $F(g|k)$ is given by Eq.~\eqref{Fdef}.
Because group $r$ is empty, $\frac{\P(A|g',k)}{\P(A|g,k)} = 1$, as the empty group does not change any of the terms in $\P(A|g',k)$ from the terms in $\P(A|g,k)$.
It thus suffices to compute $\frac{F(g'|k)}{F(g|k)}$. By Eq.~\eqref{Fdef}, we see that 
\begin{equation}\label{GroupRemovalQuotient}
	\frac{F(g'|k)}{F(g|k)} = \prod_{\ell = 2}^L \P(g^r_{(\ell)} | g^r_{(\ell - 1)}) \, .
\end{equation}
For all layers $\ell \in \{2,\ldots,L\}$, we have $n_0(g^r_{(\ell)}) = n$, $n_1(g^r_{(\ell)}) = 0$, $c^{(\ell)}_{r;00} = n$, and $c^{(\ell)}_{r;11} = 0$.
Substituting these values into Eq.~\eqref{prob} yields
\begin{equation} \label{GroupRemovalP}
	\P(g^r_{(\ell)} | g^r_{(\ell - 1)}) = J(0,0) \times J(0,n)
\end{equation}
for all $\ell \in \{2,\ldots,L\}$. We then substitute Eq.~\eqref{GroupRemovalP} into Eq.~\eqref{GroupRemovalQuotient} to obtain
\begin{equation*}
	\frac{F(g'|k)}{F(g|k)} = \left[J(0,0) \times J(0,n)\right]^{1 - L} \, ,
\end{equation*}
which requires $O(1)$ operations, as we precompute $J(k_1,k_2)$ for all $k_1, k_2 \in \{0, \ldots, n\}$ with $k_1 \leq k_2$.

Now consider a standard move that changes the group assignment of one node-layer. 
Suppose that we change $g^r_{(i,\ell)}$. 
(Specifically, we set $(g')^r_{(i,\ell)} = 0$ if $ g^r_{(i,\ell)} = 1$, and we set $(g')^r_{(i,\ell)} = 1$ if $ g^r_{(i,\ell)} = 0$.)
By Eq.~\eqref{ProbAdjacencyMarginal}, to compute $\frac{\P(A|g',k)}{\P(A|g,k)}$, it suffices to compute $m_r^{(\ell)}$ and $t_r^{(\ell)}$ for $g$ and $g'$ for each group $r \in \{1,\ldots,k - 1\}$, as $g$ and $g'$ have the same group assignments as each other in all other layers. 
Computing $m_r^{(\ell)}$ and $t_r^{(\ell)}$ naively requires $O(n^2)$ operations, as it is necessary to sum over all pairs of distinct 
node-layers $(i,\ell)$ and $(j,\ell)$ in layer $\ell$. Therefore, computing all of these quantities takes $O(kn^2)$ operations.

To compute $\frac{F(g'|k)}{F(g|k)}$, it follows directly from Eq.~\eqref{Fdef} that it suffices to compute $\P(g^r_{(\ell)} | g^r_{(\ell - 1)})$, $\P(g^r_{(\ell+1)} | g^r_{(\ell)})$, $\P((g')^r_{(\ell)} | (g')^r_{(\ell - 1)})$, and $\P((g')^r_{(\ell + 1)} | (g')^r_{(\ell)})$, 
as all other terms cancel each other.
To compute each of the four terms, it suffices to compute $n_{s}(g^r_{(\ell - 1)})$ (i.e., the number of nodes in layer $\ell - 1$ with a group-$r$ indicator variable of $s$) and $c_{r;ss}^{(\ell)}$ (i.e., the number of nodes $i$ with $g^r_{(i,\ell - 1)} = s$ and $g^r_{(i,\ell)} = s$).
Computing $n_{s}(g^r_{(\ell - 1)})$ and $c_{r;ss}^{(\ell)}$ requires $O(n)$ operations. 
Because we precompute $J(k_1,k_2)$ for all $k_1, k_2 \in \{0, \ldots, n\}$ with $k_1 \leq k_2$, computing $\frac{F(g'|k)}{F(g|k)}$ requires $O(n)$ operations.

If we accept the proposed standard move, both removing an empty group and changing the value of $g^r_{(i,\ell)}$ require $O(1)$ operations.


\subsubsection{Run-Time Complexity of Group-Addition Moves} \label{GroupAdditionComplexity}

In this subsubsection, we explain that it takes $O(1)$ operations to propose a group-addition move, compute the acceptance probability, and perform the move if it is accepted.

In Algorithm \ref{GroupAdditionProposal}, the run-time complexity of proposing a standard move is $O(1)$, as it suffices to sample $r \in \{1,\ldots,k\}$ uniformly at random.

We now discuss the run-time complexity of computing the acceptance probability of a group-addition move. Recall from Eq.~\eqref{GADiscAcceptance} that 
\begin{equation*}
	\alpha(g,k \to g',k') = \min\left\{1,\frac{\P(A|g',k)F(g'|k)}{\P(A|g,k)F(g|k)}\right\} \,,
\end{equation*}
where $\P(A|g,k)$ is given by Eq.~\eqref{ProbAdjacencyMarginal} and $F(g|k)$ is given by Eq.~\eqref{Fdef}. Because the group $r$ that we add is empty, $\frac{\P(A|g',k)}{\P(A|g,k)} = 1$, 
as the empty group does not change any of the terms in $\P(A|g',k)$ from the terms in $\P(A|g,k)$.
It thus suffices to compute $\frac{F(g'|k)}{F(g|k)}$. By Eq.~\eqref{Fdef}, we see that 
\begin{equation}\label{GroupAdditionQuotient}
	\frac{F(g'|k)}{F(g|k)} = \prod_{\ell = 2}^L \P((g')^r_{(\ell)} | (g')^r_{(\ell - 1)}) \, .
\end{equation}
For all layers $\ell \in \{2,\ldots,L\}$, we have $n_0((g')^r_{(\ell)}) = n$, $n_1((g')^r_{(\ell)}) = 0$, $c^{(\ell)}_{r;00} = n$, and $c^{(\ell)}_{r;11} = 0$. 
Substituting these values into Eq.~\eqref{prob} yields
\begin{equation} \label{GroupAdditionP}
	\P((g')^r_{(\ell)} | (g')^r_{(\ell - 1)}) = J(0,0) \times J(0,n)
\end{equation}
for all $\ell \in \{2,\ldots,L\}$. We then substitute Eq.~\eqref{GroupAdditionP} into Eq.~\eqref{GroupAdditionQuotient} to obtain
\begin{equation*}
	\frac{F(g'|k)}{F(g|k)} = \left[J(0,0) \times J(0,n)\right]^{L - 1} \, ,
\end{equation*}
which requires $O(1)$ operations, as we precompute $J(k_1,k_2)$ for all $k_1, k_2 \in \{0, \ldots, n\}$ with $k_1 \leq k_2$.

If we accept the proposed group-addition move, adding an empty group requires $O(1)$ operations. 


\subsubsection{Run-Time Complexity of Multi-Node Moves} \label{MultiNodeComplexity}

In this subsubsection, we explain that it takes $O(kn^2)$ operations to propose a multi-node move, compute the acceptance probability, and perform the move if it is accepted.

In Algorithm \ref{MultiNodeProposal}, we see that the run-time complexity of proposing a multi-node move is at most $O(kn)$, as we need to construct the set of node-layers in layer $\ell$ with group assignments $\mathcal{G}_1$ or $\mathcal{G}_2$.

We now discuss the run-time complexity of computing the acceptance probability of a multi-node move. 
Suppose that such a move changes the group assignments in layer $\ell$. 
By Eq.~\eqref{ProbAdjacencyMarginal}, to compute $\frac{\P(A|g',k)}{\P(A|g,k)}$, it suffices to compute $m_r^{(\ell)}$ and $t_r^{(\ell)}$ for each group $r \in \{1,\ldots,k - 1\}$ for both $g$ and $g'$.
The reason for this is that the group-indicator arrays $g$ and $g'$ have the same group assignments for all other layers.
Computing $m_r^{(\ell)}$ and $t_r^{(\ell)}$ requires $O(n^2)$ operations, as it is necessary to sum over all pairs of distinct node-layers $(i,\ell)$ and $(j,\ell)$ in layer $\ell$. 
Therefore, computing all of these quantities takes $O(kn^2)$ operations.

To compute $\frac{F(g'|k)}{F(g|k)}$, it follows directly from Eq.~\eqref{Fdef} that it suffices to compute $\P(g^r_{(\ell)} | g^r_{(\ell - 1)})$, $\P(g^r_{(\ell + 1)} | g^r_{(\ell)})$, $\P((g')^r_{(\ell)} | (g')^r_{(\ell - 1)})$, and $\P((g')^r_{(\ell + 1)} | (g')^r_{(\ell)})$ for each group $r \in \{1,\ldots,k - 1\}$, as all other terms cancel each other out.
To compute each of these four terms, it suffices to compute $n_{s}(g^r_{(\ell - 1)})$ (i.e., the number of nodes in layer $\ell - 1$ with a group-$r$ indicator variable of $s$) and $c_{r;ss}^{(\ell)}$ (i.e., the number of nodes $i$ with $g^r_{(i,\ell - 1)} = s$ and $g^r_{(i,\ell)} = s$).
Both of these quantities require $O(n)$ operations to compute. Because we precompute $J(k_1,k_2)$ for all $k_1, k_2 \in \{0, \ldots, n\}$ with $k_1 \leq k_2$, computing $\frac{F(g'|k)}{F(g|k)}$ requires $O(kn)$ operations.

If we accept the proposed multi-node move, constructing the new group assignments $g'$ requires at most $O(kn)$ operations.

\subsubsection{Overall Run-Time Complexity of Algorithm \ref{FinalAlg}}\label{summary-complexity}

Now that we have expressions for the run-time complexity of each type of move, we can state the run-time complexity of Algorithm \ref{FinalAlg}. 
To precompute $J(k_1,k_2)$ for all $k_1, k_2 \in \{0, \ldots, n\}$ with $k_1 \leq k_2$, it follows from Lemma A.1 in \cite{faust2026} that it takes $O(1)$ operations to compute $J(k_1,k_2)$ for one value of $(k_1,k_2)$. 
Therefore, this computation requires $O(n^2)$ operations.
Each component of Algorithm \ref{FinalAlg} takes at most $O(kn^2)$ operations, so the complete algorithm has a run-time complexity of $O(kn^2)$.


\section{Discussion of our Group-Assignment Approach}\label{GroupAssignmentDiscussion}

Now that we have discussed Algorithm \ref{FinalAlg} (see Sections \ref{MainMethod} and \ref{MNMoves2}), which we use to sample from the desired posterior distribution $\P(g,k|A)$ in 
Eq.~\eqref{bayesposteriorkvariable}, we are equipped to discuss an additional rationale behind our choice of the group-assignment probability distribution $\P(g|k)$.
Recall from Algorithm \ref{FinalAlg} and Section \ref{AcceptanceProb} that the acceptance probability $\alpha(g,k \to g',k')$ is
\begin{equation}\label{GADiscAcceptance}
	\alpha(g,k \to g',k') = \min\left\{1, \frac{\P(A|g',k')F(g',k')}{\P(A|g,k)F(g,k)}\right\}.
\end{equation}
By Eqs.~\eqref{gsubone} and \eqref{ggivenkprod}, we generate the group assignments for each group independently. Therefore, if we propose a group-addition move that adds an empty group $r$, the acceptance probability is
\begin{equation*}
	\P(g_{(i,\ell)}^r = 0 \,\, \text{ for all }\,\, i \in \{1,\ldots,n\} \,\text{ and }\, \ell \in \{1,\ldots,L\}) \, .
\end{equation*}	
Consequently, the acceptance probability \eqref{GADiscAcceptance} is very small if a generative model is biased against groups with few node-layers. This bias, in turn, causes an associated statistical-inference method to underestimate the number of groups. 

For a monolayer network (i.e., an ordinary graph), one can resolve the above issue by generating the group sizes uniformly at random.
With this choice, the probability of generating an empty group is $\frac{1}{n + 1}$.
However, for a temporal network, it is difficult to choose a generative model that is not biased against groups with few node-layers. 
Faust et al. \cite{FaustThesis,faust2026} showed that the commonly-used approach (see, e.g., \cite{Yang11, Ghasemian16, Matias17, Bazzi20}) of evolving group assignments via a discrete-time Markov process (where one bases the group assignment of a node in a given layer on its group assignment in the previous layer) leads to an increased bias against small and large groups in later layers. 
This bias, in turn, leads to a strong bias of a generative model against small and large groups in a network.
In our context of hierarchical core--periphery structure, this generative-model bias makes the acceptance probability \eqref{GADiscAcceptance} of a group-addition move prohibitively small, leading to an underestimation in the number of groups.

Our novel group-assignment approach is much less biased against small and large groups than such Markov-process approaches. The reason for this is that our approach generates the group assignments in a layer based on all of the group assignments in the previous layer, instead of using a discrete-time Markov process to evolve the group assignments of each node individually. (See \cite[Sections 4.2.3.3 and 4.3]{FaustThesis} and \cite[Sections 3.3.3 and 4]{faust2026} of \cite{FaustThesis} for more information.)
Consequently, our method is much less likely to underestimate the number of groups in a network. 

This discussion illustrates the practical importance of carefully considering potentially undesirable assumptions of generative models in statistical-inference methods. See \cite{FaustThesis,faust2026} for further discussions and examples in the context of community structure.


\section{Results for Example Networks} \label{ComparisonSICP}

In this section, we apply our hierarchical core--periphery detection method to one synthetic temporal network and two real-world temporal networks, and we discuss the structures that we identify in these networks. Statistically significant core--periphery structure is not present in many real-world networks \cite{yan2025}, 
so we examine real-world networks for which prior investigations have provided evidence of such structure or other motivation to study it. In Section \ref{synthetic}, we consider a synthetic network with a known hierarchical core--periphery structure. In Section \ref{terror}, we consider a network of terrorists in the Indian states of Jammu and Kashmir. In Section \ref{literary}, we consider a network of co-appearances of individuals in the \emph{Gospel of Luke}. 


\medskip
\medskip

\subsection{A Synthetic Temporal Network} \label{synthetic}

\begin{figure*}[h]
\centering
\includegraphics[width=0.7\textwidth]{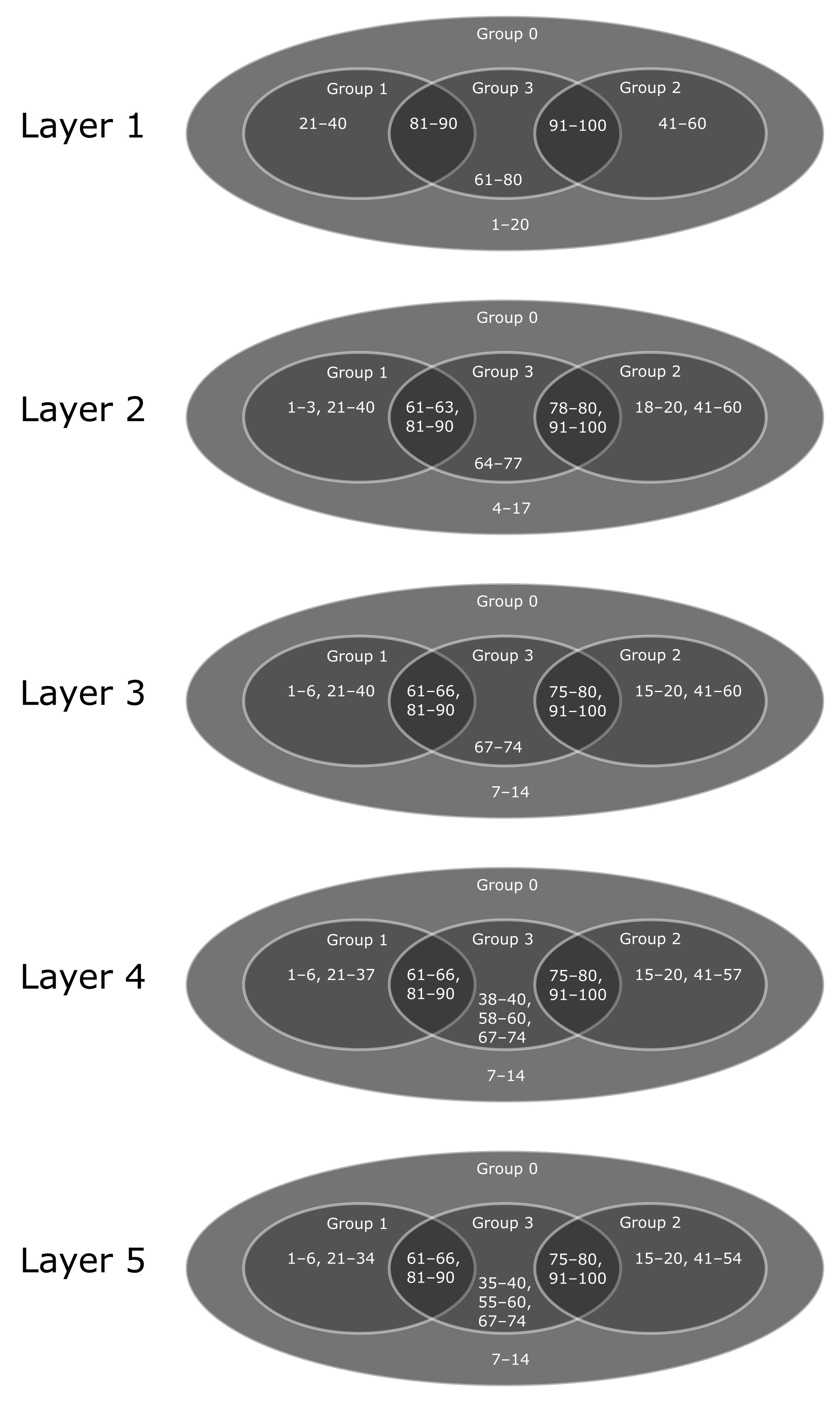}
\caption{The seeded core--periphery structure of our synthetic temporal network. When a node $i$ is in a shaded region with the label ``Group $r$'' in a layer with the label ``Layer $\ell$'', the a group-$r$ indicator variable is $g^r_{(i,\ell)} = 1$. The notation $i$--$j$ signifies the set $\{i, i + 1,\ldots, j - 1, j\}$ of nodes.}
\label{SyntheticSeeded}
\end{figure*}

We first apply our hierarchical core--periphery detection method to a synthetic network with a known hierarchical core--periphery structure. This allows us to compare the core--periphery structure that we detect using our method with a known ground-truth structure.
We generate our synthetic network, which has $n = 100$ nodes and $L = 5$ layers, using the generative model in Section \ref{GenerativeModel}. In Figure \ref{SyntheticSeeded}, we show the seeded group structure of the network. The edge-probability parameters $\omega_r^{(\ell)}$ are $(\omega_0^{(\ell)}, \omega_1^{(\ell)}, \omega_2^{(\ell)}, \omega_3^{(\ell)}) = (0.05, 0.35, 0.65, 0.95)$ for each layer $\ell \in \{1,2,\ldots,5\}$.

We choose the core--periphery structure in Figure \ref{SyntheticSeeded} for two primary reasons. First, it has a non-nested group structure. For example, groups $1$ and $3$ have nonempty intersections in each layer, but neither group completely contains the other. Additionally, as is reasonable to expect in many real-world temporal networks, the nodes in our synthetic network tend to remain in the same group across layers.

We perform $20$ independent runs of $10^6$ steps of our main MCMC algorithm (see Algorithm \ref{FinalAlg}) with $k = 4$ groups initially, a multi-node mode probability of $p = 10^{-3}$, and uniformly random initial group assignments.
We use distinct initial group assignments for each run, and we save the output group assignments every $10^4$ steps of our MCMC algorithm.
In 18 of the 20 runs, the inferred group assignments either match or almost match the ground-truth groups, aside from three specific types of discrepancies.

\begin{figure*}[h]
\centering
\includegraphics[width=0.98\textwidth]{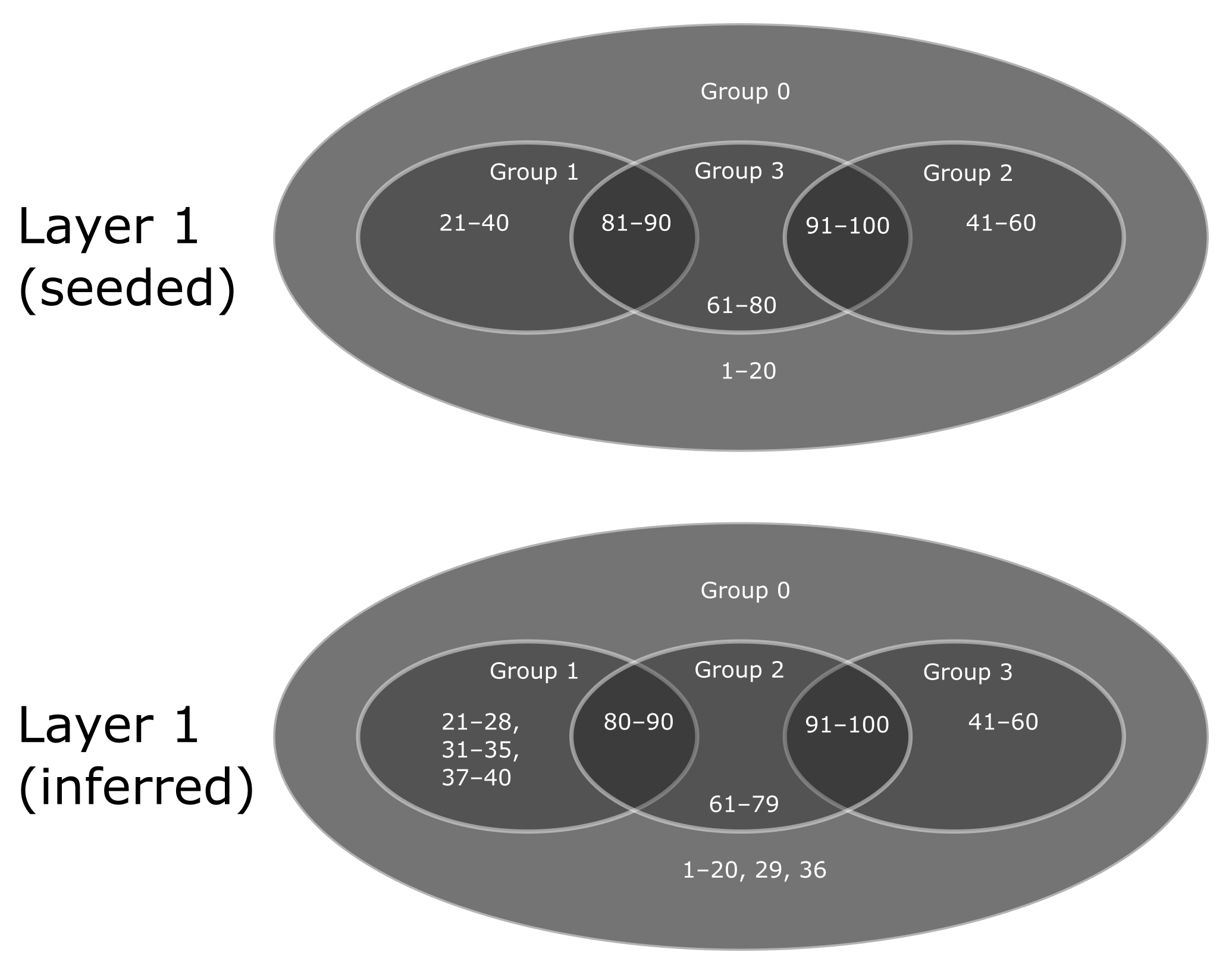}
\caption{An example of inferred group assignments with permuted groups.  When a node $i$ is in a shaded region with the label ``Group $r$'' in a layer with the label ``Layer $\ell$'', the a group-$r$ indicator variable is $g^r_{(i,\ell)} = 1$. The notation $i$--$j$ signifies the set $\{i, i + 1,\ldots, j - 1, j\}$ of nodes.}
\label{SyntheticPermutedStructure}
\end{figure*}

\begin{figure*}[h]
\centering
\includegraphics[width=0.98\textwidth]{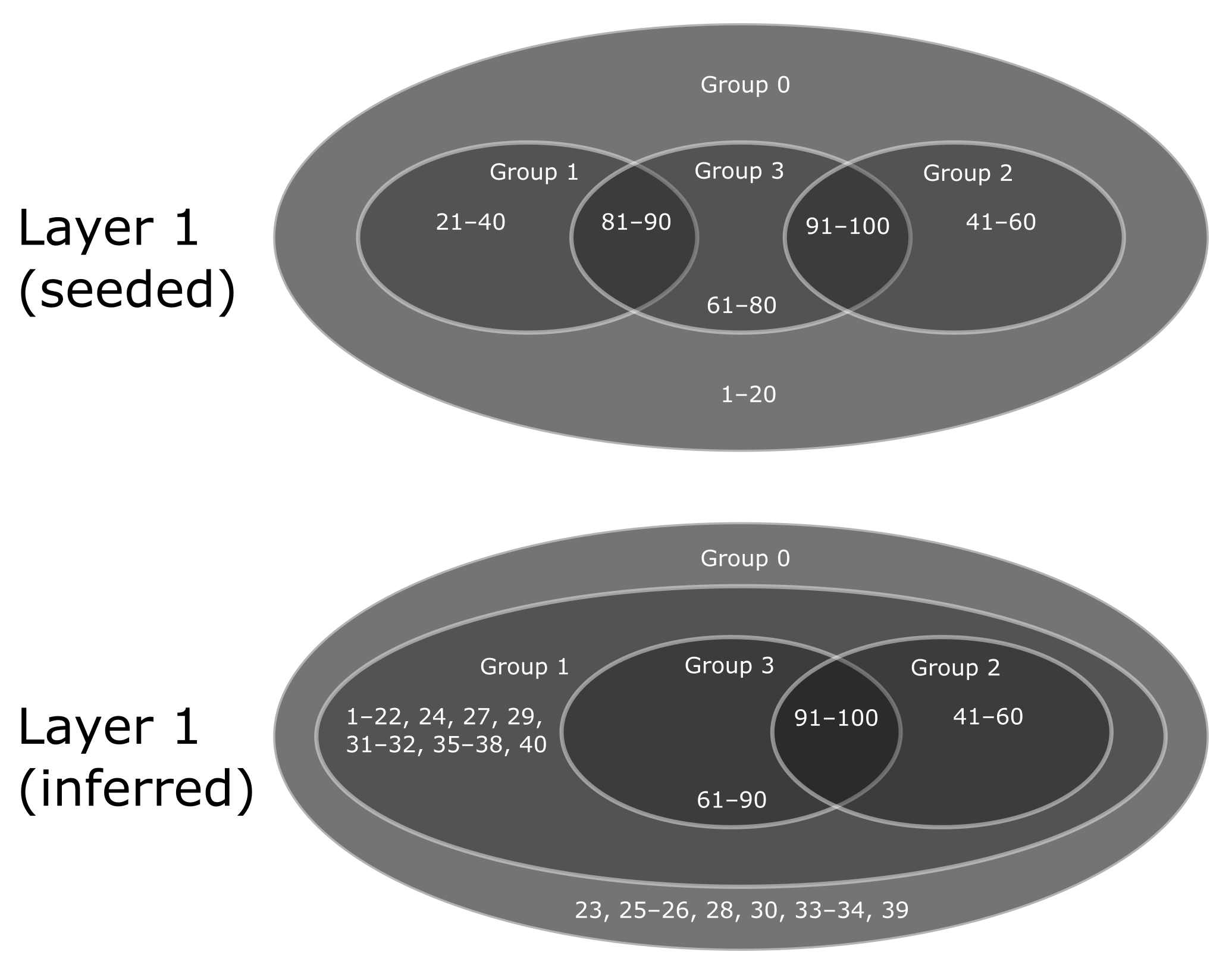}
\caption{An example of inferred group assignments with combined groups.  When a node $i$ is in a shaded region with the label ``Group $r$'' in a layer with the label ``Layer $\ell$'', the a group-$r$ indicator variable is $g^r_{(i,\ell)} = 1$. The notation $i$--$j$ signifies the set $\{i, i + 1,\ldots, j - 1, j\}$ of nodes.}
\label{SyntheticCombinedStructure}
\end{figure*}

Unsurprisingly, the inferred group assignments differ across distinct runs. There are three primary reasons for this observation. 
First, in the inferred group assignments, the group labels are often permutations of the ground-truth groups. In Figure \ref{SyntheticPermutedStructure}, we illustrate this situation with inferred group assignments that permute groups $2$ and $3$.
Second, the inferred group assignments occasionally either combine two or more ground-truth groups or separate a ground-truth group into two or more groups. 
In Figure \ref{SyntheticCombinedStructure}, we illustrate this situation with inferred group assignments in which group $1$ also includes all of the nodes from groups $2$ and $3$. 
Third, the inferred group assignments often differ across layers in many of the runs. 
In particular, both (1) permutations of group labels from ground-truth groups and (2) inferred group assignments that combine two ground-truth groups or separate a ground-truth group into two groups can occur in individual layers.
Accordingly, the inferred core--periphery structure can be correct in some layers but incorrect in these specific ways in other layers.\
Aside from the three aforementioned types of discrepancies, the inferred group assignments always match or almost match the ground-truth groups in 18 of our 20 runs. 

The permutations of ground-truth group labels and combinations and separations of ground-truth groups in the inferred groups also occur in the monolayer-network core--periphery detection method in \cite{Polanco23}. 
Additionally, recall from Section \ref{MNMoves2} that we commonly observe permutations of group assignments between layers in local maxima of the MCMC algorithm without multi-node moves (see Algorithm \ref{KVariableAlg}). 
Accordingly, incorporating multi-node moves is insufficient for our MCMC approach to entirely avoid becoming stuck at such local maxima.


\subsection{A Terrorist Network}\label{terror}

Several researchers have illustrated that terrorist groups can have core--periphery structure and/or community structure~\cite{Asal16, Bacon14, Horowitz14}.
Accordingly, we apply our core--periphery detection method to a network of ties between terrorist organizations in the Indian states of Jammu and Kashmir from 2000 to 2003 \cite{Saxena04}. This temporal network has $n = 34$ nodes (which represent terrorist organizations) and $L = 4$ layers (which represent years). 

\begin{figure*}[h]
\centering
\includegraphics[width=0.98\textwidth]{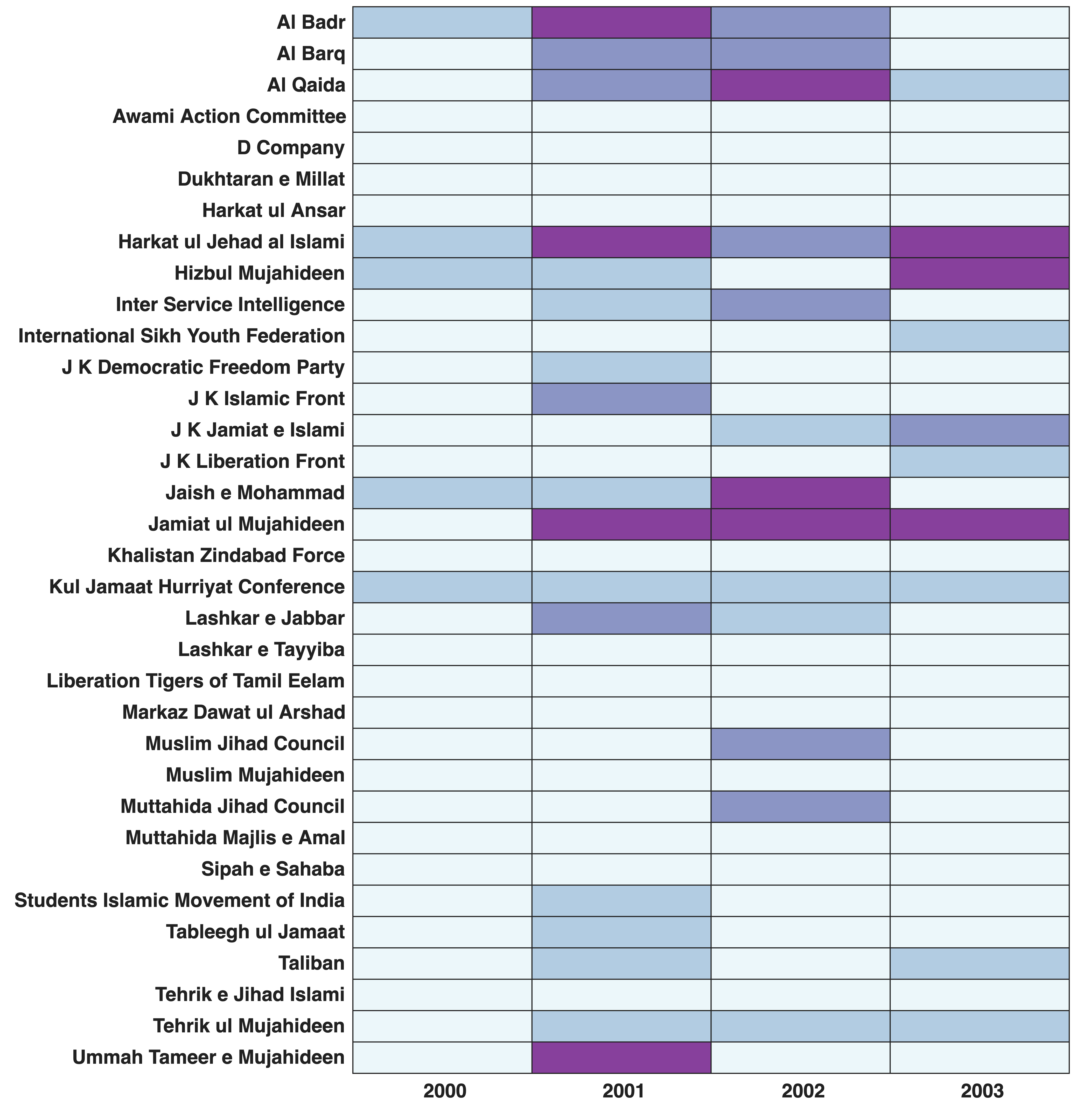}
\caption{The inferred core--periphery structure in the Jammu--Kashmir terrorist network. The light blue rectangles signify the group assignments $g^1_{(i,\ell)} = 0$ and $g^2_{(i,\ell)} = 0$, the dark blue rectangles signify the group assignments $g^1_{(i,\ell)} = 1$ and $g^2_{(i,\ell)} = 0$, the light purple rectangles signify the group assignments $g^1_{(i,\ell)} = 0$ and $g^2_{(i,\ell)} = 1$, and the dark purple rectangles signify the group assignments $g^1_{(i,\ell)} = 1$ and $g^2_{(i,\ell)} = 1$. The horizontal axis indicates years and the vertical axis indicates terrorist organizations.}
\label{SaxenaCP}
\end{figure*}

 Applying our hierarchical core--periphery detection method to this network yields $k = 3$ groups and the core--periphery structure in Figure \ref{SaxenaCP}. To obtain this core--periphery structure, we perform $5$ runs of $10^6$ steps of our main MCMC algorithm (see Algorithm \ref{FinalAlg}) with $k = 4$ groups initially, a multi-mode move probability of $p = 10^{-3}$, and uniformly random initial group assignments.
We use distinct initial group assignments for each run, and we save the output group assignments every $10^4$ steps of our MCMC algorithm.
For each node-layer $(i,\ell)$, we set the group assignment of $(i,\ell)$ to be the most frequent group assignment across the 5 MCMC runs.

\begin{figure*}[h]
\centering
\subfloat[]{\includegraphics[width=0.49\textwidth]{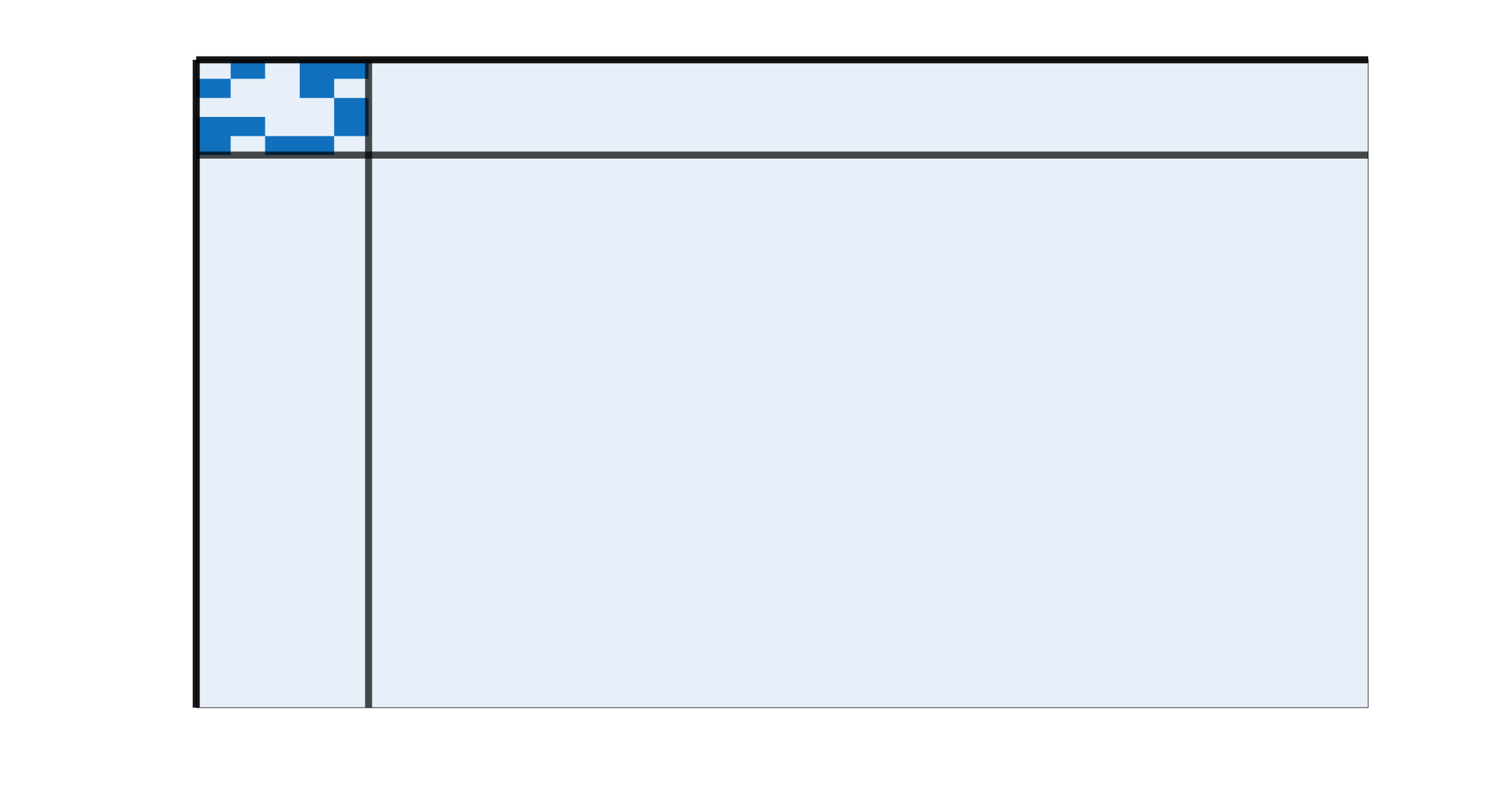}} \;
\subfloat[]{\includegraphics[width=0.49\textwidth]{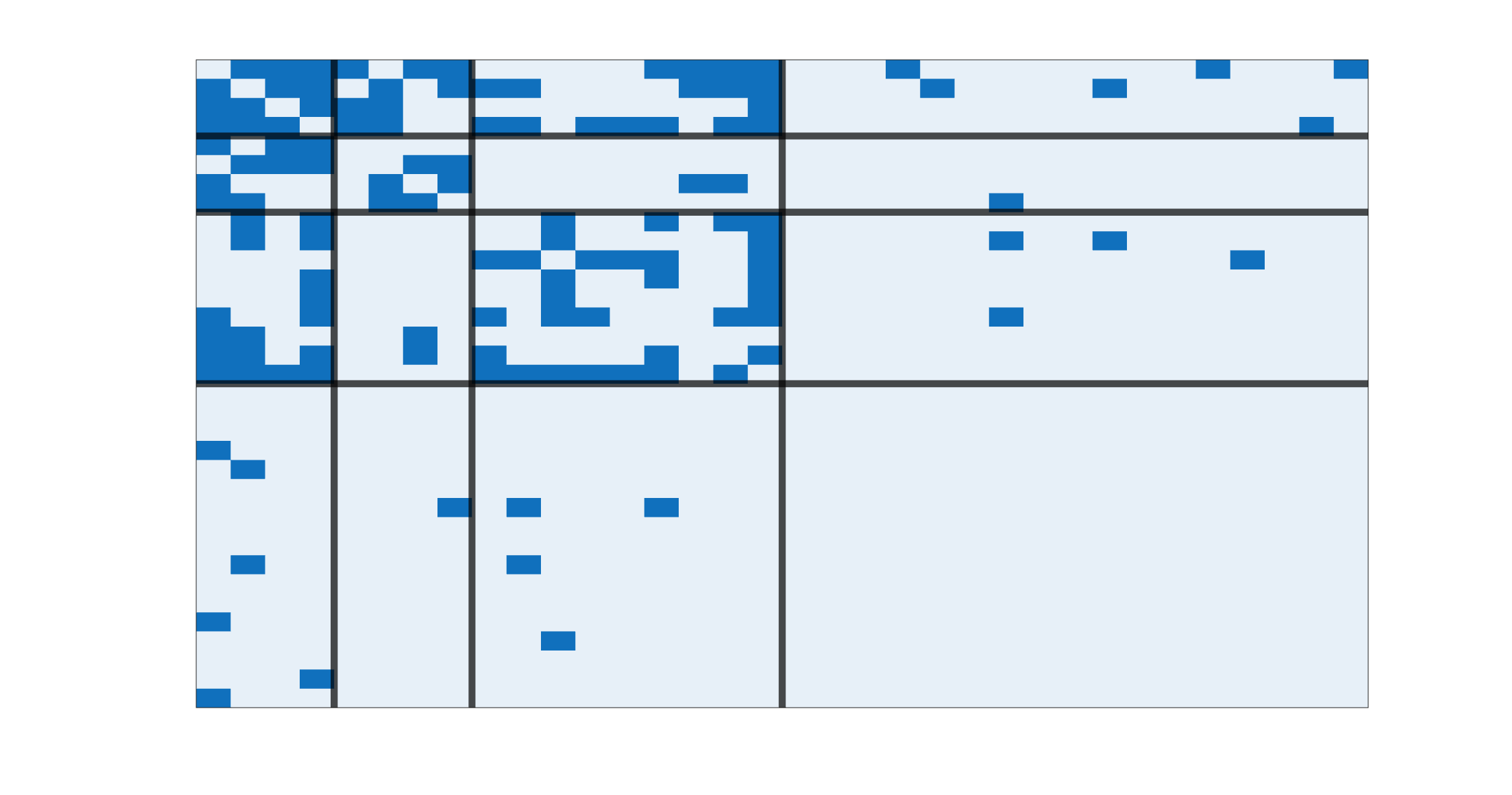}}\\
\subfloat[]{\includegraphics[width=0.49\textwidth]{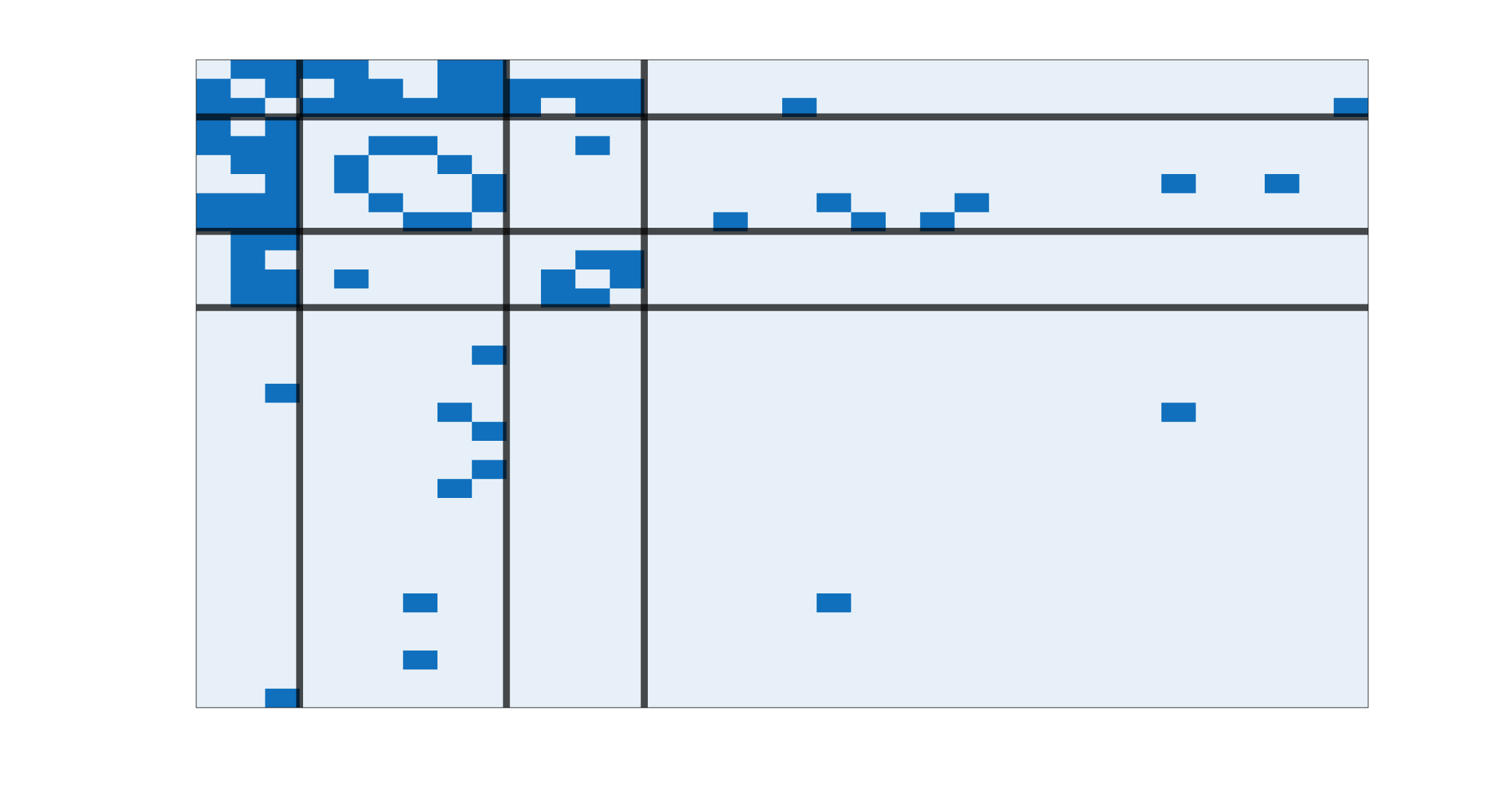}} \;
\subfloat[]{\includegraphics[width=0.49\textwidth]{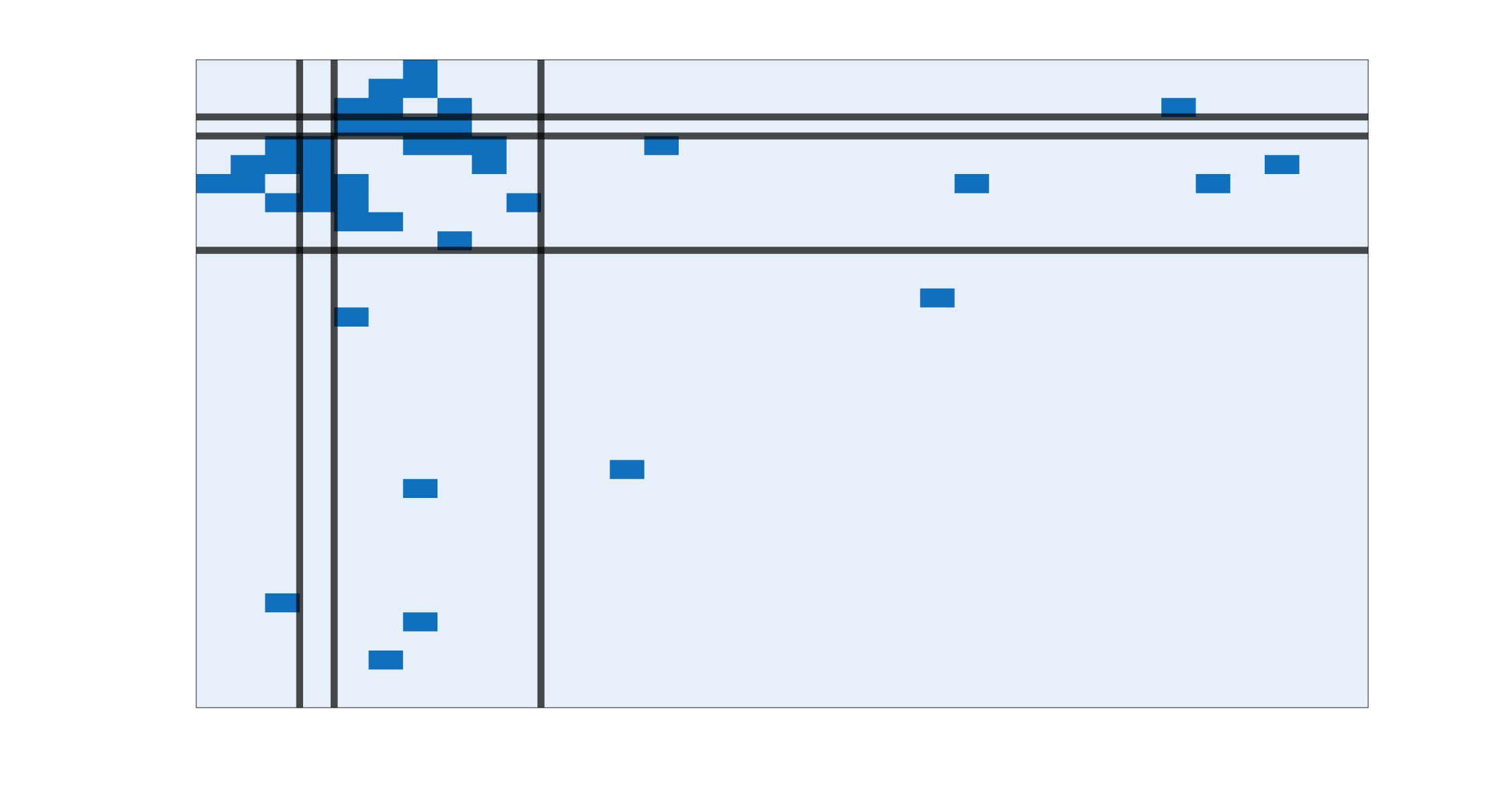}}
\caption{Permuted adjacency matrices to illustrate our inferred core--periphery structure (see Figure \ref{SaxenaCP}) of the Jammu--Kashmir terrorist network. We show (a) layer 1 (the year $2000$), (b) layer 2 (the year $2001$), (c) layer 3 (the year $2002$), and (d) layer 4 (the year $2003$).}
\label{SaxenaAdjacency}
\end{figure*}

To illustrate that the detected core--periphery structure in Figure \ref{SaxenaCP} is reasonable, we plot heat maps (see Figure \ref{SaxenaAdjacency}) of the adjacency matrices $A^{(\ell)}$ of each layer $\ell$. In these heat maps, we permute the rows and columns so that all node-layers $(i,\ell)$ with group assignments $g_{(i,\ell)}^{1} = 1$ and $g_{(i,\ell)}^{2} = 1$ appear earliest, followed by node-layers $(i,\ell)$ with group assignments $g_{(i,\ell)}^{1} = 0$ and $g_{(i,\ell)}^{2} = 1$, node-layers $(i,\ell)$ with group assignments $g_{(i,\ell)}^{1} = 1$ and $g_{(i,\ell)}^{2} = 0$, and finally node-layers $(i,\ell)$ with group assignments $g_{(i,\ell)}^{1} = 0$ and $g_{(i,\ell)}^{2} = 0$. We place dividing lines between sets of node-layers with different group assignments.
In Figure \ref{SaxenaAdjacency}, we show these heat maps. 

We observe that the densities of the edges between node-layers depend on the groups of the node-layers. For example, in layers $2$ and $3$, node-layers $(i,\ell)$ with group assignments $g_{(i,\ell)}^{1} = 1$ and $g_{(i,\ell)}^{2} = 1$ tend to be connected densely to node-layers $(j,\ell)$ with group assignments $g_{(j,\ell)}^{1} = 1$ and $g_{(j,\ell)}^{2} = 1$ but connected sparsely to node-layers $(j,\ell)$ with group assignments $g_{(j,\ell)}^{1} = 0$ and $g_{(j,\ell)}^{2} = 0$. 
To further illustrate these qualitative observations, we compute the mean edge densities of all pairs of group assignments for each layer and compile our results in Table \ref{densitytablesSaxena}. The heterogeneous edge densities indicate that the inferred core--periphery structure distinguishes between terrorist organizations with different network ``positions"~\footnote{In social-network analysis, ``positions" of nodes in a network indicate features like their structural locations~\cite{Wasserman94}.}

\begin{table*}
\subfloat[]{
\begin{tabular}{p{1.5cm}|p{1.5cm}|p{1.5cm}|p{1.5cm}|p{1.5cm}}
 & $g_{(j,\ell)}^{1} = 0$\newline$g_{(j,\ell)}^{2} = 0$ & $g_{(j,\ell)}^{1} = 1$\newline$g_{(j,\ell)}^{2} = 0$ & $g_{(j,\ell)}^{1} = 0$\newline$g_{(j,\ell)}^{2} = 1$ & $g_{(j,\ell)}^{1} = 1$\newline$g_{(i,\ell)}^{2} = 1$ \\ \hline
$g_{(i,\ell)}^{1} = 0$\newline$g_{(i,\ell)}^{2} = 0$ & 0.000 & 0.000 & N/A & N/A \\ \hline
$g_{(i,\ell)}^{1} = 1$\newline$g_{(i,\ell)}^{2} = 0$ & 0.000 & 0.600 & N/A & N/A \\ \hline
$g_{(i,\ell)}^{1} = 0$\newline$g_{(i,\ell)}^{2} = 1$ & N/A & N/A & N/A & N/A \\ \hline
$g_{(i,\ell)}^{1} = 1$\newline$g_{(i,\ell)}^{2} = 1$ & N/A & N/A & N/A & N/A \\ 
\end{tabular}}
\subfloat[]{
\begin{tabular}{p{1.5cm}|p{1.5cm}|p{1.5cm}|p{1.5cm}|p{1.5cm}}
 & $g_{(j,\ell)}^{1} = 0$\newline$g_{(j,\ell)}^{2} = 0$ & $g_{(j,\ell)}^{1} = 1$\newline$g_{(i,\ell)}^{2} = 0$ & $g_{(j,\ell)}^{1} = 0$\newline$g_{(j,\ell)}^{2} = 1$ & $g_{(j,\ell)}^{1} = 1$\newline$g_{(j,\ell)}^{2} = 1$ \\ \hline
$g_{(i,\ell)}^{1} = 0$\newline$g_{(i,\ell)}^{2} = 0$ & 0.000 & 0.026 & 0.015 & 0.088 \\ \hline
$g_{(i,\ell)}^{1} = 1$\newline$g_{(i,\ell)}^{2} = 0$ & 0.026 & 0.444 & 0.056 & 0.472 \\ \hline
$g_{(i,\ell)}^{1} = 0$\newline$g_{(i,\ell)}^{2} = 1$ & 0.015 & 0.056 & 0.500 & 0.563 \\ \hline
$g_{(i,\ell)}^{1} = 1$\newline$g_{(i,\ell)}^{2} = 1$ & 0.088 & 0.472 & 0.563 & 1.000 \\ 
\end{tabular}\label{densitytable2Saxena}}\\
\subfloat[]{
\begin{tabular}{p{1.5cm}|p{1.5cm}|p{1.5cm}|p{1.5cm}|p{1.5cm}}
 & $g_{(j,\ell)}^{1} = 0$\newline$g_{(j,\ell)}^{2} = 0$ & $g_{(j,\ell)}^{1} = 1$\newline$g_{(j,\ell)}^{2} = 0$ & $g_{(j,\ell)}^{1} = 0$\newline$g_{(j,\ell)}^{2} = 1$ & $g_{(j,\ell)}^{1} = 1$\newline$g_{(j,\ell)}^{2} = 1$ \\ \hline
$g_{(i,\ell)}^{1} = 0$\newline$g_{(i,\ell)}^{2} = 0$ & 0.005 & 0.000 & 0.056 & 0.032 \\ \hline
$g_{(i,\ell)}^{1} = 1$\newline$g_{(i,\ell)}^{2} = 0$ & 0.000 & 0.500 & 0.042 & 0.583 \\ \hline
$g_{(i,\ell)}^{1} = 0$\newline$g_{(i,\ell)}^{2} = 1$ & 0.056 & 0.042 & 0.333 & 0.778 \\ \hline
$g_{(i,\ell)}^{1} = 1$\newline$g_{(i,\ell)}^{2} = 1$ & 0.032 & 0.583 & 0.778 & 1.000 \\ 
\end{tabular}\label{densitytable3Saxena}}
\subfloat[]{
\begin{tabular}{p{1.5cm}|p{1.5cm}|p{1.5cm}|p{1.5cm}|p{1.5cm}}
 & $g_{(j,\ell)}^{1} = 0$\newline$g_{(j,\ell)}^{2} = 0$ & $g_{(j,\ell)}^{1} = 1$\newline$g_{(j,\ell)}^{2} = 0$ & $g_{(j,\ell)}^{1} = 0$\newline$g_{(j,\ell)}^{2} = 1$ & $g_{(j,\ell)}^{1} = 1$\newline$g_{(j,\ell)}^{2} = 1$ \\ \hline
$g_{(i,\ell)}^{1} = 0$\newline$g_{(i,\ell)}^{2} = 0$ & 0.004 & 0.028 & 0.000 & 0.013 \\ \hline
$g_{(i,\ell)}^{1} = 1$\newline$g_{(i,\ell)}^{2} = 0$ & 0.028 & 0.333 & 0.667 & 0.333 \\ \hline
$g_{(i,\ell)}^{1} = 0$\newline$g_{(i,\ell)}^{2} = 1$ & 0.000 & 0.667 & N/A & 0.000 \\ \hline
$g_{(i,\ell)}^{1} = 1$\newline$g_{(i,\ell)}^{2} = 1$ & 0.013 & 0.333 & 0.000 & 0.000 \\ 
\end{tabular}}\\
\caption{The densities of the edges between the node-layers of each of the groups in Figure \ref{SaxenaCP}. We show (a) layer 1 (the year $2000$), (b) layer 2 (the year $2001$), (c) layer 3 (the year $2002$), and (d) layer 4 (the year $2003$). To compute the number in each cell, we calculate the number of edges between the node-layers $(i,\ell)$ with the group assignments $g_{(i,\ell)}^{1}$ and $g_{(i,\ell)}^{2}$ in the row and the node-layers $(j,\ell)$ with the group assignments $g_{(j,\ell)}^{1}$ and $g_{(j,\ell)}^{2}$ in the column. We then divide this number by the number of possible edges between such pairs of node-layers with those group assignments. If there are no possible edges, we write ``N/A'' in the cell.}
\label{densitytablesSaxena}
\end{table*}

The densities of the edges between node-layers $(i,\ell)$ with group assignments $g_{(i,\ell)}^{1} = 1$ and $g_{(i,\ell)}^{2} = 1$ and node-layers $(j,\ell)$ with group assignments $g_{(j,\ell)}^{1} = 1$ and $g_{(j,\ell)}^{2} = 1$ are $1.000$ for both layer $2$ (see Table \ref{densitytable2Saxena}) and layer $3$ (see Table \ref{densitytable3Saxena}). By contrast, the densities of the edges between node-layers $(i,\ell)$ with group assignments $g_{(i,\ell)}^{1} = 1$ and $g_{(i,\ell)}^{2} = 1$ and node-layers $(j,\ell)$ with group assignments $g_{(j,\ell)}^{1} = 0$ and $g_{(j,\ell)}^{2} = 0$ are $0.088$ for layer $2$ and $0.032$ for layer $3$. 
Accordingly, node-layers $(i,\ell)$ with group assignments $g_{(i,\ell)}^{1} = 1$ and $g_{(i,\ell)}^{2} = 1$ tend to be connected densely to node-layers $(j,\ell)$ with group assignments $g_{(j,\ell)}^{1} = 1$ and $g_{(j,\ell)}^{2} = 1$ but connected sparsely to node-layers 
$(j,\ell)$ with group assignments $g_{(j,\ell)}^{1} = 0$ and $g_{(j,\ell)}^{2} = 0$. 
Additionally, when looking at Table \ref{densitytablesSaxena} as a whole, we also observe that each layer has rather heterogeneous edge densities.


\subsection{A Literary Co-Appearance Network}\label{literary}

We now apply our hierarchical core--periphery detection method to a network of co-appearances of individuals in the \emph{Gospel of Luke}~\cite{Holanda19}. This temporal network consists of $n = 76$ nodes (which represent characters) and $L = 5$ layers (which represent non-overlapping ranges of $4$ ordered chapters of the \emph{Gospel of Luke} \footnote{Specifically, layer 1 encodes information from chapters 1--4; layer 2 encodes information from chapters 5--8; layer 3 encodes information from chapters 9--12; layer 4 encodes information from chapters 13, 17, 18, and 19; and layer 5 encodes information from chapters 20--24. Holanda et al.~\cite{Holanda19} did not explain why their data set does not include information from chapters 14--16.}). We place an edge between two nodes in a layer if the two associated characters encounter each other in the associated chapters. 
Prior investigations of similar networks have calculated large global clustering coefficients and observed that there are individuals with large degree, closeness, and betweenness centralities~\cite{Massey19, Dorpinghaus24}, so it seems reasonable to examine core--periphery structure in such networks.

\begin{figure*}[h]
\centering
\includegraphics[width=\textwidth]{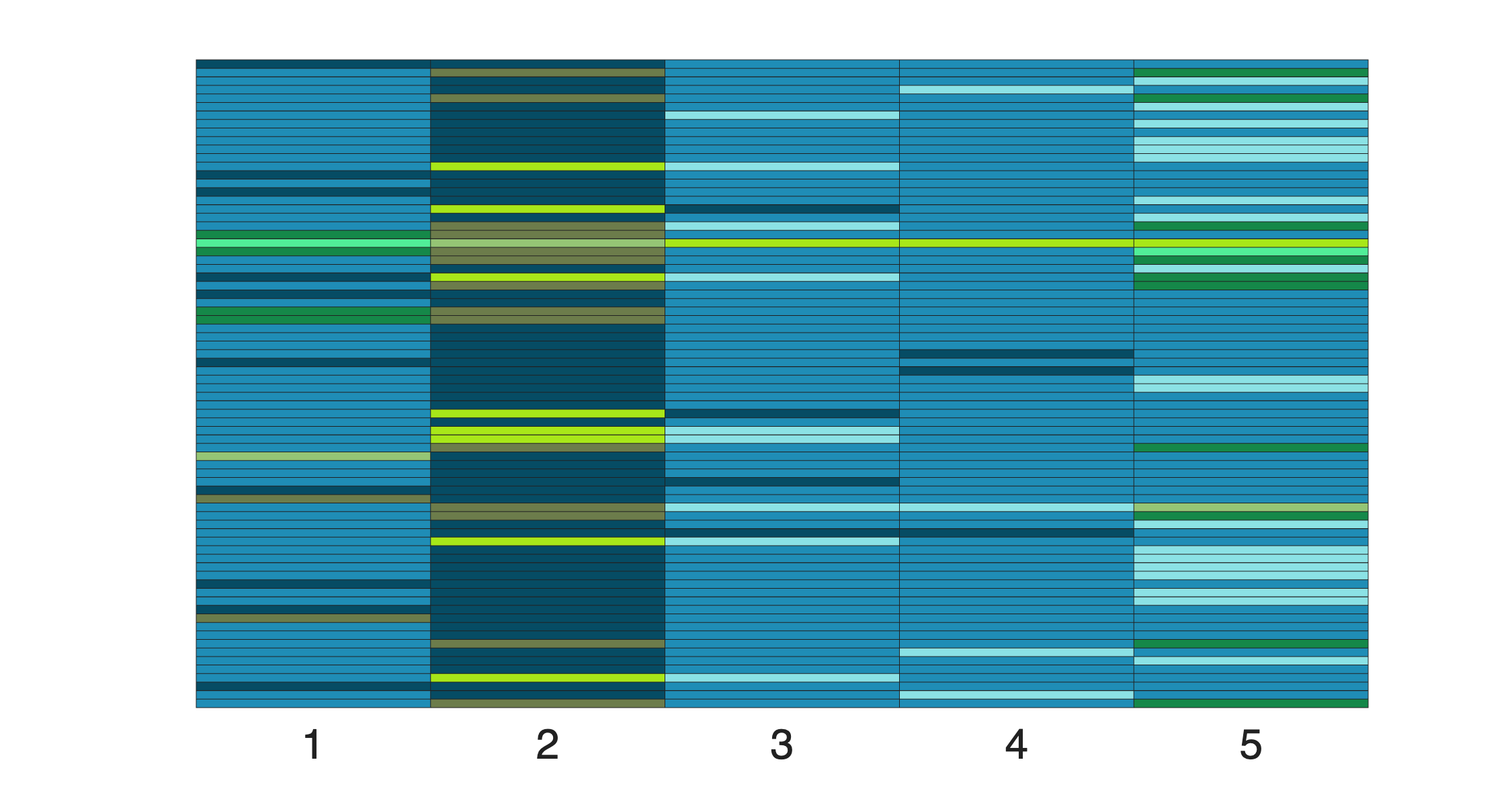}
\caption{The inferred core--periphery structure from one run of Algorithm \ref{FinalAlg} for the \emph{Gospel of Luke} co-appearance network. In Table \ref{ColorPaletteLuke}, we indicate the groups that correspond to each rectangle color. The horizontal axis indicates layers and the vertical axis indicates characters.}
\label{LukeCP}
\end{figure*}

Applying our hierarchical core--periphery detection method to the \emph{Gospel of Luke} network yields $k = 4$ groups and the core--periphery structure in Figure \ref{LukeCP}.
To obtain this core--periphery structure, we perform $5$ runs of $10^6$ steps of our main MCMC algorithm (see Algorithm \ref{FinalAlg}) with $k = 4$ groups initially, a multi-mode move probability of $p = 10^{-3}$, and uniformly random initial group assignments.
We use distinct initial group assignments for each run, and we save the output group assignments every $10^4$ steps of our MCMC algorithm.
As in the Jammu--Kashmir terrorist network, we set the group assignment of each node-layer $(i,\ell)$ to be the most frequent group assignment across the 5 MCMC runs.

\begin{table*}
\begin{tabular}{c|c}
Group assignments & Rectangle color in Figure \ref{LukeCP}
\\ \hline
$g_{(i,\ell)}^{1} = 1$ & \\ 
$g_{(i,\ell)}^{2} = 1$ & \colorbox[RGB]{169, 232, 26}{\vphantom{$g_{(i,\ell)}^{1} = 1$}\hspace{0.5in}} \\
$g_{(i,\ell)}^{3} = 1$ & \\ \hline
$g_{(i,\ell)}^{1} = 0$ & \\ 
$g_{(i,\ell)}^{2} = 1$ & \colorbox[RGB]{109, 125, 76}{\vphantom{$g_{(i,\ell)}^{1} = 1$}\hspace{0.5in}} \\
$g_{(i,\ell)}^{3} = 1$ & \\ \hline
$g_{(i,\ell)}^{1} = 1$ & \\ 
$g_{(i,\ell)}^{2} = 0$ & \colorbox[RGB]{150, 198, 118}{\vphantom{$g_{(i,\ell)}^{1} = 1$}\hspace{0.5in}} \\
$g_{(i,\ell)}^{3} = 1$ & \\ \hline
$g_{(i,\ell)}^{1} = 0$ & \\ 
$g_{(i,\ell)}^{2} = 0$ & \colorbox[RGB]{22, 137, 74}{\vphantom{$g_{(i,\ell)}^{1} = 1$}\hspace{0.5in}} \\
$g_{(i,\ell)}^{3} = 1$ & \\ \hline
$g_{(i,\ell)}^{1} = 1$ & \\ 
$g_{(i,\ell)}^{2} = 1$ & \colorbox[RGB]{82, 239, 153}{\vphantom{$g_{(i,\ell)}^{1} = 1$}\hspace{0.5in}} \\
$g_{(i,\ell)}^{3} = 0$ & \\ \hline
$g_{(i,\ell)}^{1} = 0$ & \\ 
$g_{(i,\ell)}^{2} = 1$ & \colorbox[RGB]{7, 77, 101}{\vphantom{$g_{(i,\ell)}^{1} = 1$}\hspace{0.5in}} \\
$g_{(i,\ell)}^{3} = 0$ & \\ \hline
$g_{(i,\ell)}^{1} = 1$ & \\ 
$g_{(i,\ell)}^{2} = 0$ & \colorbox[RGB]{140, 227, 230}{\vphantom{$g_{(i,\ell)}^{1} = 1$}\hspace{0.5in}} \\
$g_{(i,\ell)}^{3} = 0$ & \\ \hline
$g_{(i,\ell)}^{1} = 0$ & \\ 
$g_{(i,\ell)}^{2} = 0$ & \colorbox[RGB]{32, 142, 183}{\vphantom{$g_{(i,\ell)}^{1} = 1$}\hspace{0.5in}} \\
$g_{(i,\ell)}^{3} = 0$ & \\ \hline
\end{tabular}
\caption{The colors that we use in Figure \ref{LukeCP} to signify different group assignments $g_{(i,\ell)}^{r}$ for groups $r \in \{1,2,3\}$.}
\label{ColorPaletteLuke}
\end{table*}

\begin{figure*}[h]
\centering
\subfloat[]{\includegraphics[width=0.49\textwidth]{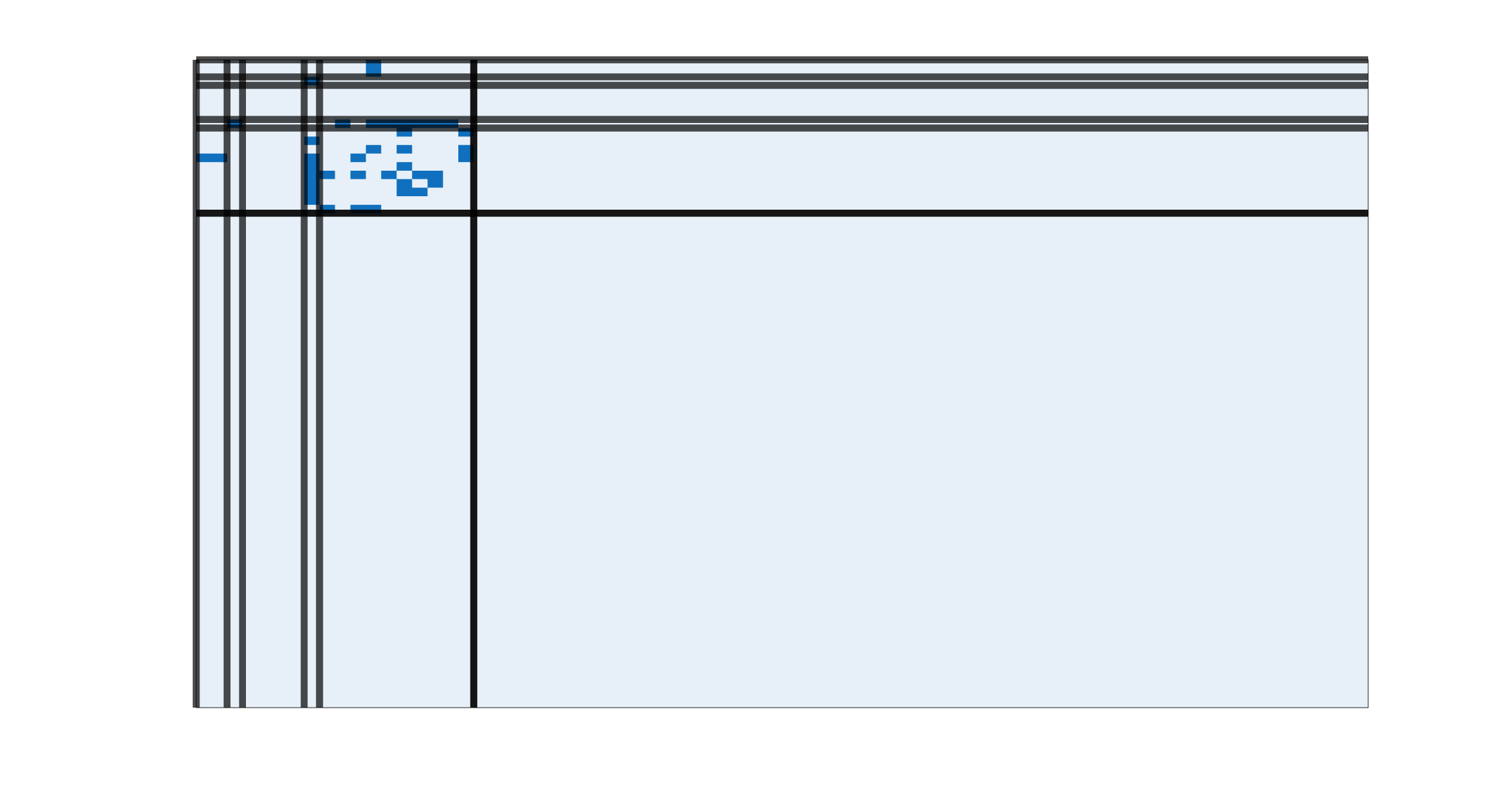}} \;
\subfloat[]{\includegraphics[width=0.49\textwidth]{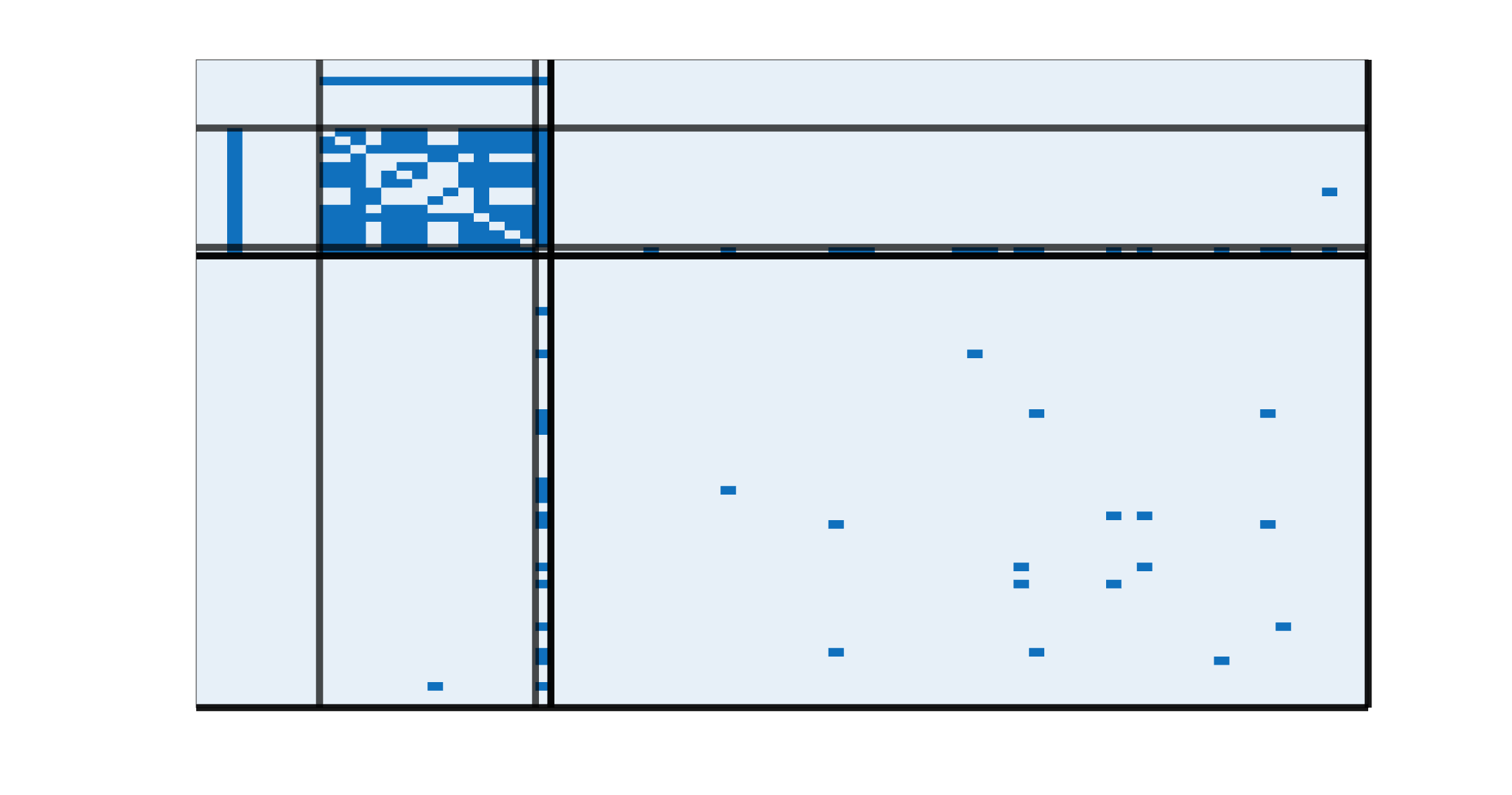}}\\
\subfloat[]{\includegraphics[width=0.49\textwidth]{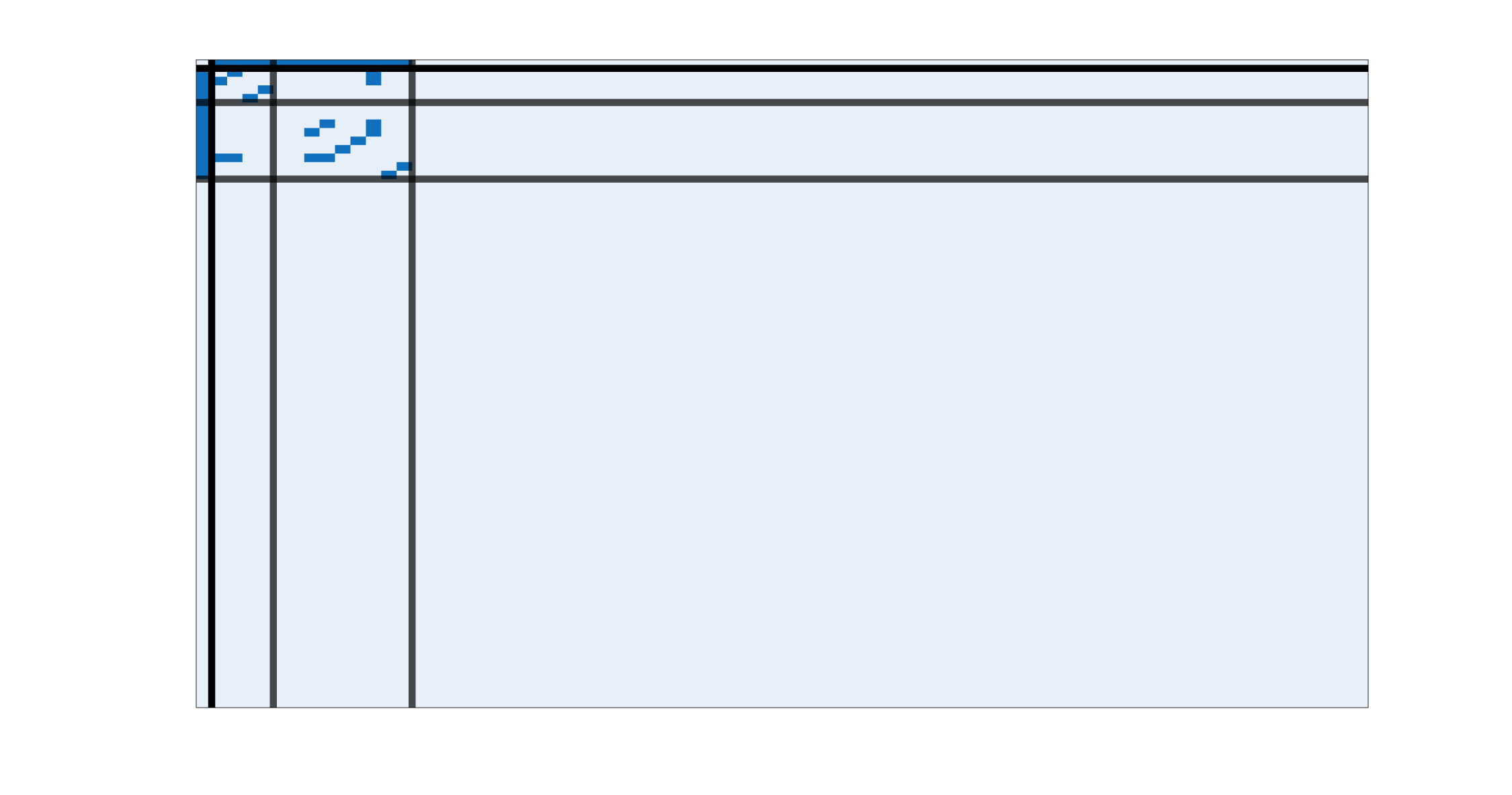}} \;
\subfloat[]{\includegraphics[width=0.49\textwidth]{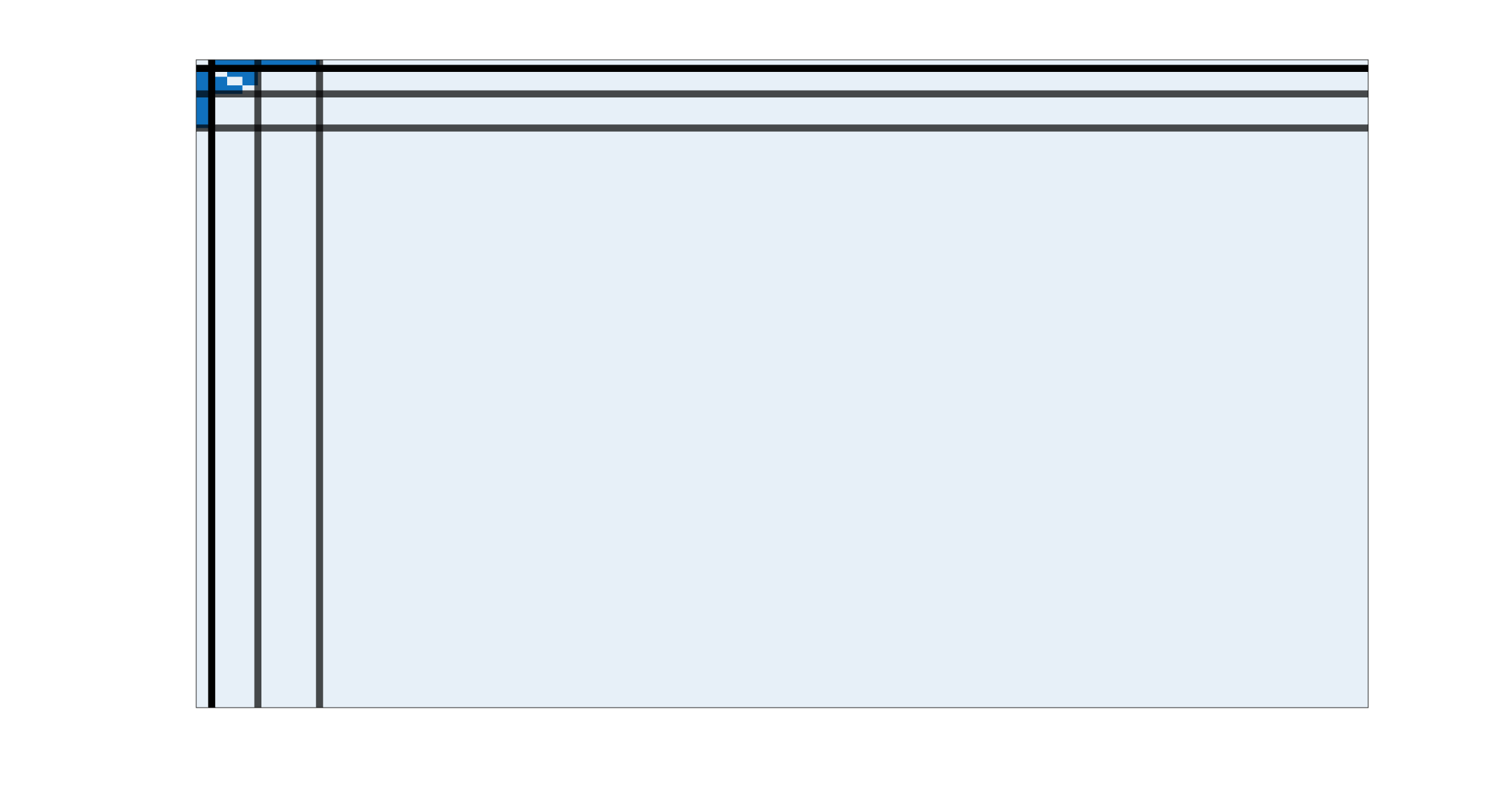}} \\
\subfloat[]{\includegraphics[width=0.49\textwidth]{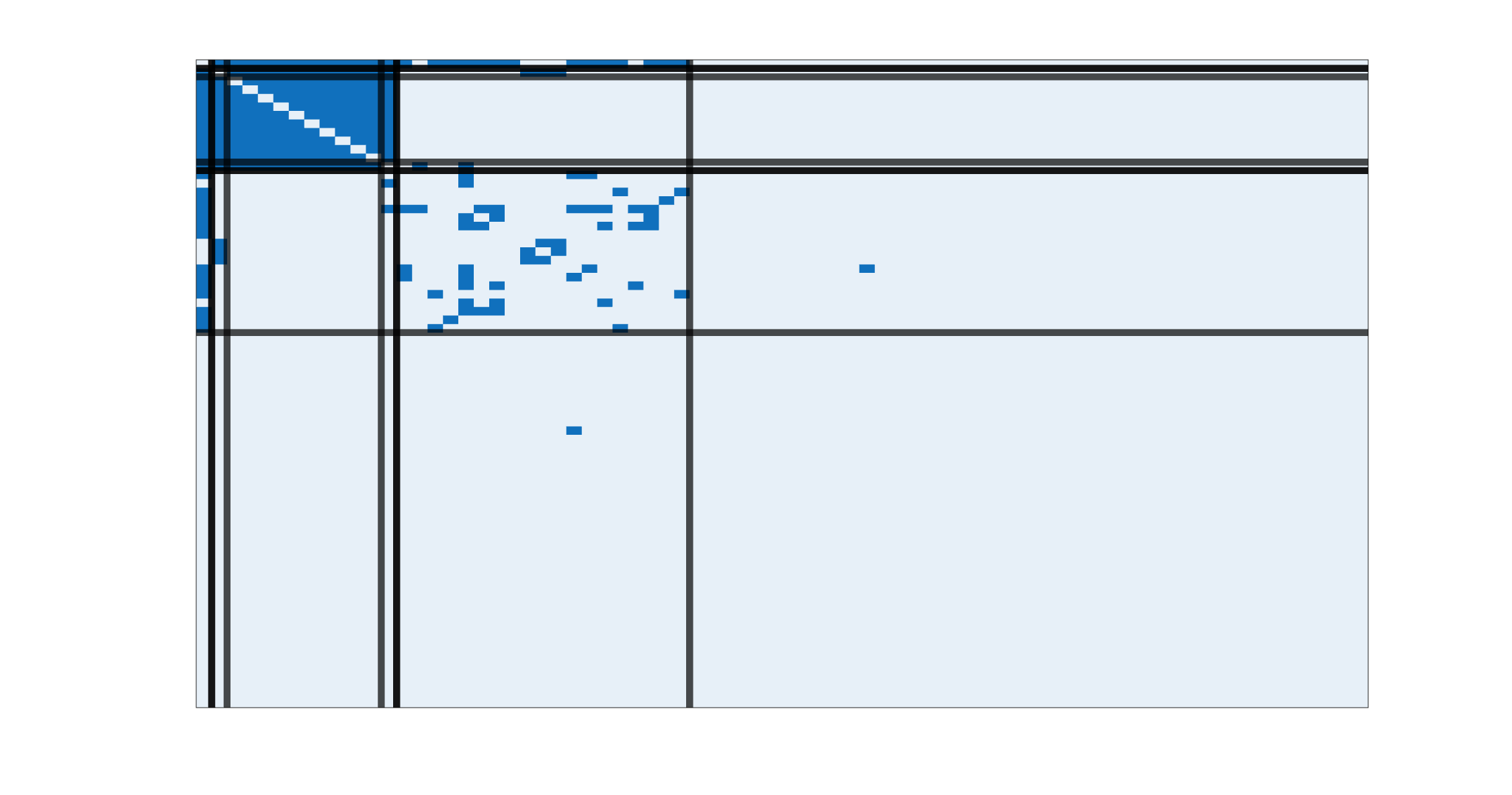}}
\caption{Permuted adjacency matrices to illustrate our inferred core--periphery structure (see Figure \ref{LukeCP}) of the \emph{Gospel of Luke} literary co-appearance network. We show (a) layer 1, (b) layer 2, (c) layer 3, (d) layer 4, and (e) layer 5.}
\label{LukeAdjacency}
\end{figure*}

To illustrate that the detected core--periphery structure in Figure \ref{LukeCP} is reasonable, we plot heat maps (see Figure \ref{LukeAdjacency}) of the adjacency matrices $A^{(\ell)}$ of each layer $\ell$. In these heat maps, we permute the rows and columns to arrange node-layers by their group assignments in the order in Table \ref{ColorPaletteLuke}. 
Node-layers $(i,\ell)$ with $(g_{(i,\ell)}^{1}, g_{(i,\ell)}^{2}, g_{(i,\ell)}^{3}) = (1,1,1)$ appear first, node-layers $(i,\ell)$ with $(g_{(i,\ell)}^{1}, g_{(i,\ell)}^{2}, g_{(i,\ell)}^{3}) = (0,1,1)$ appear next, and so on. 

As in the Jammu--Kashmir terrorist network, the densities of the edges between node-layers depend on the groups of the node-layers. For example, in layer $3$, node-layers $(i,\ell)$ with group assignments $(g_{(i,\ell)}^{1}, g_{(i,\ell)}^{2}, g_{(i,\ell)}^{3}) = (0,1,0)$ tend to be connected densely to node-layers $(j,\ell)$ with group assignments $(g_{(j,\ell)}^{1}, g_{(j,\ell)}^{2}, g_{(j,\ell)}^{3}) = (0,1,0)$ or group assignments $(g_{(j,\ell)}^{1}, g_{(j,\ell)}^{2}, g_{(j,\ell)}^{3}) = (1,1,1)$ but connected sparsely to node-layers $(j,\ell)$ with other group assignments. 
To further illustrate these qualitative observations, we compute the mean edge densities of all pairs of group assignments for each layer. In Table \ref{densitytablesLuke}, we show our results for layer $3$. Analogous results for all layers are available in our repository of code and data \cite{faust-code}.
Analogously to our observation in the Jammu--Kashmir terrorist network (see Section \ref{terror}), the heterogeneous edge densities indicate that the inferred core--periphery structure distinguishes between characters with different network positions \cite{Wasserman94}.

\begin{table*}
\begin{tabular}{p{1.5cm}|p{1.5cm}|p{1.5cm}|p{1.5cm}|p{1.5cm}|p{1.5cm}|p{1.5cm}|p{1.5cm}|p{1.5cm}}
 & $g_{(j,\ell)}^{1} = 0$\newline$g_{(j,\ell)}^{2} = 0$\newline$g_{(j,\ell)}^{3} = 0$ & $g_{(j,\ell)}^{1} = 1$\newline$g_{(j,\ell)}^{2} = 0$\newline$g_{(j,\ell)}^{3} = 0$ & $g_{(j,\ell)}^{1} = 0$\newline$g_{(j,\ell)}^{2} = 1$\newline$g_{(j,\ell)}^{3} = 0$ & $g_{(j,\ell)}^{1} = 1$\newline$g_{(i,\ell)}^{2} = 1$\newline$g_{(j,\ell)}^{3} = 0$ & $g_{(j,\ell)}^{1} = 0$\newline$g_{(j,\ell)}^{2} = 0$\newline$g_{(j,\ell)}^{3} = 1$ & $g_{(j,\ell)}^{1} = 1$\newline$g_{(j,\ell)}^{2} = 0$\newline$g_{(j,\ell)}^{3} = 1$ & $g_{(j,\ell)}^{1} = 0$\newline$g_{(j,\ell)}^{2} = 1$\newline$g_{(j,\ell)}^{3} = 1$ & $g_{(j,\ell)}^{1} = 1$\newline$g_{(i,\ell)}^{2} = 1$\newline$g_{(j,\ell)}^{3} = 1$ \\ \hline
$g_{(i,\ell)}^{1} = 0$\newline$g_{(i,\ell)}^{2} = 0$\newline$g_{(i,\ell)}^{3} = 0$ & 0.000 & 0.000 & 0.000 & N/A & N/A & N/A & N/A & 0.000 \\ \hline
$g_{(i,\ell)}^{1} = 1$\newline$g_{(i,\ell)}^{2} = 0$\newline$g_{(i,\ell)}^{3} = 0$ & 0.000 & 0.139 & 0.056 & N/A & N/A & N/A & N/A & 1.000 \\ \hline
$g_{(i,\ell)}^{1} = 0$\newline$g_{(i,\ell)}^{2} = 1$\newline$g_{(i,\ell)}^{3} = 0$ & 0.000 & 0.056 & 0.333 & N/A & N/A & N/A & N/A & 1.000 \\ \hline
$g_{(i,\ell)}^{1} = 1$\newline$g_{(i,\ell)}^{2} = 1$\newline$g_{(i,\ell)}^{3} = 0$ & N/A & N/A & N/A & N/A & N/A & N/A & N/A & N/A  \\ \hline
$g_{(i,\ell)}^{1} = 0$\newline$g_{(i,\ell)}^{2} = 0$\newline$g_{(i,\ell)}^{3} = 1$ & N/A & N/A & N/A & N/A & N/A & N/A & N/A & N/A \\ \hline
$g_{(i,\ell)}^{1} = 1$\newline$g_{(i,\ell)}^{2} = 0$\newline$g_{(i,\ell)}^{3} = 1$ & N/A & N/A & N/A & N/A & N/A & N/A & N/A & N/A \\ \hline
$g_{(i,\ell)}^{1} = 0$\newline$g_{(i,\ell)}^{2} = 1$\newline$g_{(i,\ell)}^{3} = 1$ & N/A & N/A & N/A & N/A & N/A & N/A & N/A & N/A \\ \hline
$g_{(i,\ell)}^{1} = 1$\newline$g_{(i,\ell)}^{2} = 1$\newline$g_{(i,\ell)}^{3} = 1$ & 0.000 & 1.000 & 1.000 & N/A & N/A & N/A & N/A & N/A \\ 
\end{tabular}
\caption{The densities of the edges between node-layers in layer $\ell = 3$ for each of the groups in Figure \ref{LukeAdjacency}.
 To compute the number in each cell, we calculate the number of edges between node-layers $(i,\ell)$ with group assignments $(g_{(i,\ell)}^{1}, g_{(i,\ell)}^{2}, g_{(i,\ell)}^{3})$ in the row and node-layers $(j,\ell)$ with group assignments $(g_{(j,\ell)}^{1}, g_{(j,\ell)}^{2}, g_{(j,\ell)}^{3})$ in the column.
We then divide this number by the number of possible edges between such pairs of node-layers with those group assignments.
If there are no possible edges, we write ``N/A'' in the cell. The mean edge densities in all layers $\ell$ are available in our repository of code and data \cite{faust-code}.}
\label{densitytablesLuke}
\end{table*}

As we can see in Table \ref{densitytablesLuke}, in layer $\ell = 3$, node-layers $(i,\ell)$ with group assignments $(g_{(i,\ell)}^{1}, g_{(i,\ell)}^{2}, g_{(i,\ell)}^{3}) = (0,1,0)$ tend to be connected densely to node-layers $(j,\ell)$ with group assignments  $(g_{(j,\ell)}^{1}, g_{(j,\ell)}^{2}, g_{(j,\ell)}^{3}) = (0,1,0)$ or group assignments $(g_{(j,\ell)}^{1}, g_{(j,\ell)}^{2}, g_{(j,\ell)}^{3}) = (1,1,1)$ but connected sparsely to node-layers $(j,\ell)$ with other group assignments. 
In Table \ref{densitytablesLuke}, we also observe that layer $\ell = 3$ has rather heterogeneous edge densities.
We also observe heterogeneous edge densities in the other layers $\ell$. We include density tables for all layers in our repository of code and data \cite{faust-code}.


\section{Conclusions and Discussion} \label{FinalConclusionsCP}

There are many different types of mesoscale structures, and it is important to develop methods to study such structures in temporal networks. In the present paper, we proposed and examined a method to identify hierarchical core--periphery structure in temporal networks. We also applied our method to one synthetic temporal network and two real-world temporal networks, and we discussed the structures that we identified in these networks. 

Our method uses statistical inference to identify hierarchical core--periphery structure. By using the group-evolution probability distribution from \cite{FaustThesis,faust2026} (instead of a group-evolution probability distribution that is based on a discrete-time Markov process), the Markov-chain Monte Carlo (MCMC) approach that we employed for such statistical inference is able to mitigate the underestimation of the number of groups in networks in the detection of hierarchical core--periphery structure. Additionally, by using multi-node moves, we accelerated the convergence to inferred core--periphery structure. 

There are a variety of ways to build on our work. The primary weakness of our approach is the large computational cost of our MCMC computations. Like many other similar inference-based approaches (e.g., \cite{Polanco23}), our method is quadratic in the number of nodes of a network, so it is prohibitive to use it for networks with many nodes.
It is important to develop and implement methods that can identify hierarchical core--periphery structure in temporal networks (and, more generally, in multilayer networks) that are subquadratic in the number of nodes.
Finally, we emphasize that there have been very few studies of core--periphery structure in temporal and multilayer networks --- a situation that contrasts with the voluminous analysis of temporal and multilayer community structure \cite{Rossetti18, Huang21} --- and such efforts deserve more attention. We expect that such studies will yield both theoretical and practical insights.


\section*{Acknowledgements}

We thank Arash Amini, Andrea Barreiro, Grace Li, Robert Meyers, Mark Newman, Austin Polanco, and two anonymous referees for helpful comments. TYF was supported by the National Science Foundation Graduate Research Fellowship Program (under Grant No.~DGE-2034835). 

\section*{Statement on AI Usage}

We used Claude Opus 5 and Claude Opus 5.5, respectively, to make graphical tweaks to our original hand-produced versions of Figures \ref{SyntheticSeeded}--\ref{SyntheticCombinedStructure} and Figure \ref{SaxenaCP}. We carefully checked the data in the figures before and after the changes to ensure their correctness.


\section*{Data Availability}

The code to produce our simulations, the adjacency matrices of the Jammu--Kashmir terrorist network and the \emph{Luke Gospel} literary co-appearance network, the output group assignments for each simulation run, and the mean edge densities are available at \cite{faust-code}.



\begin{thebibliography}{60}%
\makeatletter
\providecommand \@ifxundefined [1]{%
 \@ifx{#1\undefined}
}%
\providecommand \@ifnum [1]{%
 \ifnum #1\expandafter \@firstoftwo
 \else \expandafter \@secondoftwo
 \fi
}%
\providecommand \@ifx [1]{%
 \ifx #1\expandafter \@firstoftwo
 \else \expandafter \@secondoftwo
 \fi
}%
\providecommand \natexlab [1]{#1}%
\providecommand \enquote  [1]{``#1''}%
\providecommand \bibnamefont  [1]{#1}%
\providecommand \bibfnamefont [1]{#1}%
\providecommand \citenamefont [1]{#1}%
\providecommand \href@noop [0]{\@secondoftwo}%
\providecommand \href [0]{\begingroup \@sanitize@url \@href}%
\providecommand \@href[1]{\@@startlink{#1}\@@href}%
\providecommand \@@href[1]{\endgroup#1\@@endlink}%
\providecommand \@sanitize@url [0]{\catcode `\\12\catcode `\$12\catcode
  `\&12\catcode `\#12\catcode `\^12\catcode `\_12\catcode `\%12\relax}%
\providecommand \@@startlink[1]{}%
\providecommand \@@endlink[0]{}%
\providecommand \url  [0]{\begingroup\@sanitize@url \@url }%
\providecommand \@url [1]{\endgroup\@href {#1}{\urlprefix }}%
\providecommand \urlprefix  [0]{URL }%
\providecommand \Eprint [0]{\href }%
\providecommand \doibase [0]{https://doi.org/}%
\providecommand \selectlanguage [0]{\@gobble}%
\providecommand \bibinfo  [0]{\@secondoftwo}%
\providecommand \bibfield  [0]{\@secondoftwo}%
\providecommand \translation [1]{[#1]}%
\providecommand \BibitemOpen [0]{}%
\providecommand \bibitemStop [0]{}%
\providecommand \bibitemNoStop [0]{.\EOS\space}%
\providecommand \EOS [0]{\spacefactor3000\relax}%
\providecommand \BibitemShut  [1]{\csname bibitem#1\endcsname}%
\let\auto@bib@innerbib\@empty
\bibitem [{\citenamefont {Polanco}\ and\ \citenamefont
  {Newman}(2023)}]{Polanco23}%
  \BibitemOpen
  \bibfield  {author} {\bibinfo {author} {\bibfnamefont {A.}~\bibnamefont
  {Polanco}}\ and\ \bibinfo {author} {\bibfnamefont {M.~E.~J.}\ \bibnamefont
  {Newman}},\ }\bibfield  {title} {\bibinfo {title} {Hierarchical
  core--periphery structure in networks},\ }\href@noop {} {\bibfield  {journal}
  {\bibinfo  {journal} {Physical Review E}\ }\textbf {\bibinfo {volume}
  {108}},\ \bibinfo {pages} {024311} (\bibinfo {year} {2023})}\BibitemShut
  {NoStop}%
\bibitem [{\citenamefont {Newman}(2018)}]{Newman18}%
  \BibitemOpen
  \bibfield  {author} {\bibinfo {author} {\bibfnamefont {M.}~\bibnamefont
  {Newman}},\ }\href@noop {} {\emph {\bibinfo {title} {{Networks}}}},\ \bibinfo
  {edition} {{2nd}}\ ed.\ (\bibinfo  {publisher} {Oxford University Press},\
  \bibinfo {address} {Oxford, UK},\ \bibinfo {year} {2018})\BibitemShut
  {NoStop}%
\bibitem [{\citenamefont {Fortunato}\ and\ \citenamefont
  {Hric}(2016)}]{Fortunato16}%
  \BibitemOpen
  \bibfield  {author} {\bibinfo {author} {\bibfnamefont {S.}~\bibnamefont
  {Fortunato}}\ and\ \bibinfo {author} {\bibfnamefont {D.}~\bibnamefont
  {Hric}},\ }\bibfield  {title} {\bibinfo {title} {Community detection in
  networks: {A} user guide},\ }\href@noop {} {\bibfield  {journal} {\bibinfo
  {journal} {Physics Reports}\ }\textbf {\bibinfo {volume} {659}},\ \bibinfo
  {pages} {1} (\bibinfo {year} {2016})}\BibitemShut {NoStop}%
\bibitem [{\citenamefont {Rossi}\ and\ \citenamefont
  {Ahmed}(2015)}]{rossi2015}%
  \BibitemOpen
  \bibfield  {author} {\bibinfo {author} {\bibfnamefont {R.~A.}\ \bibnamefont
  {Rossi}}\ and\ \bibinfo {author} {\bibfnamefont {N.~K.}\ \bibnamefont
  {Ahmed}},\ }\bibfield  {title} {\bibinfo {title} {Role discovery in
  networks},\ }\href@noop {} {\bibfield  {journal} {\bibinfo  {journal} {IEEE
  Transactions on Knowledge and Data Engineering}\ }\textbf {\bibinfo {volume}
  {27}},\ \bibinfo {pages} {1112} (\bibinfo {year} {2015})}\BibitemShut
  {NoStop}%
\bibitem [{\citenamefont {Csermely}\ \emph {et~al.}(2013)\citenamefont
  {Csermely}, \citenamefont {London}, \citenamefont {Wu},\ and\ \citenamefont
  {Uzzi}}]{csermely2013}%
  \BibitemOpen
  \bibfield  {author} {\bibinfo {author} {\bibfnamefont {P.}~\bibnamefont
  {Csermely}}, \bibinfo {author} {\bibfnamefont {A.}~\bibnamefont {London}},
  \bibinfo {author} {\bibfnamefont {L.-Y.}\ \bibnamefont {Wu}},\ and\ \bibinfo
  {author} {\bibfnamefont {B.}~\bibnamefont {Uzzi}},\ }\bibfield  {title}
  {\bibinfo {title} {Structure and dynamics of core/periphery networks},\
  }\href@noop {} {\bibfield  {journal} {\bibinfo  {journal} {Journal of Complex
  Networks}\ }\textbf {\bibinfo {volume} {1}},\ \bibinfo {pages} {93} (\bibinfo
  {year} {2013})}\BibitemShut {NoStop}%
\bibitem [{\citenamefont {Rombach}\ \emph {et~al.}(2017)\citenamefont
  {Rombach}, \citenamefont {Porter}, \citenamefont {Fowler},\ and\
  \citenamefont {Mucha}}]{Rombach17}%
  \BibitemOpen
  \bibfield  {author} {\bibinfo {author} {\bibfnamefont {P.}~\bibnamefont
  {Rombach}}, \bibinfo {author} {\bibfnamefont {M.}~\bibnamefont {Porter}},
  \bibinfo {author} {\bibfnamefont {J.}~\bibnamefont {Fowler}},\ and\ \bibinfo
  {author} {\bibfnamefont {P.}~\bibnamefont {Mucha}},\ }\bibfield  {title}
  {\bibinfo {title} {Core-periphery structure in networks (revisited)},\
  }\href@noop {} {\bibfield  {journal} {\bibinfo  {journal} {SIAM Review}\
  }\textbf {\bibinfo {volume} {59}},\ \bibinfo {pages} {619} (\bibinfo {year}
  {2017})}\BibitemShut {NoStop}%
\bibitem [{\citenamefont {Yanchenko}\ and\ \citenamefont
  {Sengupta}(2023)}]{yan2022}%
  \BibitemOpen
  \bibfield  {author} {\bibinfo {author} {\bibfnamefont {E.}~\bibnamefont
  {Yanchenko}}\ and\ \bibinfo {author} {\bibfnamefont {S.}~\bibnamefont
  {Sengupta}},\ }\bibfield  {title} {\bibinfo {title} {Core-periphery structure
  in networks: {A} statistical exposition},\ }\href@noop {} {\bibfield
  {journal} {\bibinfo  {journal} {Statistics Surveys}\ }\textbf {\bibinfo
  {volume} {17}},\ \bibinfo {pages} {42} (\bibinfo {year} {2023})}\BibitemShut
  {NoStop}%
\bibitem [{\citenamefont {Granovetter}(1983)}]{Granovetter83}%
  \BibitemOpen
  \bibfield  {author} {\bibinfo {author} {\bibfnamefont {M.}~\bibnamefont
  {Granovetter}},\ }\bibfield  {title} {\bibinfo {title} {The strength of weak
  ties: A network theory revisited},\ }\href@noop {} {\bibfield  {journal}
  {\bibinfo  {journal} {Sociological Theory}\ }\textbf {\bibinfo {volume}
  {1}},\ \bibinfo {pages} {201} (\bibinfo {year} {1983})}\BibitemShut {NoStop}%
\bibitem [{\citenamefont {White}\ \emph {et~al.}(1976)\citenamefont {White},
  \citenamefont {Boorman},\ and\ \citenamefont {Breiger}}]{White76}%
  \BibitemOpen
  \bibfield  {author} {\bibinfo {author} {\bibfnamefont {H.~C.}\ \bibnamefont
  {White}}, \bibinfo {author} {\bibfnamefont {S.~A.}\ \bibnamefont {Boorman}},\
  and\ \bibinfo {author} {\bibfnamefont {R.~L.}\ \bibnamefont {Breiger}},\
  }\bibfield  {title} {\bibinfo {title} {Social structure from multiple
  networks. {I. B}lockmodels of roles and positions},\ }\href@noop {}
  {\bibfield  {journal} {\bibinfo  {journal} {American Journal of Sociology}\
  }\textbf {\bibinfo {volume} {81}},\ \bibinfo {pages} {730} (\bibinfo {year}
  {1976})}\BibitemShut {NoStop}%
\bibitem [{\citenamefont {Doreian}(1985)}]{Doreian85}%
  \BibitemOpen
  \bibfield  {author} {\bibinfo {author} {\bibfnamefont {P.}~\bibnamefont
  {Doreian}},\ }\bibfield  {title} {\bibinfo {title} {Structural equivalence in
  a psychology journal network},\ }\href@noop {} {\bibfield  {journal}
  {\bibinfo  {journal} {Journal of the American Society for Information
  Science}\ }\textbf {\bibinfo {volume} {36}},\ \bibinfo {pages} {411}
  (\bibinfo {year} {1985})}\BibitemShut {NoStop}%
\bibitem [{\citenamefont {Willis}\ and\ \citenamefont
  {McNamee}(1990)}]{Willis90}%
  \BibitemOpen
  \bibfield  {author} {\bibinfo {author} {\bibfnamefont {C.~L.}\ \bibnamefont
  {Willis}}\ and\ \bibinfo {author} {\bibfnamefont {S.~J.}\ \bibnamefont
  {McNamee}},\ }\bibfield  {title} {\bibinfo {title} {Social networks of
  science and patterns of publication in leading sociology journals, 1960 to
  1985},\ }\href@noop {} {\bibfield  {journal} {\bibinfo  {journal}
  {Knowledge}\ }\textbf {\bibinfo {volume} {11}},\ \bibinfo {pages} {363}
  (\bibinfo {year} {1990})}\BibitemShut {NoStop}%
\bibitem [{\citenamefont {Roy}(1983)}]{Roy83}%
  \BibitemOpen
  \bibfield  {author} {\bibinfo {author} {\bibfnamefont {W.~G.}\ \bibnamefont
  {Roy}},\ }\bibfield  {title} {\bibinfo {title} {The unfolding of the
  interlocking directorate structure of the {United States}},\ }\href@noop {}
  {\bibfield  {journal} {\bibinfo  {journal} {American Sociological Review}\
  }\textbf {\bibinfo {volume} {48}},\ \bibinfo {pages} {248} (\bibinfo {year}
  {1983})}\BibitemShut {NoStop}%
\bibitem [{\citenamefont {Williams}(1977)}]{Williams77}%
  \BibitemOpen
  \bibfield  {author} {\bibinfo {author} {\bibfnamefont {S.~W.}\ \bibnamefont
  {Williams}},\ }\bibfield  {title} {\bibinfo {title} {Internal colonialism,
  core-periphery contrasts and devolution: An integrative comment},\
  }\href@noop {} {\bibfield  {journal} {\bibinfo  {journal} {Area}\ }\textbf
  {\bibinfo {volume} {9}},\ \bibinfo {pages} {272} (\bibinfo {year}
  {1977})}\BibitemShut {NoStop}%
\bibitem [{\citenamefont {Lee}\ \emph {et~al.}(2014)\citenamefont {Lee},
  \citenamefont {Cucuringu},\ and\ \citenamefont {Porter}}]{lee2014}%
  \BibitemOpen
  \bibfield  {author} {\bibinfo {author} {\bibfnamefont {S.~H.}\ \bibnamefont
  {Lee}}, \bibinfo {author} {\bibfnamefont {M.}~\bibnamefont {Cucuringu}},\
  and\ \bibinfo {author} {\bibfnamefont {M.~A.}\ \bibnamefont {Porter}},\
  }\bibfield  {title} {\bibinfo {title} {Density-based and transport-based
  core-periphery structures in networks},\ }\href@noop {} {\bibfield  {journal}
  {\bibinfo  {journal} {Physical Review E}\ }\textbf {\bibinfo {volume} {89}},\
  \bibinfo {pages} {032810} (\bibinfo {year} {2014})}\BibitemShut {NoStop}%
\bibitem [{\citenamefont {Fournet}\ and\ \citenamefont
  {Barrat}(2014)}]{Fournet14}%
  \BibitemOpen
  \bibfield  {author} {\bibinfo {author} {\bibfnamefont {J.}~\bibnamefont
  {Fournet}}\ and\ \bibinfo {author} {\bibfnamefont {A.}~\bibnamefont
  {Barrat}},\ }\bibfield  {title} {\bibinfo {title} {Contact patterns among
  high school students},\ }\href@noop {} {\bibfield  {journal} {\bibinfo
  {journal} {PLoS ONE}\ }\textbf {\bibinfo {volume} {9}},\ \bibinfo {eid}
  {e107878} (\bibinfo {year} {2014})}\BibitemShut {NoStop}%
\bibitem [{\citenamefont {Morer}\ \emph {et~al.}(2020)\citenamefont {Morer},
  \citenamefont {Cardillo}, \citenamefont {D\'{\i}az-Guilera}, \citenamefont
  {Prignano},\ and\ \citenamefont {Lozano}}]{Morer20}%
  \BibitemOpen
  \bibfield  {author} {\bibinfo {author} {\bibfnamefont {I.}~\bibnamefont
  {Morer}}, \bibinfo {author} {\bibfnamefont {A.}~\bibnamefont {Cardillo}},
  \bibinfo {author} {\bibfnamefont {A.}~\bibnamefont {D\'{\i}az-Guilera}},
  \bibinfo {author} {\bibfnamefont {L.}~\bibnamefont {Prignano}},\ and\
  \bibinfo {author} {\bibfnamefont {S.}~\bibnamefont {Lozano}},\ }\bibfield
  {title} {\bibinfo {title} {Comparing spatial networks: {A} one-size-fits-all
  efficiency-driven approach},\ }\href@noop {} {\bibfield  {journal} {\bibinfo
  {journal} {Physical Review E}\ }\textbf {\bibinfo {volume} {101}},\ \bibinfo
  {pages} {042301} (\bibinfo {year} {2020})}\BibitemShut {NoStop}%
\bibitem [{\citenamefont {Lee}\ \emph {et~al.}(2016)\citenamefont {Lee},
  \citenamefont {Magallanes},\ and\ \citenamefont {Porter}}]{lee2016}%
  \BibitemOpen
  \bibfield  {author} {\bibinfo {author} {\bibfnamefont {S.~H.}\ \bibnamefont
  {Lee}}, \bibinfo {author} {\bibfnamefont {J.~M.}\ \bibnamefont
  {Magallanes}},\ and\ \bibinfo {author} {\bibfnamefont {M.~A.}\ \bibnamefont
  {Porter}},\ }\bibfield  {title} {\bibinfo {title} {Time-dependent community
  structure in legislation cosponsorship networks in the {Congress} of the
  {Republic of Peru}},\ }\href@noop {} {\bibfield  {journal} {\bibinfo
  {journal} {Journal of Complex Networks}\ }\textbf {\bibinfo {volume} {5}},\
  \bibinfo {pages} {127} (\bibinfo {year} {2016})}\BibitemShut {NoStop}%
\bibitem [{\citenamefont {Neal}(2020)}]{Neal20}%
  \BibitemOpen
  \bibfield  {author} {\bibinfo {author} {\bibfnamefont {Z.~P.}\ \bibnamefont
  {Neal}},\ }\bibfield  {title} {\bibinfo {title} {A sign of the times? {Weak}
  and strong polarization in the {U.S. Congress}, 1973--2016},\ }\href@noop {}
  {\bibfield  {journal} {\bibinfo  {journal} {Social Networks}\ }\textbf
  {\bibinfo {volume} {60}},\ \bibinfo {pages} {103} (\bibinfo {year}
  {2020})}\BibitemShut {NoStop}%
\bibitem [{\citenamefont {Holme}\ and\ \citenamefont
  {Saram{\"a}ki}(2012)}]{Holme12}%
  \BibitemOpen
  \bibfield  {author} {\bibinfo {author} {\bibfnamefont {P.}~\bibnamefont
  {Holme}}\ and\ \bibinfo {author} {\bibfnamefont {J.}~\bibnamefont
  {Saram{\"a}ki}},\ }\bibfield  {title} {\bibinfo {title} {Temporal networks},\
  }\href@noop {} {\bibfield  {journal} {\bibinfo  {journal} {Physics Reports}\
  }\textbf {\bibinfo {volume} {519}},\ \bibinfo {pages} {97} (\bibinfo {year}
  {2012})}\BibitemShut {NoStop}%
\bibitem [{\citenamefont {Holme}(2015)}]{Holme15}%
  \BibitemOpen
  \bibfield  {author} {\bibinfo {author} {\bibfnamefont {P.}~\bibnamefont
  {Holme}},\ }\bibfield  {title} {\bibinfo {title} {Modern temporal network
  theory: {A} colloquium},\ }\href@noop {} {\bibfield  {journal} {\bibinfo
  {journal} {The European Physical Journal B}\ }\textbf {\bibinfo {volume}
  {88}},\ \bibinfo {pages} {234} (\bibinfo {year} {2015})}\BibitemShut
  {NoStop}%
\bibitem [{\citenamefont {Holme}\ and\ \citenamefont
  {Saram{\"a}ki}(2023)}]{Holme19}%
  \BibitemOpen
  \bibfield  {author} {\bibinfo {author} {\bibfnamefont {P.}~\bibnamefont
  {Holme}}\ and\ \bibinfo {author} {\bibfnamefont {J.}~\bibnamefont
  {Saram{\"a}ki}},\ }\href@noop {} {\emph {\bibinfo {title} {Temporal Network
  Theory}}},\ \bibinfo {edition} {2nd}\ ed.\ (\bibinfo  {publisher}
  {Springer},\ \bibinfo {address} {Cham, Switzerland},\ \bibinfo {year}
  {2023})\BibitemShut {NoStop}%
\bibitem [{\citenamefont {Palla}\ \emph {et~al.}(2007)\citenamefont {Palla},
  \citenamefont {Barab{\'a}si},\ and\ \citenamefont {Vicsek}}]{Palla07}%
  \BibitemOpen
  \bibfield  {author} {\bibinfo {author} {\bibfnamefont {G.}~\bibnamefont
  {Palla}}, \bibinfo {author} {\bibfnamefont {A.-L.}\ \bibnamefont
  {Barab{\'a}si}},\ and\ \bibinfo {author} {\bibfnamefont {T.}~\bibnamefont
  {Vicsek}},\ }\bibfield  {title} {\bibinfo {title} {Quantifying social group
  evolution},\ }\href@noop {} {\bibfield  {journal} {\bibinfo  {journal}
  {Nature}\ }\textbf {\bibinfo {volume} {446}},\ \bibinfo {pages} {664}
  (\bibinfo {year} {2007})}\BibitemShut {NoStop}%
\bibitem [{\citenamefont {Peixoto}\ and\ \citenamefont
  {Rosvall}(2017)}]{Peixoto17}%
  \BibitemOpen
  \bibfield  {author} {\bibinfo {author} {\bibfnamefont {T.~P.}\ \bibnamefont
  {Peixoto}}\ and\ \bibinfo {author} {\bibfnamefont {M.}~\bibnamefont
  {Rosvall}},\ }\bibfield  {title} {\bibinfo {title} {Modelling sequences and
  temporal networks with dynamic community structures},\ }\href@noop {}
  {\bibfield  {journal} {\bibinfo  {journal} {Nature Communications}\ }\textbf
  {\bibinfo {volume} {8}},\ \bibinfo {pages} {582} (\bibinfo {year}
  {2017})}\BibitemShut {NoStop}%
\bibitem [{\citenamefont {Gauvin}\ \emph {et~al.}(2014)\citenamefont {Gauvin},
  \citenamefont {Panisson},\ and\ \citenamefont {Cattuto}}]{Gauvin14}%
  \BibitemOpen
  \bibfield  {author} {\bibinfo {author} {\bibfnamefont {L.}~\bibnamefont
  {Gauvin}}, \bibinfo {author} {\bibfnamefont {A.}~\bibnamefont {Panisson}},\
  and\ \bibinfo {author} {\bibfnamefont {C.}~\bibnamefont {Cattuto}},\
  }\bibfield  {title} {\bibinfo {title} {Detecting the community structure and
  activity patterns of temporal networks: A non-negative tensor factorization
  approach},\ }\href@noop {} {\bibfield  {journal} {\bibinfo  {journal} {PLoS
  ONE}\ }\textbf {\bibinfo {volume} {9}},\ \bibinfo {pages} {e86028} (\bibinfo
  {year} {2014})}\BibitemShut {NoStop}%
\bibitem [{\citenamefont {Mucha}\ \emph {et~al.}(2010)\citenamefont {Mucha},
  \citenamefont {Richardson}, \citenamefont {Macon}, \citenamefont {Porter},\
  and\ \citenamefont {Onnela}}]{Mucha10}%
  \BibitemOpen
  \bibfield  {author} {\bibinfo {author} {\bibfnamefont {P.~J.}\ \bibnamefont
  {Mucha}}, \bibinfo {author} {\bibfnamefont {T.}~\bibnamefont {Richardson}},
  \bibinfo {author} {\bibfnamefont {K.}~\bibnamefont {Macon}}, \bibinfo
  {author} {\bibfnamefont {M.~A.}\ \bibnamefont {Porter}},\ and\ \bibinfo
  {author} {\bibfnamefont {J.-P.}\ \bibnamefont {Onnela}},\ }\bibfield  {title}
  {\bibinfo {title} {Community structure in time-dependent, multiscale, and
  multiplex networks},\ }\href@noop {} {\bibfield  {journal} {\bibinfo
  {journal} {Science}\ }\textbf {\bibinfo {volume} {328}},\ \bibinfo {pages}
  {876} (\bibinfo {year} {2010})}\BibitemShut {NoStop}%
\bibitem [{\citenamefont {Rossetti}\ and\ \citenamefont
  {Cazabet}(2018)}]{Rossetti18}%
  \BibitemOpen
  \bibfield  {author} {\bibinfo {author} {\bibfnamefont {G.}~\bibnamefont
  {Rossetti}}\ and\ \bibinfo {author} {\bibfnamefont {R.}~\bibnamefont
  {Cazabet}},\ }\bibfield  {title} {\bibinfo {title} {Community discovery in
  dynamic networks: {A} survey},\ }\href@noop {} {\bibfield  {journal}
  {\bibinfo  {journal} {ACM Computing Surveys}\ }\textbf {\bibinfo {volume}
  {51}},\ \bibinfo {pages} {35} (\bibinfo {year} {2018})}\BibitemShut {NoStop}%
\bibitem [{\citenamefont {Kivel{\"a}}\ \emph {et~al.}(2014)\citenamefont
  {Kivel{\"a}}, \citenamefont {Arenas}, \citenamefont {Barthelemy},
  \citenamefont {Gleeson}, \citenamefont {Moreno},\ and\ \citenamefont
  {Porter}}]{kivela2014}%
  \BibitemOpen
  \bibfield  {author} {\bibinfo {author} {\bibfnamefont {M.}~\bibnamefont
  {Kivel{\"a}}}, \bibinfo {author} {\bibfnamefont {A.}~\bibnamefont {Arenas}},
  \bibinfo {author} {\bibfnamefont {M.}~\bibnamefont {Barthelemy}}, \bibinfo
  {author} {\bibfnamefont {J.~P.}\ \bibnamefont {Gleeson}}, \bibinfo {author}
  {\bibfnamefont {Y.}~\bibnamefont {Moreno}},\ and\ \bibinfo {author}
  {\bibfnamefont {M.~A.}\ \bibnamefont {Porter}},\ }\bibfield  {title}
  {\bibinfo {title} {{Multilayer Networks}},\ }\href@noop {} {\bibfield
  {journal} {\bibinfo  {journal} {{Journal of Complex Networks}}\ }\textbf
  {\bibinfo {volume} {2}},\ \bibinfo {pages} {203} (\bibinfo {year}
  {2014})}\BibitemShut {NoStop}%
\bibitem [{\citenamefont {Aleta}\ \emph {et~al.}(2026)\citenamefont {Aleta},
  \citenamefont {Teixeira}, \citenamefont {de~Arruda}, \citenamefont
  {Baronchelli}, \citenamefont {Barrat}, \citenamefont {Kert\'{e}sz},
  \citenamefont {D\'{i}az-Guilera}, \citenamefont {Artime}, \citenamefont
  {Starnini}, \citenamefont {Petri}, \citenamefont {Karsai}, \citenamefont
  {Patwardhan}, \citenamefont {Vespignani}, \citenamefont {Moreno},\ and\
  \citenamefont {Fortunato}}]{aleta2026}%
  \BibitemOpen
  \bibfield  {author} {\bibinfo {author} {\bibfnamefont {A.}~\bibnamefont
  {Aleta}}, \bibinfo {author} {\bibfnamefont {A.~S.}\ \bibnamefont {Teixeira}},
  \bibinfo {author} {\bibfnamefont {G.~F.}\ \bibnamefont {de~Arruda}}, \bibinfo
  {author} {\bibfnamefont {A.}~\bibnamefont {Baronchelli}}, \bibinfo {author}
  {\bibfnamefont {A.}~\bibnamefont {Barrat}}, \bibinfo {author} {\bibfnamefont
  {J.}~\bibnamefont {Kert\'{e}sz}}, \bibinfo {author} {\bibfnamefont
  {A.}~\bibnamefont {D\'{i}az-Guilera}}, \bibinfo {author} {\bibfnamefont
  {O.}~\bibnamefont {Artime}}, \bibinfo {author} {\bibfnamefont
  {M.}~\bibnamefont {Starnini}}, \bibinfo {author} {\bibfnamefont
  {G.}~\bibnamefont {Petri}}, \bibinfo {author} {\bibfnamefont
  {M.}~\bibnamefont {Karsai}}, \bibinfo {author} {\bibfnamefont
  {S.}~\bibnamefont {Patwardhan}}, \bibinfo {author} {\bibfnamefont
  {A.}~\bibnamefont {Vespignani}}, \bibinfo {author} {\bibfnamefont
  {Y.}~\bibnamefont {Moreno}},\ and\ \bibinfo {author} {\bibfnamefont
  {S.}~\bibnamefont {Fortunato}},\ }\bibfield  {title} {\bibinfo {title}
  {Multilayer network science: {T}heory, methods, and applications},\
  }\href@noop {} {\bibfield  {journal} {\bibinfo  {journal} {Journal of Complex
  Networks}\ }\textbf {\bibinfo {volume} {14}},\ \bibinfo {pages} {cnag007}
  (\bibinfo {year} {2026})}\BibitemShut {NoStop}%
\bibitem [{\citenamefont {Zhang}\ \emph {et~al.}(2015)\citenamefont {Zhang},
  \citenamefont {Martin},\ and\ \citenamefont {Newman}}]{Zhang15}%
  \BibitemOpen
  \bibfield  {author} {\bibinfo {author} {\bibfnamefont {X.}~\bibnamefont
  {Zhang}}, \bibinfo {author} {\bibfnamefont {T.}~\bibnamefont {Martin}},\ and\
  \bibinfo {author} {\bibfnamefont {M.~E.~J.}\ \bibnamefont {Newman}},\
  }\bibfield  {title} {\bibinfo {title} {Identification of core-periphery
  structure in networks},\ }\href@noop {} {\bibfield  {journal} {\bibinfo
  {journal} {Physical Review E}\ }\textbf {\bibinfo {volume} {91}},\ \bibinfo
  {pages} {032803} (\bibinfo {year} {2015})}\BibitemShut {NoStop}%
\bibitem [{\citenamefont {Gallagher}\ \emph {et~al.}(2021)\citenamefont
  {Gallagher}, \citenamefont {Young},\ and\ \citenamefont
  {Welles}}]{Gallagher21}%
  \BibitemOpen
  \bibfield  {author} {\bibinfo {author} {\bibfnamefont {R.~J.}\ \bibnamefont
  {Gallagher}}, \bibinfo {author} {\bibfnamefont {J.-G.}\ \bibnamefont
  {Young}},\ and\ \bibinfo {author} {\bibfnamefont {B.~F.}\ \bibnamefont
  {Welles}},\ }\bibfield  {title} {\bibinfo {title} {A clarified typology of
  core-periphery structure in networks},\ }\href@noop {} {\bibfield  {journal}
  {\bibinfo  {journal} {Science Advances}\ }\textbf {\bibinfo {volume} {7}},\
  \bibinfo {pages} {eabc9800} (\bibinfo {year} {2021})}\BibitemShut {NoStop}%
\bibitem [{\citenamefont {Bergermann}\ \emph {et~al.}(2024)\citenamefont
  {Bergermann}, \citenamefont {Stoll},\ and\ \citenamefont
  {Tudisco}}]{Bergermann1024}%
  \BibitemOpen
  \bibfield  {author} {\bibinfo {author} {\bibfnamefont {K.}~\bibnamefont
  {Bergermann}}, \bibinfo {author} {\bibfnamefont {M.}~\bibnamefont {Stoll}},\
  and\ \bibinfo {author} {\bibfnamefont {F.}~\bibnamefont {Tudisco}},\
  }\bibfield  {title} {\bibinfo {title} {A nonlinear spectral core--periphery
  detection method for multiplex networks},\ }\href@noop {} {\bibfield
  {journal} {\bibinfo  {journal} {Proceedings of the Royal Society A:
  Mathematical, Physical and Engineering Sciences}\ }\textbf {\bibinfo {volume}
  {480}},\ \bibinfo {pages} {20230914} (\bibinfo {year} {2024})}\BibitemShut
  {NoStop}%
\bibitem [{\citenamefont {Bergermann}\ and\ \citenamefont
  {Tudisco}(2025)}]{Bergermann24}%
  \BibitemOpen
  \bibfield  {author} {\bibinfo {author} {\bibfnamefont {K.}~\bibnamefont
  {Bergermann}}\ and\ \bibinfo {author} {\bibfnamefont {F.}~\bibnamefont
  {Tudisco}},\ }\bibfield  {title} {\bibinfo {title} {Core-periphery detection
  in multilayer networks},\ }\href@noop {} {\bibfield  {journal} {\bibinfo
  {journal} {Physical Review Letters}\ }\textbf {\bibinfo {volume} {135}},\
  \bibinfo {pages} {087402} (\bibinfo {year} {2025})}\BibitemShut {NoStop}%
\bibitem [{\citenamefont {Nie}\ \emph {et~al.}(2025)\citenamefont {Nie},
  \citenamefont {Xuan}, \citenamefont {Gao},\ and\ \citenamefont
  {Ruan}}]{Nie25}%
  \BibitemOpen
  \bibfield  {author} {\bibinfo {author} {\bibfnamefont {J.}~\bibnamefont
  {Nie}}, \bibinfo {author} {\bibfnamefont {Q.}~\bibnamefont {Xuan}}, \bibinfo
  {author} {\bibfnamefont {D.}~\bibnamefont {Gao}},\ and\ \bibinfo {author}
  {\bibfnamefont {Z.}~\bibnamefont {Ruan}},\ }\bibfield  {title} {\bibinfo
  {title} {An effective method for profiling core--periphery structures in
  complex networks},\ }\href@noop {} {\bibfield  {journal} {\bibinfo  {journal}
  {Physica A: Statistical Mechanics and its Applications}\ }\textbf {\bibinfo
  {volume} {669}},\ \bibinfo {pages} {130618} (\bibinfo {year}
  {2025})}\BibitemShut {NoStop}%
\bibitem [{\citenamefont {Hashemi}\ and\ \citenamefont
  {Behrouz}(2024)}]{Hashemi24}%
  \BibitemOpen
  \bibfield  {author} {\bibinfo {author} {\bibfnamefont {F.}~\bibnamefont
  {Hashemi}}\ and\ \bibinfo {author} {\bibfnamefont {A.}~\bibnamefont
  {Behrouz}},\ }\bibfield  {title} {\bibinfo {title} {A unified core structure
  in multiplex networks: From finding the densest subgraph to modeling user
  engagement},\ }in\ \href@noop {} {\emph {\bibinfo {booktitle} {Proceedings of
  the 30th ACM SIGKDD Conference on Knowledge Discovery and Data Mining}}},\
  \bibinfo {series and number} {KDD '24}\ (\bibinfo  {publisher} {Association
  for Computing Machinery},\ \bibinfo {address} {New York, NY, USA},\ \bibinfo
  {year} {2024})\ pp.\ \bibinfo {pages} {1028--1039}\BibitemShut {NoStop}%
\bibitem [{\citenamefont {Mosca}\ and\ \citenamefont
  {Mazzarisi}(2026)}]{mosca2026}%
  \BibitemOpen
  \bibfield  {author} {\bibinfo {author} {\bibfnamefont {A.}~\bibnamefont
  {Mosca}}\ and\ \bibinfo {author} {\bibfnamefont {P.}~\bibnamefont
  {Mazzarisi}},\ }\href@noop {} {\bibinfo {title} {Penalized maximum likelihood
  inference of core-periphery networks}} (\bibinfo {year} {2026}),\ \bibinfo
  {note} {https://arxiv.org/abs/2609.15796}\BibitemShut {NoStop}%
\bibitem [{Note1()}]{Note1}%
  \BibitemOpen
  \bibinfo {note} {For illustrations of the types of structures that one can
  generate using this hierarchical notion of core--periphery structure, see
  Figure 1 of \cite {Polanco23}.}\BibitemShut {Stop}%
\bibitem [{\citenamefont {Peixoto}(2023)}]{peixoto2023}%
  \BibitemOpen
  \bibfield  {author} {\bibinfo {author} {\bibfnamefont {T.~P.}\ \bibnamefont
  {Peixoto}},\ }\href@noop {} {\emph {\bibinfo {title} {Descriptive vs.
  Inferential Community Detection in Networks: Pitfalls, Myths and
  Half-Truths}}},\ Elements in the Structure and Dynamics of Complex Networks\
  (\bibinfo  {publisher} {Cambridge University Press},\ \bibinfo {address}
  {Cambridge, UK},\ \bibinfo {year} {2023})\BibitemShut {NoStop}%
\bibitem [{\citenamefont {Saxena}\ \emph {et~al.}(2004)\citenamefont {Saxena},
  \citenamefont {Santhanam},\ and\ \citenamefont {Basu}}]{Saxena04}%
  \BibitemOpen
  \bibfield  {author} {\bibinfo {author} {\bibfnamefont {S.}~\bibnamefont
  {Saxena}}, \bibinfo {author} {\bibfnamefont {K.}~\bibnamefont {Santhanam}},\
  and\ \bibinfo {author} {\bibfnamefont {A.}~\bibnamefont {Basu}},\ }\bibfield
  {title} {\bibinfo {title} {Application of {Social Network Analysis} ({SNA})
  to terrorist networks in {Jammu} \& {Kashmir}},\ }\href@noop {} {\bibfield
  {journal} {\bibinfo  {journal} {Strategic Analysis}\ }\textbf {\bibinfo
  {volume} {28}},\ \bibinfo {pages} {84} (\bibinfo {year} {2004})}\BibitemShut
  {NoStop}%
\bibitem [{\citenamefont {Holanda}\ \emph {et~al.}(2019)\citenamefont
  {Holanda}, \citenamefont {Matias}, \citenamefont {Ferreira}, \citenamefont
  {Benevides},\ and\ \citenamefont {Kinouchi}}]{Holanda19}%
  \BibitemOpen
  \bibfield  {author} {\bibinfo {author} {\bibfnamefont {A.~J.}\ \bibnamefont
  {Holanda}}, \bibinfo {author} {\bibfnamefont {M.}~\bibnamefont {Matias}},
  \bibinfo {author} {\bibfnamefont {S.~M. S.~P.}\ \bibnamefont {Ferreira}},
  \bibinfo {author} {\bibfnamefont {G.~M.~L.}\ \bibnamefont {Benevides}},\ and\
  \bibinfo {author} {\bibfnamefont {O.}~\bibnamefont {Kinouchi}},\ }\bibfield
  {title} {\bibinfo {title} {Character networks and book genre
  classification},\ }\href@noop {} {\bibfield  {journal} {\bibinfo  {journal}
  {International Journal of Modern Physics C}\ }\textbf {\bibinfo {volume}
  {30}},\ \bibinfo {pages} {1950058} (\bibinfo {year} {2019})}\BibitemShut
  {NoStop}%
\bibitem [{\citenamefont {Faust}(2026)}]{faust-code}%
  \BibitemOpen
  \bibfield  {author} {\bibinfo {author} {\bibfnamefont {T.~Y.}\ \bibnamefont
  {Faust}},\ }\href@noop {} {\bibinfo {title} {Repository for ``{I}nference of
  hierarchical core--periphery structure in temporal networks"}} (\bibinfo
  {year} {2026}),\ \bibinfo {note}
  {\url{https://github.com/tfaust0196/mhCorePeriphery}}\BibitemShut {NoStop}%
\bibitem [{\citenamefont {Bauz{\'a}~Mingueza}\ \emph
  {et~al.}(2023)\citenamefont {Bauz{\'a}~Mingueza}, \citenamefont
  {Flor{\'\i}a}, \citenamefont {G{\'o}mez-Garde{\~n}es}, \citenamefont
  {Arenas},\ and\ \citenamefont {Cardillo}}]{Mingueza23}%
  \BibitemOpen
  \bibfield  {author} {\bibinfo {author} {\bibfnamefont {F.}~\bibnamefont
  {Bauz{\'a}~Mingueza}}, \bibinfo {author} {\bibfnamefont {M.}~\bibnamefont
  {Flor{\'\i}a}}, \bibinfo {author} {\bibfnamefont {J.}~\bibnamefont
  {G{\'o}mez-Garde{\~n}es}}, \bibinfo {author} {\bibfnamefont {A.}~\bibnamefont
  {Arenas}},\ and\ \bibinfo {author} {\bibfnamefont {A.}~\bibnamefont
  {Cardillo}},\ }\bibfield  {title} {\bibinfo {title} {Characterization of
  interactions' persistence in time-varying networks},\ }\href@noop {}
  {\bibfield  {journal} {\bibinfo  {journal} {Scientific Reports}\ }\textbf
  {\bibinfo {volume} {13}},\ \bibinfo {pages} {765} (\bibinfo {year}
  {2023})}\BibitemShut {NoStop}%
\bibitem [{\citenamefont {Faust}(2025)}]{FaustThesis}%
  \BibitemOpen
  \bibfield  {author} {\bibinfo {author} {\bibfnamefont {T.~Y.}\ \bibnamefont
  {Faust}},\ }\emph {\bibinfo {title} {Inference and Size Localization of
  Mesoscale Structures in Temporal Networks}},\ \href@noop {} {Ph.D. thesis},\
  \bibinfo  {school} {University of California, Los Angeles}, \bibinfo
  {address} {https://escholarship.org/uc/item/8q68n9v4} (\bibinfo {year}
  {2025})\BibitemShut {NoStop}%
\bibitem [{\citenamefont {Faust}\ \emph {et~al.}(2026)\citenamefont {Faust},
  \citenamefont {Amini},\ and\ \citenamefont {Porter}}]{faust2026}%
  \BibitemOpen
  \bibfield  {author} {\bibinfo {author} {\bibfnamefont {T.~Y.}\ \bibnamefont
  {Faust}}, \bibinfo {author} {\bibfnamefont {A.~A.}\ \bibnamefont {Amini}},\
  and\ \bibinfo {author} {\bibfnamefont {M.~A.}\ \bibnamefont {Porter}},\
  }\href@noop {} {\bibinfo {title} {Community-size biases in statistical
  inference in temporal networks}} (\bibinfo {year} {2026}),\ \bibinfo {note}
  {https://arxiv.org/abs/2601.15635}\BibitemShut {NoStop}%
\bibitem [{Note2()}]{Note2}%
  \BibitemOpen
  \bibinfo {note} {For a complete discussion of the approach that we use to
  generate $g^r_{(\ell )}$ from $g^r_{(\ell - 1)}$, see \cite [Section 4.2.3.3
  and Appendix B.2]{FaustThesis} and \cite [Section 3.3.3 and Appendices A.2
  and A.3]{faust2026}.}\BibitemShut {Stop}%
\bibitem [{\citenamefont {Robert}\ and\ \citenamefont
  {Casella}(2004)}]{Robert04}%
  \BibitemOpen
  \bibfield  {author} {\bibinfo {author} {\bibfnamefont {C.~P.}\ \bibnamefont
  {Robert}}\ and\ \bibinfo {author} {\bibfnamefont {G.}~\bibnamefont
  {Casella}},\ }\bibfield  {title} {\bibinfo {title} {The
  {Metropolis--Hastings} algorithm},\ }in\ \href@noop {} {\emph {\bibinfo
  {booktitle} {{Monte Carlo} Statistical Methods}}}\ (\bibinfo  {publisher}
  {Springer},\ \bibinfo {address} {Heidelberg, Germany},\ \bibinfo {year}
  {2004})\ pp.\ \bibinfo {pages} {267--320},\ \bibinfo {note} {chapter
  7}\BibitemShut {NoStop}%
\bibitem [{Note3()}]{Note3}%
  \BibitemOpen
  \bibinfo {note} {If $n$ is sufficiently large, the integral $J(k_1,k_2)$ can
  become very small, which causes finite-precision issues and thereby leads to
  inaccurate results when computing $J(k_1,k_2)$ using the procedure in \cite
  [Appendix B.2.1]{FaustThesis} and \cite [Appendix A.3]{faust2026}. To
  mitigate this problem, we use the approximation \begin {equation*} \protect
  \frac {J(k_1 + 1,k_2)}{J(k_1,k_2)} \approx 1 \end {equation*} for large $k_1$
  and $k_2$. In particular, for fixed $k_2$, we set the computed values of
  $J(k_1,k_2)$ to $\exp (-16)$ for all $k_1 \ge k_1'$, where $k_1'$ is the
  smallest $k_1'$ such that $J(k_1',k_2) < \exp (-16)$.}\BibitemShut {Stop}%
\bibitem [{\citenamefont {Yang}\ \emph {et~al.}(2011)\citenamefont {Yang},
  \citenamefont {Chi}, \citenamefont {Zhu}, \citenamefont {Gong},\ and\
  \citenamefont {Jin}}]{Yang11}%
  \BibitemOpen
  \bibfield  {author} {\bibinfo {author} {\bibfnamefont {T.}~\bibnamefont
  {Yang}}, \bibinfo {author} {\bibfnamefont {Y.}~\bibnamefont {Chi}}, \bibinfo
  {author} {\bibfnamefont {S.}~\bibnamefont {Zhu}}, \bibinfo {author}
  {\bibfnamefont {Y.}~\bibnamefont {Gong}},\ and\ \bibinfo {author}
  {\bibfnamefont {R.}~\bibnamefont {Jin}},\ }\bibfield  {title} {\bibinfo
  {title} {Detecting communities and their evolutions in dynamic social
  networks---{A} {Bayesian} approach},\ }\href@noop {} {\bibfield  {journal}
  {\bibinfo  {journal} {Machine Learning}\ }\textbf {\bibinfo {volume} {82}},\
  \bibinfo {pages} {157} (\bibinfo {year} {2011})}\BibitemShut {NoStop}%
\bibitem [{\citenamefont {Ghasemian}\ \emph {et~al.}(2016)\citenamefont
  {Ghasemian}, \citenamefont {Zhang}, \citenamefont {Clauset}, \citenamefont
  {Moore},\ and\ \citenamefont {Peel}}]{Ghasemian16}%
  \BibitemOpen
  \bibfield  {author} {\bibinfo {author} {\bibfnamefont {A.}~\bibnamefont
  {Ghasemian}}, \bibinfo {author} {\bibfnamefont {P.}~\bibnamefont {Zhang}},
  \bibinfo {author} {\bibfnamefont {A.}~\bibnamefont {Clauset}}, \bibinfo
  {author} {\bibfnamefont {C.}~\bibnamefont {Moore}},\ and\ \bibinfo {author}
  {\bibfnamefont {L.}~\bibnamefont {Peel}},\ }\bibfield  {title} {\bibinfo
  {title} {Detectability thresholds and optimal algorithms for community
  structure in dynamic networks},\ }\href@noop {} {\bibfield  {journal}
  {\bibinfo  {journal} {Physical Review X}\ }\textbf {\bibinfo {volume} {6}},\
  \bibinfo {pages} {031005} (\bibinfo {year} {2016})}\BibitemShut {NoStop}%
\bibitem [{\citenamefont {Matias}\ and\ \citenamefont
  {Miele}(2016)}]{Matias17}%
  \BibitemOpen
  \bibfield  {author} {\bibinfo {author} {\bibfnamefont {C.}~\bibnamefont
  {Matias}}\ and\ \bibinfo {author} {\bibfnamefont {V.}~\bibnamefont {Miele}},\
  }\bibfield  {title} {\bibinfo {title} {Statistical clustering of temporal
  networks through a dynamic stochastic block model},\ }\href@noop {}
  {\bibfield  {journal} {\bibinfo  {journal} {Journal of the Royal Statistical
  Society Series B: Statistical Methodology}\ }\textbf {\bibinfo {volume}
  {79}},\ \bibinfo {pages} {1119} (\bibinfo {year} {2016})}\BibitemShut
  {NoStop}%
\bibitem [{\citenamefont {Bazzi}\ \emph {et~al.}(2020)\citenamefont {Bazzi},
  \citenamefont {Jeub}, \citenamefont {Arenas}, \citenamefont {Howison},\ and\
  \citenamefont {Porter}}]{Bazzi20}%
  \BibitemOpen
  \bibfield  {author} {\bibinfo {author} {\bibfnamefont {M.}~\bibnamefont
  {Bazzi}}, \bibinfo {author} {\bibfnamefont {L.~G.~S.}\ \bibnamefont {Jeub}},
  \bibinfo {author} {\bibfnamefont {A.}~\bibnamefont {Arenas}}, \bibinfo
  {author} {\bibfnamefont {S.~D.}\ \bibnamefont {Howison}},\ and\ \bibinfo
  {author} {\bibfnamefont {M.~A.}\ \bibnamefont {Porter}},\ }\bibfield  {title}
  {\bibinfo {title} {A framework for the construction of generative models for
  mesoscale structure in multilayer networks},\ }\href@noop {} {\bibfield
  {journal} {\bibinfo  {journal} {Physical Review Research}\ }\textbf {\bibinfo
  {volume} {2}},\ \bibinfo {eid} {023100} (\bibinfo {year} {2020})}\BibitemShut
  {NoStop}%
\bibitem [{\citenamefont {Yanchenko}\ \emph {et~al.}(2025)\citenamefont
  {Yanchenko}, \citenamefont {Sengupta},\ and\ \citenamefont
  {Mukherjee}}]{yan2025}%
  \BibitemOpen
  \bibfield  {author} {\bibinfo {author} {\bibfnamefont {E.}~\bibnamefont
  {Yanchenko}}, \bibinfo {author} {\bibfnamefont {S.}~\bibnamefont
  {Sengupta}},\ and\ \bibinfo {author} {\bibfnamefont {D.}~\bibnamefont
  {Mukherjee}},\ }\href@noop {} {\bibinfo {title} {Statistical inference for
  core-periphery structures}} (\bibinfo {year} {2025}),\ \bibinfo {note}
  {https://arxiv.org/abs/2508.04730}\BibitemShut {NoStop}%
\bibitem [{\citenamefont {Asal}\ \emph {et~al.}(2016)\citenamefont {Asal},
  \citenamefont {Park}, \citenamefont {Rethemeyer},\ and\ \citenamefont
  {Ackerman}}]{Asal16}%
  \BibitemOpen
  \bibfield  {author} {\bibinfo {author} {\bibfnamefont {V.~H.}\ \bibnamefont
  {Asal}}, \bibinfo {author} {\bibfnamefont {H.~H.}\ \bibnamefont {Park}},
  \bibinfo {author} {\bibfnamefont {R.~K.}\ \bibnamefont {Rethemeyer}},\ and\
  \bibinfo {author} {\bibfnamefont {G.}~\bibnamefont {Ackerman}},\ }\bibfield
  {title} {\bibinfo {title} {With friends like these {\ldots} {Why} terrorist
  organizations ally},\ }\href@noop {} {\bibfield  {journal} {\bibinfo
  {journal} {International Public Management Journal}\ }\textbf {\bibinfo
  {volume} {19}},\ \bibinfo {pages} {1} (\bibinfo {year} {2016})}\BibitemShut
  {NoStop}%
\bibitem [{\citenamefont {Bacon}(2014)}]{Bacon14}%
  \BibitemOpen
  \bibfield  {author} {\bibinfo {author} {\bibfnamefont {T.}~\bibnamefont
  {Bacon}},\ }\bibfield  {title} {\bibinfo {title} {Alliance hubs: Focal points
  in the international terrorist landscape},\ }\href@noop {} {\bibfield
  {journal} {\bibinfo  {journal} {Perspectives on Terrorism}\ }\textbf
  {\bibinfo {volume} {8}},\ \bibinfo {pages} {4} (\bibinfo {year}
  {2014})}\BibitemShut {NoStop}%
\bibitem [{\citenamefont {Horowitz}\ and\ \citenamefont
  {Potter}(2014)}]{Horowitz14}%
  \BibitemOpen
  \bibfield  {author} {\bibinfo {author} {\bibfnamefont {M.~C.}\ \bibnamefont
  {Horowitz}}\ and\ \bibinfo {author} {\bibfnamefont {P.~B.~K.}\ \bibnamefont
  {Potter}},\ }\bibfield  {title} {\bibinfo {title} {Allying to kill: Terrorist
  intergroup cooperation and the consequences for lethality},\ }\href@noop {}
  {\bibfield  {journal} {\bibinfo  {journal} {The Journal of Conflict
  Resolution}\ }\textbf {\bibinfo {volume} {58}},\ \bibinfo {pages} {199}
  (\bibinfo {year} {2014})}\BibitemShut {NoStop}%
\bibitem [{Note4()}]{Note4}%
  \BibitemOpen
  \bibinfo {note} {In social-network analysis, ``positions}\BibitemShut
  {NoStop}%
\bibitem [{Note5()}]{Note5}%
  \BibitemOpen
  \bibinfo {note} {Specifically, layer 1 encodes information from chapters
  1--4; layer 2 encodes information from chapters 5--8; layer 3 encodes
  information from chapters 9--12; layer 4 encodes information from chapters
  13, 17, 18, and 19; and layer 5 encodes information from chapters 20--24.
  Holanda et al.~\cite {Holanda19} did not explain why their data set does not
  include information from chapters 14--16.}\BibitemShut {Stop}%
\bibitem [{\citenamefont {Massey}(2019)}]{Massey19}%
  \BibitemOpen
  \bibfield  {author} {\bibinfo {author} {\bibfnamefont {S.~E.}\ \bibnamefont
  {Massey}},\ }\bibfield  {title} {\bibinfo {title} {Form and relationship of
  the social networks of the {New Testament}},\ }\href@noop {} {\bibfield
  {journal} {\bibinfo  {journal} {Social Network Analysis and Mining}\ }\textbf
  {\bibinfo {volume} {9}},\ \bibinfo {pages} {32} (\bibinfo {year}
  {2019})}\BibitemShut {NoStop}%
\bibitem [{\citenamefont {D{\"o}rpinghaus}\ \emph {et~al.}(2024)\citenamefont
  {D{\"o}rpinghaus}, \citenamefont {Weil},\ and\ \citenamefont
  {Sommer}}]{Dorpinghaus24}%
  \BibitemOpen
  \bibfield  {author} {\bibinfo {author} {\bibfnamefont {J.}~\bibnamefont
  {D{\"o}rpinghaus}}, \bibinfo {author} {\bibfnamefont {V.}~\bibnamefont
  {Weil}},\ and\ \bibinfo {author} {\bibfnamefont {M.~W.}\ \bibnamefont
  {Sommer}},\ }\bibfield  {title} {\bibinfo {title} {Towards modeling and
  analysis of longitudinal social networks},\ }\href@noop {} {\bibfield
  {journal} {\bibinfo  {journal} {Applied Network Science}\ }\textbf {\bibinfo
  {volume} {9}},\ \bibinfo {pages} {52} (\bibinfo {year} {2024})}\BibitemShut
  {NoStop}%
\bibitem [{\citenamefont {Wasserman}\ and\ \citenamefont
  {Faust}(1994)}]{Wasserman94}%
  \BibitemOpen
  \bibfield  {author} {\bibinfo {author} {\bibfnamefont {S.}~\bibnamefont
  {Wasserman}}\ and\ \bibinfo {author} {\bibfnamefont {K.}~\bibnamefont
  {Faust}},\ }\bibinfo {title} {Network positions and roles},\ in\ \href@noop
  {} {\emph {\bibinfo {booktitle} {Social Network Analysis: Methods and
  Applications}}},\ \bibinfo {series and number} {Structural Analysis in the
  Social Sciences}\ (\bibinfo  {publisher} {Cambridge University Press},\
  \bibinfo {address} {Cambridge, UK},\ \bibinfo {year} {1994})\ Chap.~\bibinfo
  {chapter} {12}, pp.\ \bibinfo {pages} {461--502}\BibitemShut {NoStop}%
\bibitem [{\citenamefont {Huang}\ \emph {et~al.}(2021)\citenamefont {Huang},
  \citenamefont {Chen}, \citenamefont {Ren},\ and\ \citenamefont
  {Wang}}]{Huang21}%
  \BibitemOpen
  \bibfield  {author} {\bibinfo {author} {\bibfnamefont {X.}~\bibnamefont
  {Huang}}, \bibinfo {author} {\bibfnamefont {D.}~\bibnamefont {Chen}},
  \bibinfo {author} {\bibfnamefont {T.}~\bibnamefont {Ren}},\ and\ \bibinfo
  {author} {\bibfnamefont {D.}~\bibnamefont {Wang}},\ }\bibfield  {title}
  {\bibinfo {title} {A survey of community detection methods in multilayer
  networks},\ }\href@noop {} {\bibfield  {journal} {\bibinfo  {journal} {Data
  Mining and Knowledge Discovery}\ }\textbf {\bibinfo {volume} {35}},\ \bibinfo
  {pages} {1} (\bibinfo {year} {2021})}\BibitemShut {NoStop}%
\end{thebibliography}


%


\end{document}